\documentclass[onecolumn,english]{revtex4}

\pdfoutput=1

\usepackage{natbib,hyperref}

\usepackage[T1]{fontenc}
\usepackage[latin9]{inputenc}
\usepackage{color}
\usepackage{graphicx}
\usepackage{amssymb,amsmath}
\usepackage{babel}
\usepackage{float}

\usepackage[normalem]{ulem}

\def\be{\begin{equation}}
\def\ee{\end{equation}}

\makeatletter

\@ifundefined{definecolor}
{\usepackage{color}}{}

\newcommand{\beq}{\begin{equation}}
\newcommand{\eeq}{\end{equation}}  
\newcommand{\ba}{\begin{eqnarray}}
\newcommand{\ea}{\end{eqnarray}}
\newcommand{\bef}{\begin{figure}}
	\newcommand{\eef}{\end{figure}}


\begin{document}

\title{COVID-19 transmission risk factors}


	\author{Alessio Notari$^{1}$}
	\email{notari@fqa.ub.edu}
	\author{Giorgio Torrieri$^{2}$}
	\email{torrieri@ifi.unicamp.br}
	
	\affiliation{$^{1}$ Departament de F\'isica Qu\`antica i Astrofis\'ica \& Institut de Ci\`encies del Cosmos (ICCUB), Universitat de Barcelona, Mart\'i i Franqu\`es 1, 08028 Barcelona, Spain}
	\affiliation{$^{2}$ Instituto de Fisica Gleb Wataghin (IFGW), Unicamp, Campinas, SP, Brazil}

\begin{abstract}
\begin{center}
\textbf{Abstract}
\end{center}
We analyze risk factors correlated with the initial transmission growth rate of the recent COVID-19 pandemic in different countries.  The number of cases follows in its early stages an almost exponential expansion; we chose as a starting point in each country the first day $d_i$ with 30 cases and we fitted for 12 days, capturing thus the early exponential growth. We looked then for linear correlations of the exponents $\alpha$ with other variables, for a sample of 126 countries. 
We find a positive correlation, {\it i.e. faster spread of COVID-19}, with high confidence level with the following variables, with respective $p$-value: low Temperature ($4\cdot10^{-7}$), high ratio of old vs.~working-age people ($3\cdot10^{-6}$), life expectancy ($8\cdot10^{-6}$),  number of international tourists ($1\cdot10^{-5}$), earlier epidemic starting date $d_i$ ($2\cdot10^{-5}$), high level of physical contact in greeting habits ($6 \cdot 10^{-5}$), lung cancer prevalence ($6 \cdot 10^{-5}$), obesity in males ($1 \cdot 10^{-4}$), share of population in urban areas ($2\cdot10^{-4}$),  cancer prevalence ($3 \cdot 10^{-4}$),  alcohol consumption ($0.0019$),  daily smoking prevalence ($0.0036$), UV index ($0.004$, 73 countries). We also find a correlation with low Vitamin D serum levels ($0.002-0.006$), but on a smaller sample, $\sim 50$ countries, to be confirmed on a larger sample. There is highly significant correlation also with blood types: positive correlation with types RH- ($3\cdot10^{-5}$) and A+ ($3\cdot10^{-3}$), negative correlation with B+ ($2\cdot10^{-4}$).
We also find positive correlation with moderate confidence level ($p$-value of $0.02\sim0.03$) with: CO$_2$/SO emissions,  type-1 diabetes in children, low vaccination coverage for Tuberculosis (BCG). Several of the above variables are correlated with each other and so they are likely to have common interpretations.   We thus performed a Principal Component Analysis, in order to find the significant independent linear combinations of such variables. 
We also analyzed the possible existence of a bias: countries with low GDP-per capita might have less intense testing and we discuss correlation with the above variables.

\end{abstract}

\maketitle

\section{ Introduction}

The recent coronavirus (COVID-19) pandemic is now spreading essentially everywhere in our planet. The growth rate of the contagion has however  a very high variability among different countries, even in its very early stages, when government intervention is still almost negligible. Any factor contributing to a faster or slower spread needs to be identified and understood with the highest degree of scrutiny. In~\cite{Notari} the early growth rate of the contagion has been found to be correlated at high significance with temperature T.  In this work we extend a similar analysis to many other variables. This correlational study could help further investigation in order to establish causal factors and it can help policy makers in their decisions.

Some factors are intuitive and have been found in other studies, such as temperature~\cite{Notari, frenchstudy, wang,Araujo2020.03.12.20034728,Bukhari2020,Wang2020,Sajadi2020,Luo2020.02.12.20022467} (see also~\cite{brazilstudy} for a different conclusion)  and air travel~\cite{gamstudy,brazilstudy}; we aim here at being more exhaustive and at finding also factors which are not ``obvious'' and have a potential biological origin, or correlation with one. 

The paper is organized as follows. In section~\ref{methods} we explain our methods, in section~\ref{results} we show our main results, in  section~\ref{resultseach} we show the detailed results for each individual variable of our analysis, in section~\ref{cross} we discuss correlations among variables and in section~\ref{conclusions} we draw our conclusions.

\section{Method} \label{methods}
As in~\cite{Notari}, we use the empirical observation that the number of COVID-19  positive cases follows a common pattern in the majority of countries: once the number of confirmed cases reaches order 10 there is a very rapid  growth, which is typically well approximated for a few weeks by an exponential. Subsequently the exponential growth typically gradually slows down, probably due to other effects, such as: lockdown policies from governments, a higher degree of awareness in the population or the tracking and isolation of the positive cases. The growth is then typically stopped and reaches a peak in countries with a strong lockdown/tracking policy.

Our aim is to find which factors correlate  with the speed of contagion, in its  first stage of {\it free} propagation. 
For this purpose we analyzed a datasets of 126 countries taken from~\cite{ref3} on April 15th. 
We have chosen our sample using the following rules:
\begin{itemize} 
\item We start analyzing data from the first day $d_i$ in which the number of cases in a given country reaches a reference number $N_i$, which we choose to be $N_i=30$~\footnote{In practice we choose, as the first day, the one in which the number of cases $N_i$ is closest to 30. In some countries, such a number $N_i$ is repeated for several days; in such cases we choose the last of such days as the starting point. For the particular case of China, we started from January 16th, with 59 cases, since the number before that day was essentially frozen.};
\item We include only countries with at least 12 days of data, after this starting point;
\item We excluded countries with too small total population (less than 300 thousands inhabitants).
\end{itemize}
We then fit the data for each country with a simple exponential curve $N(t)=N_0 \, e^{\alpha t}$, with 2 parameters, $N_0$ and $\alpha$; here $t$ is in units of days. 


Note that the statistical errors on the exponents $\alpha$
 are typically only a few percent of the spread of the values of $\alpha$ among the various countries. For this reason we disregarded statistical errors on $\alpha$. The analysis was done using  the software {\it Mathematica}, from Wolfram Research, Inc..

\section{Main Results} \label{results}

We first look for correlations with several individual variables. Most variables are taken from~\cite{worlddata}, while for a few of them have been collected from other sources, as commented below.

\subsubsection{Non-significant variables}
We find {\it no} significant correlation of the COVID-19 transmission in our set of countries with many variables, including the following ones:
\begin{enumerate}
\item Number of inhabitants;
\item Asthma-prevalence;
\item  Participation time in leisure, social and associative life per day;
\item Population density;
\item Average precipitation per year;
\item Vaccinations coverage for: Polio, Diphteria, Tetanus, Pertussis, Hepatitis B;
\item Share of men with high-blood-pressure;
\item Diabetes prevalence (type 1 and 2, together);
\item Air pollution (``Suspended particulate matter (SPM), in micrograms per cubic metre'').
\end{enumerate}

\subsubsection{Significant variables, strong evidence}

We find {\it strong} evidence for correlation with:
\begin{enumerate}
\item Temperature (negative correlation, $p$-value $4.4\cdot10^{-7}$); 
\item Old-age dependency ratio: ratio of the number of people older than 64 relative to the number of people in the
working-age (15-64 years) (positive correlation, $p$-value $3.3\cdot10^{-6}$); 
\item Life expectancy  (positive correlation  $p$-value $8.1\cdot10^{-6}$);
\item International tourism: number of arrivals (positive correlation  $p$-value $9.6\cdot10^{-6}$);
\item Starting day $d_i$ of the epidemic (negative correlation, $p$-value $1.7\cdot 10^{-5}$);
\item Amount of contact in greeting habits (positive correlation, $p$-value $5.0\cdot 10^{-5}$);
\item Lung cancer death rates (positive correlation, $p$-value $6.3\cdot 10^{-5}$);
\item Obesity in males (positive correlation,  $p$-value $1.2\cdot10^{-4}$);
\item Share of population in urban areas (positive correlation,  $p$-value $1.7\cdot10^{-4}$);
\item Share of population with cancer (positive correlation,  $p$-value $2.8\cdot10^{-4}$);
\item Alcohol consumption (positive correlation, $p$-value $0.0019$);
\item Daily smoking prevalence (positive correlation, $p$-value $0.0036$);
\item UV index (negative correlation, $p$-value $0.004$; smaller sample, 73 countries);
\item Vitamin D serum levels (negative correlation, annual values $p$-value $0.006$, seasonal values $0.002$; smaller sample, $\sim 50$ countries).
\end{enumerate}

\subsubsection{Blood types, strong evidence}
We also find strong evidence for correlation with blood types:
\begin{enumerate}
\item  RH + blood group system (negative correlation, $p$-value $3\cdot 10^{-5}$);
\item  A+ (positive correlation, $p$-value $3\cdot 10^{-3}$);
\item B + (negative correlation, $p$-value $2\cdot 10^{-4}$);
\item A-  (positive correlation, $p$-value $3\cdot 10^{-5}$);
\item 0-  (positive correlation, $p$-value $8\cdot 10^{-4}$);
\item AB-  (positive correlation, $p$-value $0.028$).
\end{enumerate}
We find moderate evidence for correlation with:
\begin{enumerate}
\item B-  (positive correlation, $p$-value $0.013$).
\end{enumerate}
We find no significant correlation with:
\begin{enumerate}
\item  0+ ;
\item AB+. 
\end{enumerate}

\subsubsection{Significant variables, moderate evidence}
We find moderate evidence for correlation with:
\begin{enumerate}
\item  CO$_2$ (and SO) emissions  (positive correlation, $p$-value $0.015$);
\item Type-1 diabetes in children (positive correlation, $p$-value $0.023$);
\item Vaccination coverage for Tuberculosis (BCG)  (negative correlation, $p$-value $0.028$).

\end{enumerate}

\subsubsection{Significant variables, counterintuitive}
Counterintuitively we also find correlations in a direction opposite to a naive expectation:
\begin{enumerate}
\item Death-rate-from-air-pollution (negative correlation, $p$-value $3.5\cdot10^{-5}$);
\item Prevalence of anemia, adults and children, (negative correlation, $p$-value $1.4\cdot10^{-4}$ and $7.\times 10^{-6}$, respectively);
\item Share of women with high-blood-pressure (negative correlation, $p$-value $1.6\cdot10^{-4}$);
\item Incidence of Hepatitis B (negative correlation, $p$-value $2.4\cdot10^{-4}$);
\item PM2.5 air pollution (negative correlation, $p$-value $0.029$).

\end{enumerate}

\subsection{Bias due to GDP: lack of testing?}
We also find a correlation with GDP per capita, which  should be an indicator of lack of testing capabilities. Note however that GDP per capita is also quite highly correlated with another important variable, life expectancy, as we will show in section~\ref{cross}: high GDP per capita is related to an older population, which is correlated with faster contagion. 

Note also that correlation of contagion with GDP disappears when excluding very poor countries, approximately below 5 thousand \$ GDP per capita: this is likely due to the fact that only below a given threshold the capability of testing becomes insufficient. 

\begin{figure}[H]
\begin{center}
\vspace*{3mm}
\includegraphics[scale=0.6]{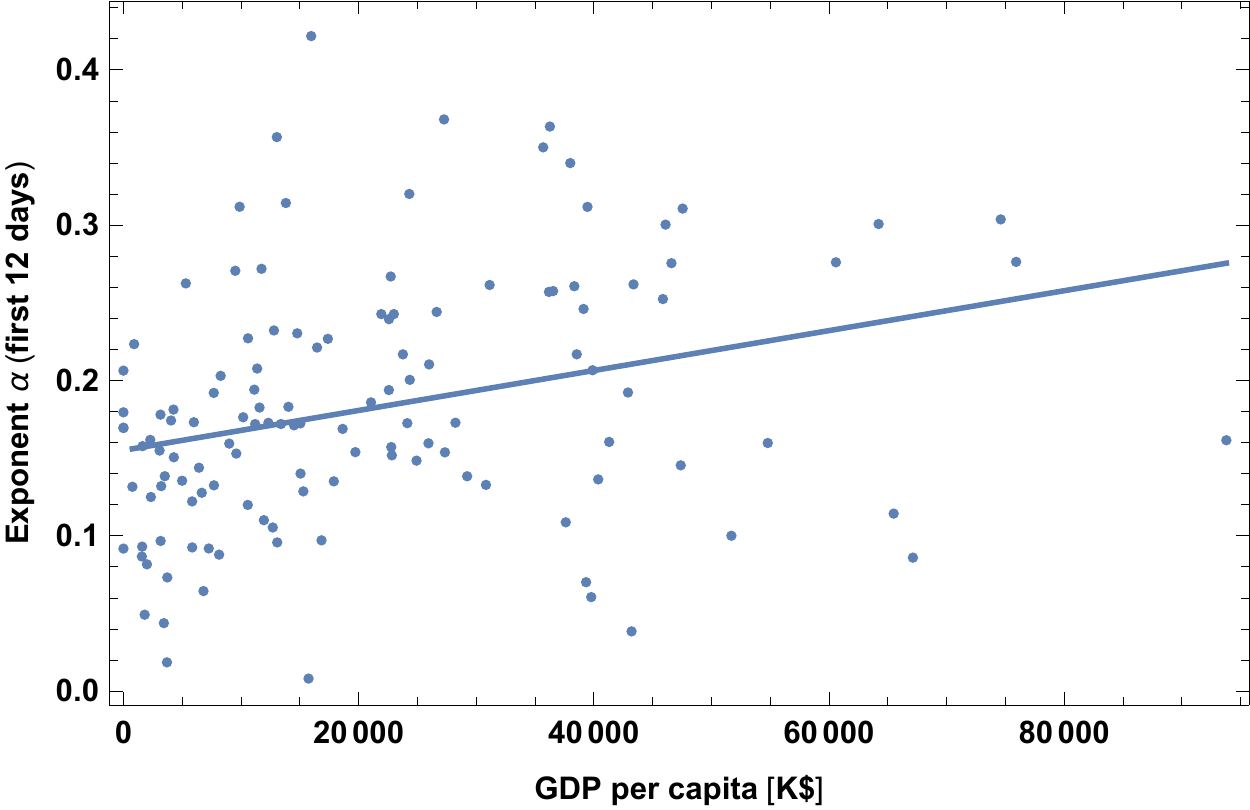} 
\caption{Exponent $\alpha$ for each country vs.~GDP per capita. We show the data points and the best-fit for the linear interpolation. \label{figT}}
\end{center}
\end{figure}

\begin{table}[H]

\begin{tabular}{cc}
    \begin{minipage}{.5\linewidth}
       \begin{tabular}{|l|c|c|c|c|c|}
       \hline
\text{} & \text{Estimate} & \text{Standard Error} & \text{t-Statistic} & \text{$p$-value} \\
\hline
 1 & 0.155 & 0.0111 & 14. & $6.79\times 10^{-27}$ \\
 \text{\it GDP} & $1.28\times 10^{-6}$ & $3.81\times 10^{-7}$ & 3.37 & 0.001 \\
\hline
     \end{tabular}
     \end{minipage} 
     &
     \hspace*{4em}
    \begin{minipage}{.5\linewidth}
             \begin{tabular}{|l|c|c|c|c|}
  \hline
$R^2$ &  $0.087$ \\ \hline
$N$ & 121 \\ \hline

 \end{tabular}
    \end{minipage} 

\end{tabular}

    \caption{In the left panel: best-estimate, standard error ($\sigma$), t-statistic and $p$-value for the parameters of the linear interpolation,  for correlation of $\alpha$ with GDP per capita ({\it GDP}). In the right panel: $R^2$ for the  best-estimate and number of countries $N$.}
    \label{tabTUB}
\end{table}

We performed 2-variables fits, including GDP and each of the above significant variables, in order to check if they remain still significant. In section~\ref{resultseach} we will show the results of such fits, and also the result of individual one variable fits excluding countries below the threshold of  5 thousand \$ GDP per capita.
We list here below the variables that are still significant even when fitting together with GDP.

\subsubsection{Significant variables, strong evidence}

In a 2-variable fit, including GDP per capita, we find strong evidence for correlation with:
\begin{enumerate}
\item Amount of contact in greeting habits (positive correlation, $p$-value  $1.5\cdot10^{-5}$);
\item Temperature (negative correlation, $p$-value  $2.3\cdot10^{-5}$);
\item International tourism: number of arrivals (positive correlation,  $p$-value $2.6\cdot10^{-4}$);
\item Old-age dependency ratio: ratio of the number of people older than 64 relative to the number of people in the working-age (15-64 years) (positive correlation, $p$-value $5.5\cdot10^{-4}$);

\item Vitamin D serum levels (negative correlation, annual values $p$-value $0.0032$, seasonal values $0.0024$; smaller sample, $\sim 50$ countries). To be confirmed on a larger sample.
\item Starting day of the epidemic (negative correlation, $p$-value $0.0037$);
\item Lung cancer death rates (positive correlation, $p$-value $0.0039$);
\item Life expectancy (positive correlation, $p$-value $0.0048$);

\end{enumerate}

\subsubsection{Blood types, strong evidence}
We still find strong evidence for correlation with blood types:
\begin{enumerate}
\item  RH + blood group system (negative correlation, $p$-value $1\cdot 10^{-3}$);
\item B + (negative correlation, $p$-value $2\cdot 10^{-3}$);
\item A-  (positive correlation, $p$-value $1\cdot 10^{-3}$);
\item 0-  (positive correlation, $p$-value $3\cdot 10^{-3}$);
\end{enumerate}
We find moderate evidence for correlation with:
\begin{enumerate}
\item  A+ (positive correlation, $p$-value $0.028$);
\item B-  (positive correlation, $p$-value $0.039$).
\item AB-  (positive correlation, $p$-value $0.012$).
\end{enumerate}
%

\subsubsection{Significant variables, moderate evidence}
We find moderate evidence for:
\begin{enumerate}
\item UV index (negative correlation, $p$-value $0.01$; smaller sample, 73 countries);
\item Type-I diabetes in children, 0-19 years-old (negative correlation, $p$-value $0.01$);
\item Vaccination coverage for Tuberculosis (BCG)  (negative correlation, $p$-value $0.023$);
\item Obesity in males (positive correlation,  $p$-value $0.02$);
\item  $\text{CO}_2$ emissions  (positive correlation, $p$-value $0.02$);
\item Alcohol consumption (positive correlation, $p$-value $0.03$);
\item Daily smoking prevalence (positive correlation, $p$-value $0.03$);
\item Share of population in urban areas (positive correlation,  $p$-value $0.04$);
\end{enumerate}

\subsubsection{Significant variables, counterintuitive}
Counterintuitively we still find correlations with:
\begin{enumerate}
\item Death rate from air pollution (negative correlation, $p$-value $0.002$);
\item Prevalence of anemia, adults and children, (negative correlation, $p$-value $0.023$ and $0.005$);
\item Incidence of Hepatitis B (negative correlation, $p$-value $0.01$);
\item Share of women with high-blood-pressure (negative correlation, $p$-value $0.03$).
\end{enumerate}

In the next section we analyze in more detail the significant variables, one by one (except for those which are not significant anymore after taking into account of GDP per capita).
In section~\ref{cross} we will analyze cross-correlations among such variables  and this will also give a plausible interpretation for the existence of the ``counterintuitive'' variables.

\section{Results for each variable} \label{resultseach}
We first analyze each individual variable, then we analyze variables that have a ``counterintuitive'' correlation and finally we separately analyze blood types.
\subsubsection{Temperature}

The average temperature T has been collected for the relevant period of time, ranging from January to mid April, weighted among the main cities of a given country, see~\cite{Notari} for details. Results are shown in fig.~\ref{figT} and Table~\ref{tabT}. 

We also found that another variable, the absolute value of latitude, has a  similar amount of correlation as for the case of T. However the two variables have a very high correlation (about 0.91) and we do not show results for latitude here. Another variable which is also very highly correlated is UV index and it is shown later.
\begin{figure}[H]
\begin{center}
\vspace*{3mm}
\includegraphics[scale=0.6]{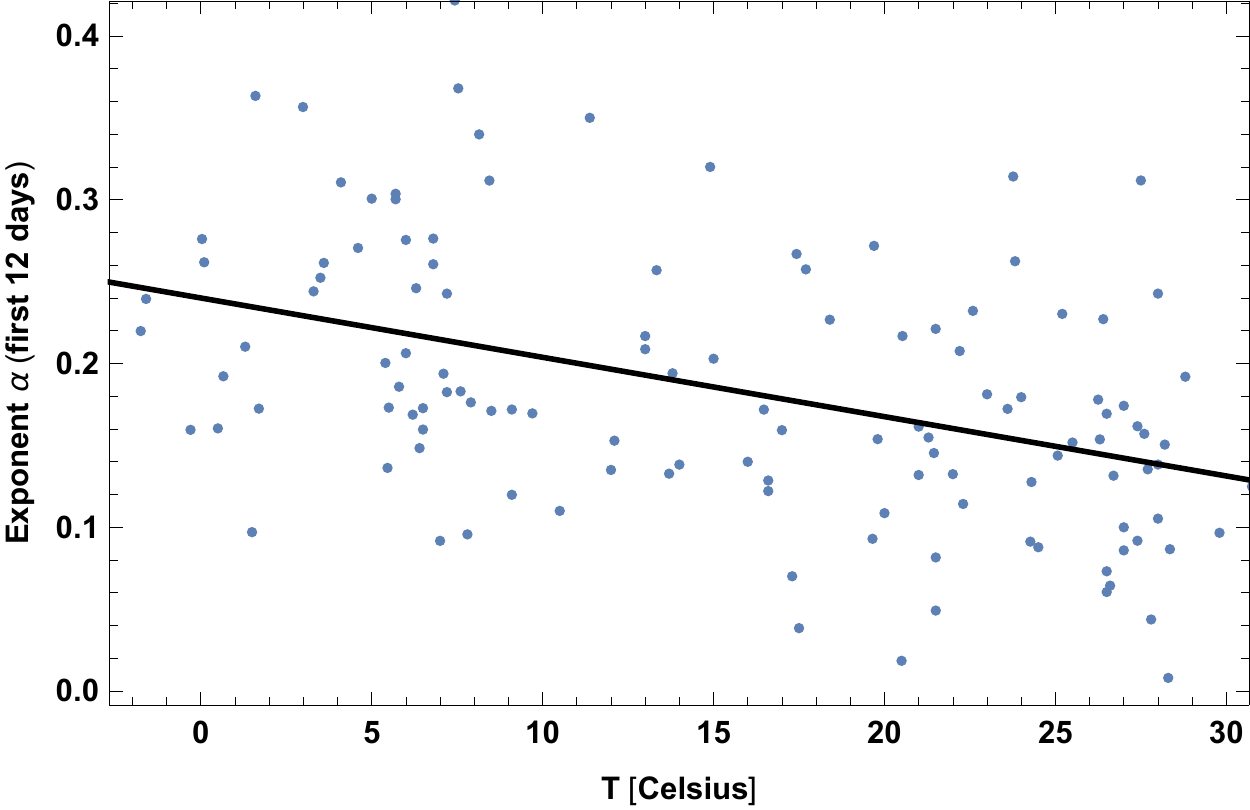} 
\caption{Exponent $\alpha$ for each country vs.~average temperature T, for the relevant period of time, as defined in~\cite{Notari}. We show the data points and the best-fit for the linear interpolation. \label{figT}}
\end{center}
\end{figure}

\begin{table}[H]

\begin{tabular}{cc}
    \begin{minipage}{.5\linewidth}
       \begin{tabular}{|l|c|c|c|c|c|}
       \hline
\text{} & \text{Estimate} & \text{Standard Error} & \text{t-Statistic} & \text{$p$-value} &  \text{$p$-value}, GDP>5K\$ \\
\hline
 1 & 0.239 & 0.0124 & 19.2 & $4.64\times 10^{-39}$ & \\
 T & -0.00359 & 0.000676 & -5.32 & $4.73\times 10^{-7}$ & $7.1 \times 10^{-5}$\\
 \hline
     \end{tabular}
     \end{minipage} 
     &
     \hspace*{12em}
    \begin{minipage}{.5\linewidth}
             \begin{tabular}{|l|c|c|c|c|}
  \hline
$R^2$ &  $0.186$ \\ \hline
$N$ & 126 \\ \hline

 \end{tabular}
    \end{minipage} 

\end{tabular}

     \vspace*{2em}

\begin{tabular}{cc}
    \begin{minipage}{.5\linewidth}
       \begin{tabular}{|l|c|c|c|c|c|}
       \hline
\text{} & \text{Estimate} & \text{Standard Error} & \text{t-Statistic} & \text{$p$-value}  \\
\hline
1 & 0.223 & 0.0188 & 11.9 & $7.45\times 10^{-22}$ \\
 \text{GDP} & $5.62\times 10^{-7}$ & $3.93\times 10^{-7}$ & 1.43 & 0.155 \\
 T & -0.0033 & 0.000761 & -4.33 & 0.0000311 \\
 \hline
     \end{tabular}
     \end{minipage} 
     &
     \hspace*{6em}
    \begin{minipage}{.5\linewidth}
             \begin{tabular}{|l|c|c|c|c|}

\hline
$R^2 $ &  $0.212$ \\ \hline
$N$ & 121 \\ \hline
{\text Cross-correlation} &     0.425\\ \hline
 
 \end{tabular}
    \end{minipage} 

\end{tabular}

    \caption{In the left top panel: best-estimate, standard error ($\sigma$), t-statistic and $p$-value for the parameters of the linear interpolation,  for correlation of $\alpha$ with temperature T. We also show the $p$-value, excluding countries below 5 thousand \$ GDP per capita. In the left bottom panel: same quantities for correlation of $\alpha$ with temperature T and GDP per capita. In the right panels: $R^2$ for the  best-estimate and number of countries $N$. We also show the correlation coefficient between the 2 variables in the two-variable fit.}
    \label{tabT}
\end{table}

\subsubsection{Old-age dependency ratio}
\begin{figure}[H]
\begin{center}
\vspace*{3mm}
\includegraphics[scale=0.6]{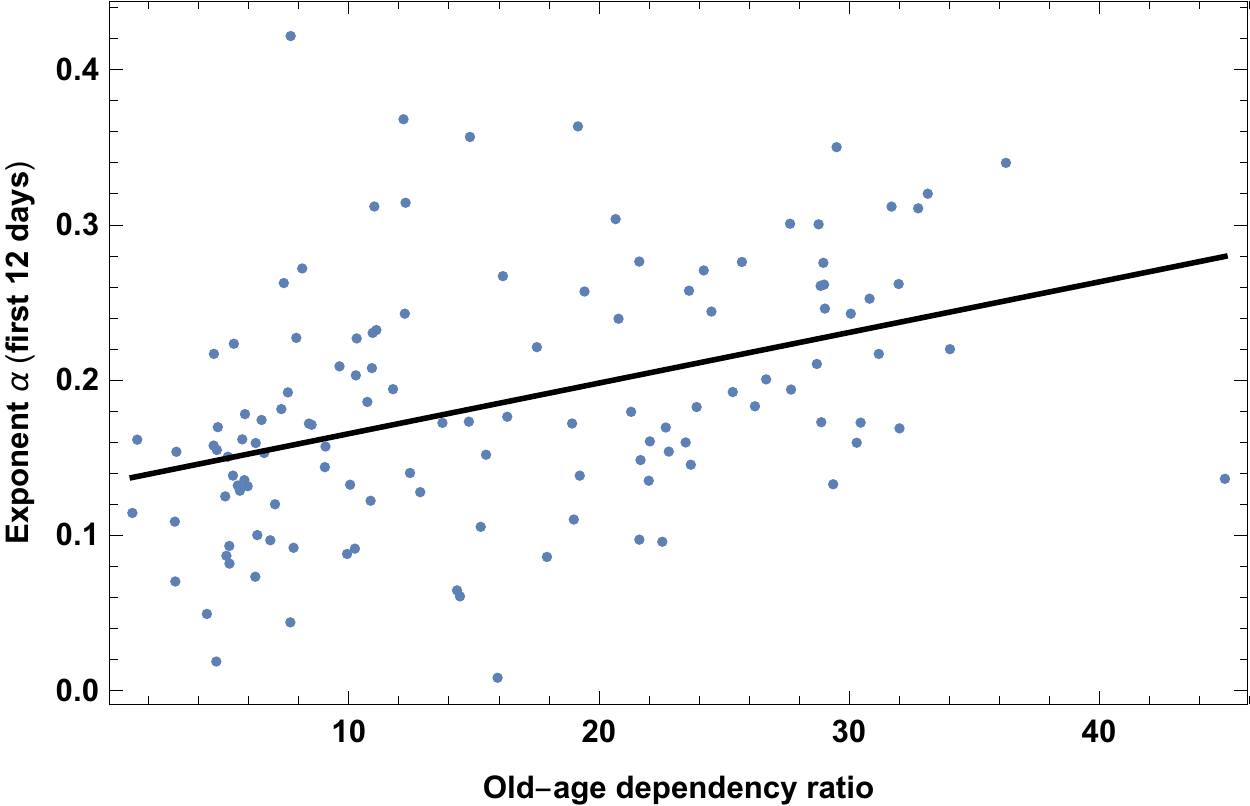} 
\caption{Exponent $\alpha$ for each country vs.~old-age dependency ratio, as defined in the text. We show the data points and the best-fit for the linear interpolation. \label{figOLD}}
\end{center}
\end{figure}

This is the ratio of the number of people older than 64 relative to the number of people in the
working-age (15-64 years). Data are shown as the proportion of dependents per 100 working-age
population, for the year 2017. Results are shown in fig.~\ref{figOLD} and Table~\ref{tabOLD}. This is an interesting finding, since it may suggest that old people are not only subject to higher mortality, but also more likely to be contagious. This could be either because they are more likely to become sick, or because their state of sickness is longer and more contagious, or because many of them live together in nursing homes, or all such reasons together.

Note that a similar variable is life expectancy (which we analyze later); other variables are also highly correlated, such as median age and child dependency ratio, which we do not show here. In analogy to the previous interpretation, data indicate that a younger population, including countries with high percentage of children, is more immune to COVID-19, or less contagious.

\begin{table}[H]

\begin{tabular}{cc}
    \begin{minipage}{.5\linewidth}
       \begin{tabular}{|l|c|c|c|c|c|}
       \hline
\text{} & \text{Estimate} & \text{Standard Error} & \text{t-Statistic} & \text{$p$-value} &  \text{$p$-value}, GDP>5K\$ \\
\hline
1 & 0.132 & 0.0126 & 10.5 & $9.03\times 10^{-19}$ & \\
 \text{OLD} & 0.00326 & 0.000669 & 4.87 & $3.37\times 10^{-6}$ & 0.0025 \\
  \hline
     \end{tabular}
     \end{minipage} 
     &
     \hspace*{12em}
    \begin{minipage}{.5\linewidth}
             \begin{tabular}{|l|c|c|c|c|}
  \hline
$R^2$ &  $0.164$ \\ \hline
$N$ &123 \\ \hline

 \end{tabular}
    \end{minipage} 

\end{tabular}

     \vspace*{2em}

\begin{tabular}{cc}
    \begin{minipage}{.5\linewidth}
       \begin{tabular}{|l|c|c|c|c|c|}
       \hline
\text{} & \text{Estimate} & \text{Standard Error} & \text{t-Statistic} & \text{$p$-value} \\
\hline
 1 & 0.126 & 0.0132 & 9.56 & $2.42\times 10^{-16}$ \\
 \text{GDP} & $7.18\times 10^{-7}$ & $4.04\times 10^{-7}$ & 1.78 & 0.0778 \\
 \text{OLD} & 0.0027 & 0.00076 & 3.55 & 0.000557 \\
 \hline
     \end{tabular}
     \end{minipage} 
     &
     \hspace*{5em}
    \begin{minipage}{.5\linewidth}
             \begin{tabular}{|l|c|c|c|c|}

\hline
$R^2 $ &  $0.188$ \\ \hline
$N$ & 120 \\ \hline
{\text Cross-correlation} &    -0.4561\\ \hline

 \end{tabular}
    \end{minipage} 

\end{tabular}

    \caption{In the left top panel: best-estimate, standard error ($\sigma$), t-statistic and $p$-value for the parameters of the linear interpolation,  for correlation of $\alpha$ with old-age dependency ratio, OLD. We also show the $p$-value, excluding countries below 5 thousand \$ GDP per capita. In the left bottom panel: same quantities for correlation of $\alpha$ with OLD and GDP per capita. In the right panels: $R^2$ for the  best-estimate and number of countries $N$. We also show the correlation coefficient between the 2 variables in the two-variable fit.}
    \label{tabOLD}
\end{table}

\subsubsection{Life expectancy}
\begin{figure}[H]
\begin{center}
\vspace*{3mm}
\includegraphics[scale=0.6]{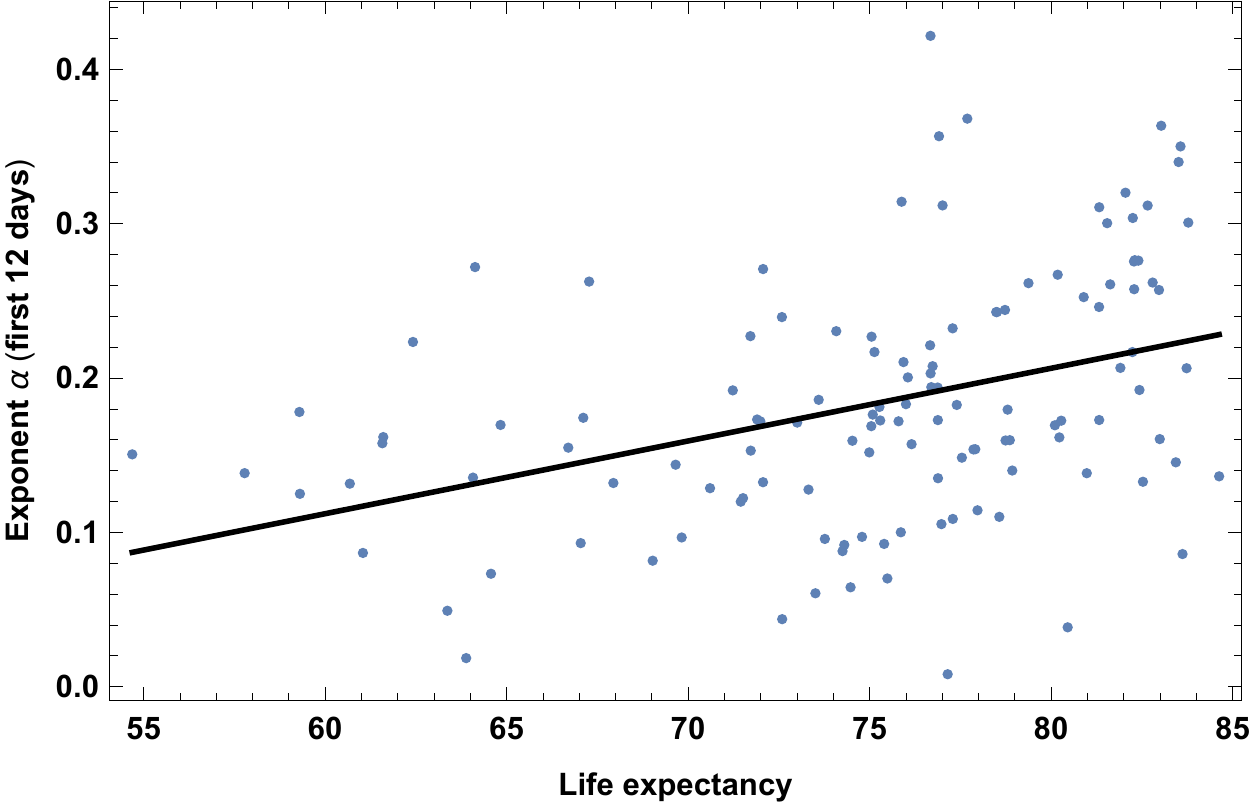} 
\caption{Exponent $\alpha$ for each country vs.~life expectancy. We show the data points and the best-fit for the linear interpolation. \label{figLIFE}}
\end{center}
\end{figure}

This dataset is for year 2016. It has high correlation with old-age dependency ratio. It also has high correlations with other datasets in~\cite{worlddata} that we do not show here, such as median age and child dependency ratio (the ratio between under-19-year-olds and 20-to-69-year-olds). Results are shown in fig.~\ref{figLIFE} and Table~\ref{tabLIFE}.

\begin{table}[H]

\begin{tabular}{cc}
    \begin{minipage}{.5\linewidth}
       \begin{tabular}{|l|c|c|c|c|c|}
       \hline
\text{} & \text{Estimate} & \text{Standard Error} & \text{t-Statistic} & \text{$p$-value} &  \text{$p$-value}, GDP>5K\$ \\
\hline
  1 & -0.147 & 0.0716 & -2.05 & 0.0424 & \\
 \text{LIFE} & 0.00446 & 0.00096 & 4.65 & $8.56\times 10^{-6}$ & 0.0041 \\
\hline
     \end{tabular}
     \end{minipage} 
     &
     \hspace*{12em}
    \begin{minipage}{.5\linewidth}
             \begin{tabular}{|l|c|c|c|c|}
  \hline
$R^2$ &  $0.151$ \\ \hline
$N$ &125 \\ \hline

 \end{tabular}
    \end{minipage} 

\end{tabular}

     \vspace*{2em}

\begin{tabular}{cc}
    \begin{minipage}{.5\linewidth}
       \begin{tabular}{|l|c|c|c|c|c|}
       \hline
\text{} & \text{Estimate} & \text{Standard Error} & \text{t-Statistic} & \text{$p$-value} \\
\hline
 1 & -0.11 & 0.0929 & -1.19 & 0.237 \\
 \text{LIFE} & 0.00386 & 0.00134 & 2.87 & 0.00485 \\
 \text{GDP} & $3.99\times 10^{-7}$ & $4.98\times 10^{-7}$ & 0.801 & 0.424 \\
 \hline
      \end{tabular}
     \end{minipage} 
     &
     \hspace*{5.5em}
    \begin{minipage}{.5\linewidth}
             \begin{tabular}{|l|c|c|c|c|}

\hline
$R^2 $ &  $0.160$ \\ \hline
$N$ & 120 \\ \hline
{\text Cross-correlation} &   -0.679\\ \hline

 \end{tabular}
    \end{minipage} 

\end{tabular}

    \caption{In the left top panel: best-estimate, standard error ($\sigma$), t-statistic and $p$-value for the parameters of the linear interpolation,  for correlation of $\alpha$ with life expectancy, LIFE. We also show the $p$-value, excluding countries below 5 thousand \$ GDP per capita. In the left bottom panel: same quantities for correlation of $\alpha$ with  LIFE and GDP per capita. In the right panels: $R^2$ for the  best-estimate and number of countries $N$. We also show the correlation coefficient between the 2 variables in the two-variable fit.}
    \label{tabLIFE}
\end{table}

\subsubsection{International tourism: number of arrivals}
\begin{figure}[H]
\begin{center}
\vspace*{3mm}
\includegraphics[scale=0.6]{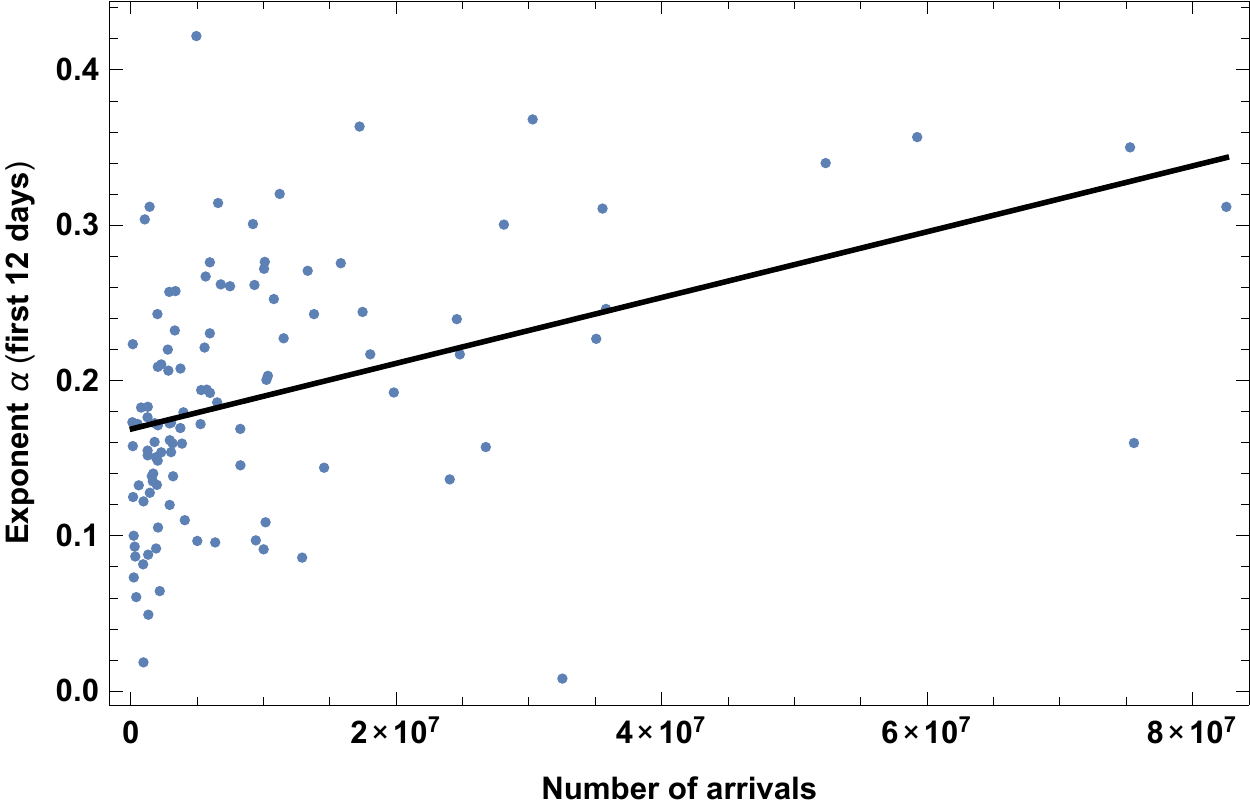} 
\caption{Exponent $\alpha$ for each country vs.~number of tourist arrivals. We show the data points and the best-fit for the linear interpolation. \label{figARR}}
\end{center}
\end{figure}

The dataset  is for year 2016. Results are shown in fig.~\ref{figARR} and Table~\ref{tabARR}. As expected, more tourists correlate with higher speed of contagion. This is in agreement with~\cite{gamstudy,brazilstudy}, that  found air travel to be an important factor, which will appear here as the number of tourists as well as a correlation with GDP.

\begin{table}[H]

\begin{tabular}{cc}
    \begin{minipage}{.5\linewidth}
       \begin{tabular}{|l|c|c|c|c|c|}
       \hline
\text{} & \text{Estimate} & \text{Standard Error} & \text{t-Statistic} & \text{$p$-value} &  \text{$p$-value}, GDP>5K\$\\
\hline
1 & 0.168 & 0.00846 & 19.9 & $7.38\times 10^{-38}$ & \\
 \text{ARR} & $2.13\times 10^{-9}$ & $4.55\times 10^{-10}$ & 4.69 & $8.12\times 10^{-6}$ & 0.0002 \\
\hline
     \end{tabular}
     \end{minipage} 
     &
     \hspace*{12em}
    \begin{minipage}{.5\linewidth}
             \begin{tabular}{|l|c|c|c|c|}
  \hline
$R^2$ &  $0.169$ \\ \hline
$N$ &110 \\ \hline

 \end{tabular}
    \end{minipage} 

\end{tabular}

     \vspace*{2em}

\begin{tabular}{cc}
    \begin{minipage}{.5\linewidth}
       \begin{tabular}{|l|c|c|c|c|c|}
       \hline
\text{} & \text{Estimate} & \text{Standard Error} & \text{t-Statistic} & \text{$p$-value} \\
\hline
1 & 0.146 & 0.0116 & 12.6 & $1.48\times 10^{-22}$ \\
 \text{GDP} & $1.11\times 10^{-6}$ & $4.02\times 10^{-7}$ & 2.76 & 0.00691 \\
 \text{ARR} & $1.77\times 10^{-9}$ & $4.68\times 10^{-10}$ & 3.78 & 0.000265 \\ \hline
     \end{tabular}
     \end{minipage} 
     &
     \hspace*{5.5em}
    \begin{minipage}{.5\linewidth}
             \begin{tabular}{|l|c|c|c|c|}

\hline
$R^2 $ &  $0.226$ \\ \hline
$N$ & 107 \\ \hline
{\text Cross-correlation} &   -0.288\\ \hline

 \end{tabular}
    \end{minipage} 

\end{tabular}

    \caption{In the left top panel: best-estimate, standard error ($\sigma$), t-statistic and $p$-value for the parameters of the linear interpolation,  for correlation of $\alpha$ with number of tourist arrivals, ARR. We also show the $p$-value, excluding countries below 5 thousand \$ GDP per capita. In the left bottom panel: same quantities for correlation of $\alpha$ with ARR and GDP per capita. In the right panels: $R^2$ for the  best-estimate and number of countries $N$. We also show the correlation coefficient between the 2 variables in the two-variable fit.}
    \label{tabARR}
\end{table}

\subsubsection{Starting date of the epidemic}

\begin{figure}[H]
\begin{center}
\vspace*{3mm}
\includegraphics[scale=0.6]{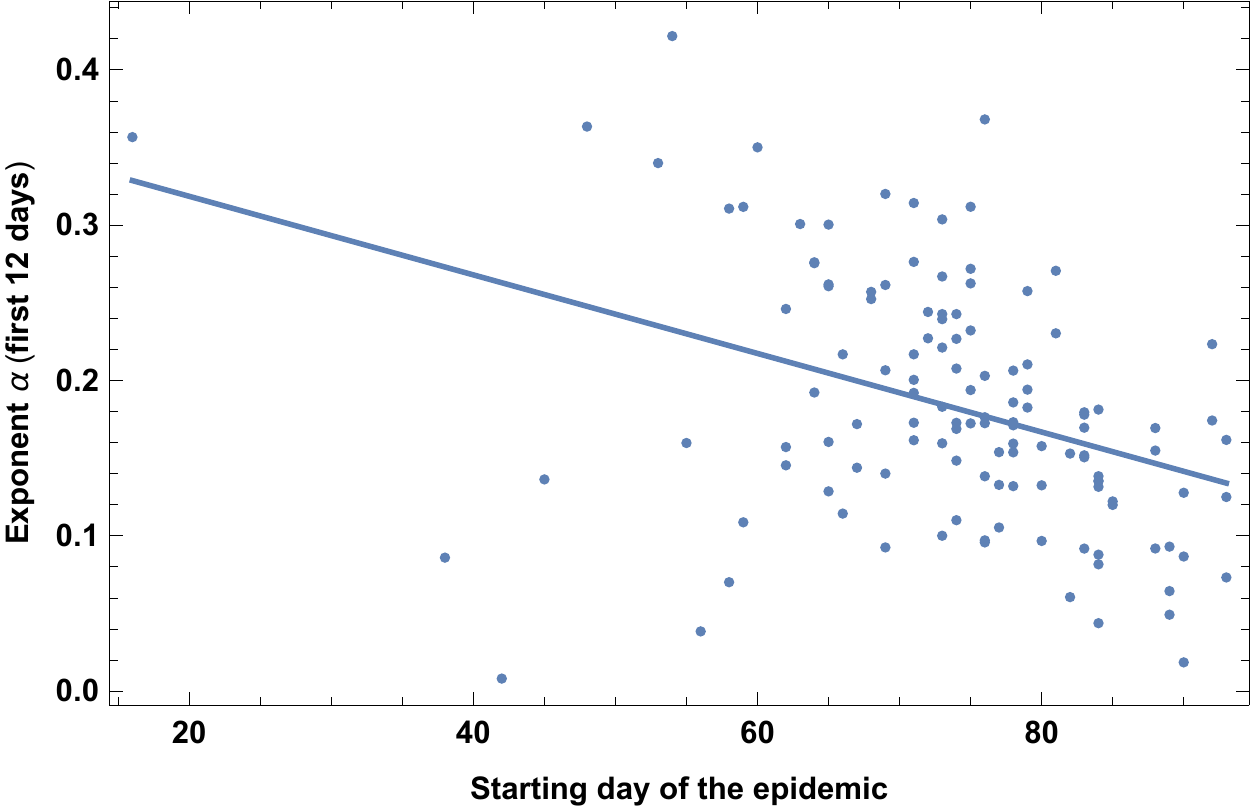} 
\caption{Exponent $\alpha$ for each country vs.~starting date of the analysis of the epidemic, DATE, defined as the day when the positive cases reached $N=30$. Days are counted from Dec 31st 2019. We show the data points and the best-fit for the linear interpolation. \label{figDATE}}
\end{center}
\end{figure}

This refers to the day $d_i$ chosen as a starting point, counted from December 31st 2019.  Results are shown in fig.~\ref{figDATE} and Table~\ref{tabDATE}, which shows that earlier contagion is correlated with faster contagion. One possible interpretation is that countries which are affected later are already more aware of the pandemic and therefore have a larger amount of social distancing, which makes the growth rate smaller. Another possible interpretation is that there is some other underlying factor that protects against contagion, and therefore epidemics spreads both later {\it and} slower. 

\begin{table}[H]

\begin{tabular}{cc}
    \begin{minipage}{.5\linewidth}
       \begin{tabular}{|l|c|c|c|c|c|}
       \hline
\text{} & \text{Estimate} & \text{Standard Error} & \text{t-Statistic} & \text{$p$-value} &  \text{$p$-value}, GDP>5K\$  \\
\hline
 1 & 0.369 & 0.0421 & 8.76 & $1.21\times 10^{-14}$ &\\
 \text{DATE} & -0.00253 & 0.000565 & -4.48 & 0.0000168 & 0.011\\
 \hline
     \end{tabular}
     \end{minipage} 
     &
     \hspace*{12em}
    \begin{minipage}{.5\linewidth}
             \begin{tabular}{|l|c|c|c|c|}
  \hline
$R^2$ &  $0.139$ \\ \hline
$N$ & 126 \\ \hline

 \end{tabular}
    \end{minipage} 

\end{tabular}

     \vspace*{2em}

\begin{tabular}{cc}
    \begin{minipage}{.5\linewidth}
       \begin{tabular}{|l|c|c|c|c|c|}
       \hline
\text{} & \text{Estimate} & \text{Standard Error} & \text{t-Statistic} & \text{$p$-value} \\
\hline
1 & 0.32 & 0.0568 & 5.64 & $1.18\times 10^{-7}$ \\
 \text{GDP} & $5.85\times 10^{-7}$ & $4.38\times 10^{-7}$ & 1.33 & 0.185 \\
 \text{DATE} & -0.00204 & 0.000689 & -2.96 & 0.00367 \\
  \hline
     \end{tabular}
     \end{minipage} 
     &
     \hspace*{5.5em}
    \begin{minipage}{.5\linewidth}
             \begin{tabular}{|l|c|c|c|c|}

\hline
$R^2 $ &  $0.151$ \\ \hline
$N$ & 121 \\ \hline
{\text Cross-correlation} &  0.539\\ \hline
 \end{tabular}
    \end{minipage} 
\\

\end{tabular}

    \caption{In the left top panel: best-estimate, standard error ($\sigma$), t-statistic and $p$-value for the parameters of the linear interpolation,  for correlation of $\alpha$ with vs.~starting date of the analysis of the epidemic, DATE, as defined in the text. We also show the $p$-value, excluding countries below 5 thousand \$ GDP per capita. In the left bottom panel: same quantities for correlation of $\alpha$ with  DATE and GDP per capita. In the right panels: $R^2$ for the  best-estimate and number of countries $N$. We also show the correlation coefficient between the 2 variables in the two-variable fit.}
    \label{tabDATE}
\end{table}

\subsubsection{Greeting habits}
\begin{figure}[t!]
\begin{center}
\vspace*{3mm}
\includegraphics[scale=0.6]{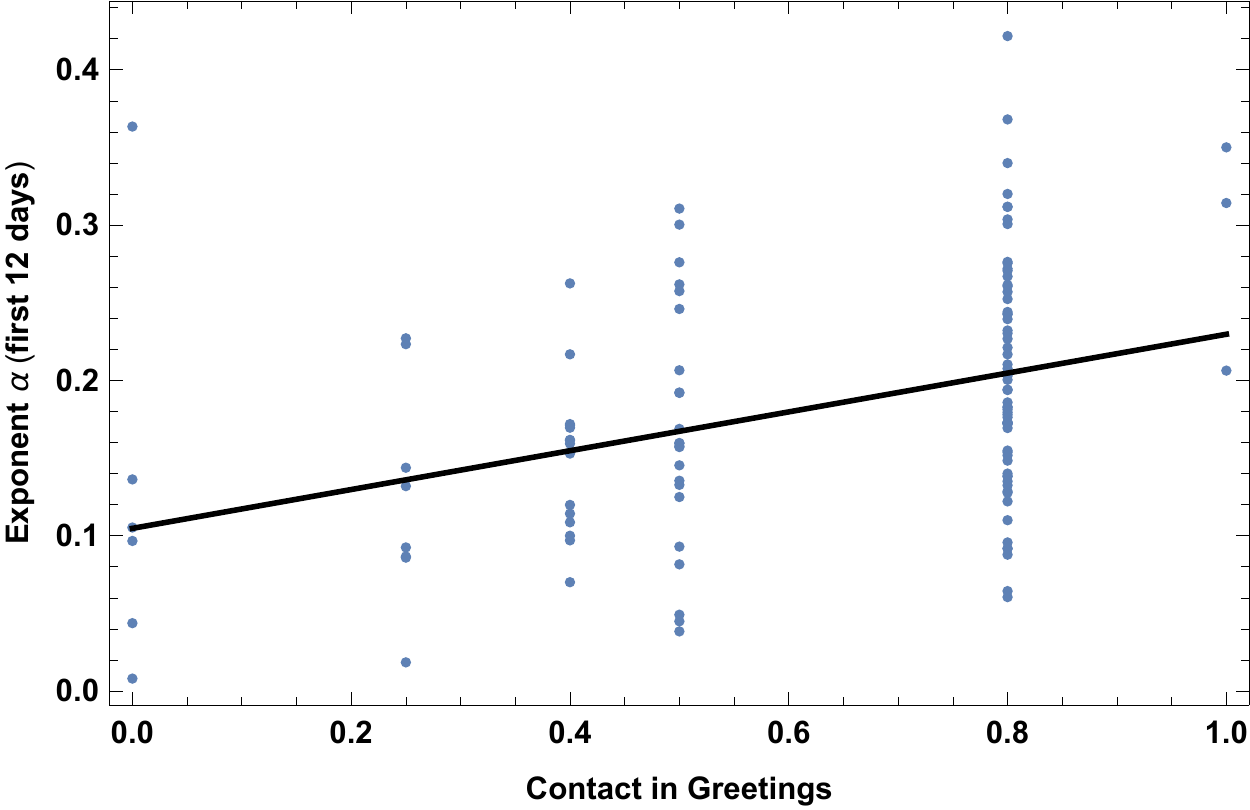} 
\caption{Exponent $\alpha$ for each country vs.~level of contact in greeting habits, $GRE$, as defined in the text. We show the data points and the best-fit for the linear interpolation. \label{figGRE}}
\end{center}
\end{figure}

A relevant variable is the level of contact in greeting habits in each country.
We have subdivided the countries in groups according to the physical contact in greeting habits; information has been taken from \cite{culture}. 
\begin{enumerate}
\item No or little physical contact, bowing. In this group we have: {Bangladesh, Cambodia, Japan, Korea South,  Sri Lanka, Thailand}.
\item Handshaking between man-man and woman-woman. No or little contact man-woman.   In this group we have: India, Indonesia, Niger, Senegal, Singapore, Togo, Vietnam, Zambia
\item  Handshaking. In this group we have: Australia, Austria, Bulgaria, Burkina Faso, Canada, China, Estonia, \
Finland, Germany, Ghana, Madagascar, Malaysia, Mali, Malta, New \
Zealand, Norway, Philippines, Rwanda, Sweden, Taiwan, Uganda, Kingdom \
United, States United.
  \item Handshaking, plus kissing among friends and relatives, but only man-man and woman-woman. No or little contact man-woman. In this group we have:  
  Afghanistan, Azerbaijan, Bahrain, Belarus, Brunei, Egypt, Guinea, \
Jordan, Kuwait, Kyrgyzstan, Oman, Pakistan, Qatar, Arabia Saudi, Arab \
Emirates United, Uzbekistan.
 \item Handshaking, plus kissing among friends and relatives.  In this group we have:  Albania, Algeria, Argentina, Armenia, Belgium, Bolivia, and Bosnia 
Herzegovina, Cameroon, Chile, Colombia, Costa Rica, C\^ote d'Ivoire, Croatia, Cuba, Cyprus, Czech Republic, Denmark, 
Dominican Republic, Ecuador, El Salvador, France, Georgia, Greece, 
Guatemala, Honduras, Hungary, Iran, Iraq, Ireland, Israel, Italy, 
Jamaica, Kazakhstan, Kenya, Kosovo, Latvia, Lebanon, Lithuania, 
Luxembourg, Macedonia, Mauritius, Mexico, Moldova, Montenegro, 
Morocco, Netherlands, Panama, Paraguay, Peru, Poland, Portugal, 
Puerto Rico, Romania, Russia, Serbia, Slovakia, Slovenia, Africa 
South, Switzerland, Trinidad and Tobago, Tunisia, Turkey, Ukraine, 
Uruguay, Venezuela.
\item Handshaking and kissing. In this group we have: Andorra, Brazil, Spain.
\end{enumerate}
We have arbitrarily assigned a variable, named GRE, from 0 to 1 to each group, namely GRE $=0, 0.25, 0.5, 0.4, 0.8$ and $1$, respectively. We have chosen a ratio of 2 between group 2 and 3 and between group 4 and 5, based on the fact that the only difference is that about half of the possible interactions (men-women) are without contact. Results are shown in Fig.~\ref{figGRE} and Table~\ref{tabGRE}.
 
%

Note also that an outlier is visible in the plot, which corresponds to South Korea. The early outbreak of the disease in this particular case was strongly affected by the Shincheonji Church, which included mass prayer and worship sessions. By excluding South Korea from the dataset one finds an even larger significance, $p$-value$= 6.4\cdot 10^{-8}$ and $R^2\approx 0.23$.


\begin{table}[H]

\begin{tabular}{cc}
    \begin{minipage}{.5\linewidth}
       \begin{tabular}{|l|c|c|c|c|c|}
       \hline
\text{} & \text{Estimate} & \text{Standard Error} & \text{t-Statistic} & \text{$p$-value} &  \text{$p$-value}, GDP>5K\$\\
\hline
 1 & 0.111 & 0.019 & 5.8 & $5.47\times 10^{-8}$ & \\
 \text{GRE} & 0.12 & 0.0287 & 4.2 & 0.000051 & 0.0029\\
\hline
     \end{tabular}
     \end{minipage} 
     &
     \hspace*{12em}
    \begin{minipage}{.5\linewidth}
             \begin{tabular}{|l|c|c|c|c|}
  \hline
$R^2$ &  $0.129$ \\ \hline
$N$ & 121 \\ \hline

 \end{tabular}
    \end{minipage} 

\end{tabular}

     \vspace*{2em}

\begin{tabular}{cc}
    \begin{minipage}{.5\linewidth}
       \begin{tabular}{|l|c|c|c|c|c|}
       \hline
\text{} & \text{Estimate} & \text{Standard Error} & \text{t-Statistic} & \text{$p$-value} \\
\hline
 1 & 0.079 & 0.0204 & 3.87 & 0.000184 \\
 \text{GDP} & $1.26\times 10^{-6}$ & $3.65\times 10^{-7}$ & 3.45 & 0.000794 \\
 \text{GRE} & 0.128 & 0.0282 & 4.53 & 0.0000149 \\
  \hline
     \end{tabular}
     \end{minipage} 
     &
     \hspace*{5em}
    \begin{minipage}{.5\linewidth}
             \begin{tabular}{|l|c|c|c|c|}

\hline
$R^2 $ &  $0.22$ \\ \hline
$N$ & 116 \\ \hline
{\text Cross-correlation} &   0.00560\\ \hline
 \end{tabular}
    \end{minipage} 
\\

\end{tabular}

    \caption{In the left top panel: best-estimate, standard error ($\sigma$), t-statistic and $p$-value for the parameters of the linear interpolation,  for correlation of $\alpha$ with level of contact in greeting habits, GRE, as defined in the text. We also show the $p$-value, excluding countries below 5 thousand \$ GDP per capita. In the left bottom panel: same quantities for correlation of $\alpha$ with  GRE and GDP per capita. In the right panels: $R^2$ for the  best-estimate and number of countries $N$. We also show the correlation coefficient between the 2 variables in the two-variable fit.}
    \label{tabGRE}
\end{table}

\subsubsection{Lung cancer death rates}
\begin{figure}[H]
\begin{center}
\vspace*{3mm}
\includegraphics[scale=0.6]{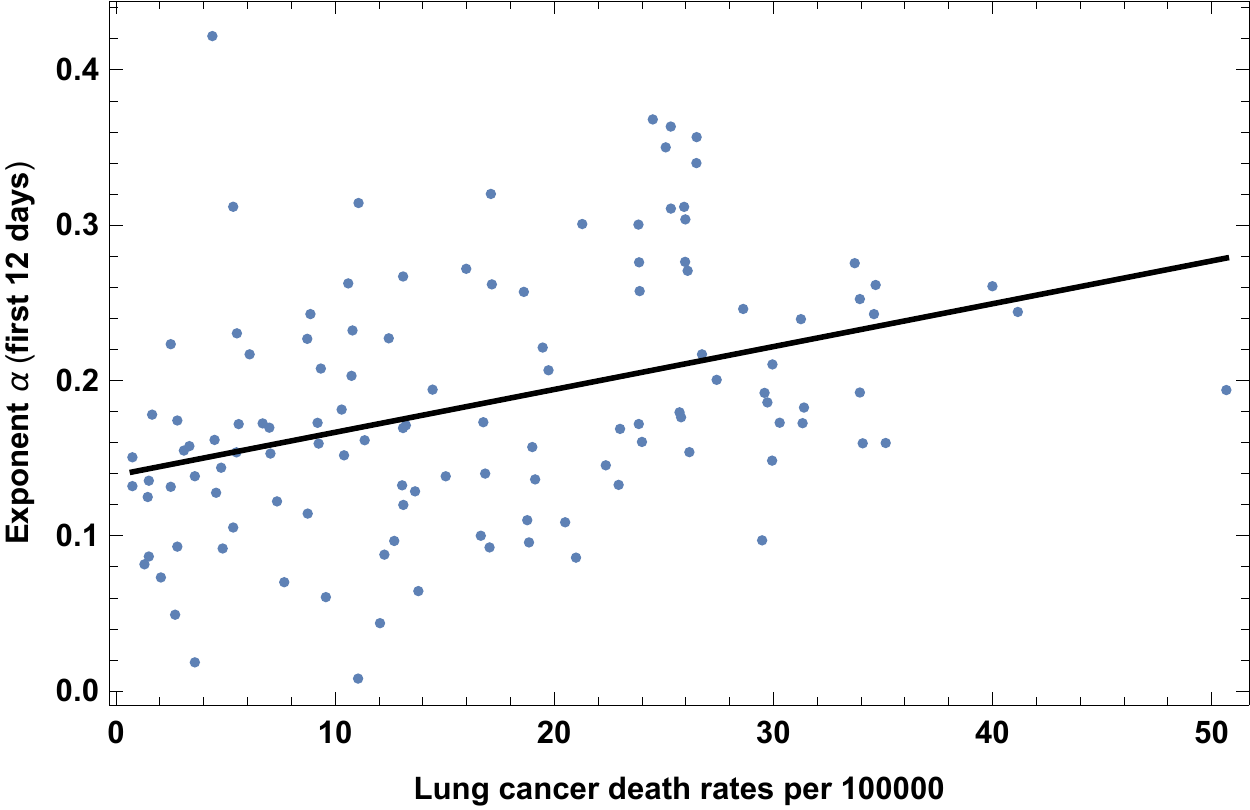} 
\caption{Exponent $\alpha$ for each country vs.~lung cancer death rates. We show the data points and the best-fit for the linear interpolation. \label{figLUNG}}
\end{center}
\end{figure}

This dataset refers to year 2002.  Results are shown in Fig.~\ref{figLUNG} and Table~\ref{tabLUNG}.
Such results are interesting and could be interpreted a priori in two ways. A first interpretation is that COVID-19 contagion might correlate to lung cancer, simply due to the fact that lung cancer is more prevalent in countries with more old people. Such a simplistic interpretation is somehow contradicted by the case of generic cancer death rates, discussed in section~\ref{results}, which is indeed less significant than lung cancer. A better interpretation is therefore that lung cancer may be a specific risk factor for COVID-19 contagion. This is supported also by the observation of high rates of  lung cancer in COVID-19 patients~\cite{lungcancer}.

\begin{table}[H]

\begin{tabular}{cc}
    \begin{minipage}{.5\linewidth}
       \begin{tabular}{|l|c|c|c|c|c|}
       \hline
\text{} & \text{Estimate} & \text{Standard Error} & \text{t-Statistic} & \text{$p$-value}  &  \text{$p$-value}, GDP>5K\$ \\
\hline
 1 & 0.143 & 0.0121 & 11.8 & $9.96\times 10^{-22}$ & \\
 \text{LUNG} & 0.00159 & 0.000381 & 4.17 & 0.0000572 & 0.024\\ \hline
     \end{tabular}
     \end{minipage} 
     &
     \hspace*{12em}
    \begin{minipage}{.5\linewidth}
             \begin{tabular}{|l|c|c|c|c|}
  \hline
$R^2$ &  $0.127$ \\ \hline
$N$ &121 \\ \hline

 \end{tabular}
    \end{minipage} 

\end{tabular}

     \vspace*{2em}

\begin{tabular}{cc}
    \begin{minipage}{.5\linewidth}
       \begin{tabular}{|l|c|c|c|c|c|}
       \hline
\text{} & \text{Estimate} & \text{Standard Error} & \text{t-Statistic} & \text{$p$-value} \\
\hline
 1 & 0.133 & 0.013 & 10.2 & $7.48\times 10^{-18}$ \\
 \text{GDP} & $8.93\times 10^{-7}$ & $4.02\times 10^{-7}$ & 2.22 & 0.0282 \\
 \text{LUNG} & 0.00122 & 0.000416 & 2.94 & 0.00396 \\
 \hline
      \end{tabular}
     \end{minipage} 
     &
     \hspace*{5.5em}
    \begin{minipage}{.5\linewidth}
             \begin{tabular}{|l|c|c|c|c|}

\hline
$R^2 $ &  $0.164$ \\ \hline
$N$ & 118 \\ \hline
{\text Cross-correlation} &   -0.403\\ \hline

 \end{tabular}
    \end{minipage} 

\end{tabular}

    \caption{In the left top panel: best-estimate, standard error ($\sigma$), t-statistic and $p$-value for the parameters of the linear interpolation,  for correlation of $\alpha$ with lung cancer death rates, LUNG. We also show the $p$-value, excluding countries below 5 thousand \$ GDP per capita. In the left bottom panel: same quantities for correlation of $\alpha$ with LUNG and GDP per capita.  In the right panels: $R^2$ for the  best-estimate and number of countries $N$. We also show the correlation coefficient between the 2 variables in the two-variable fit.}
    \label{tabLUNG}
\end{table}

\subsubsection{Obesity in males}

\begin{figure}[H]
\begin{center}
\vspace*{3mm}
\includegraphics[scale=0.6]{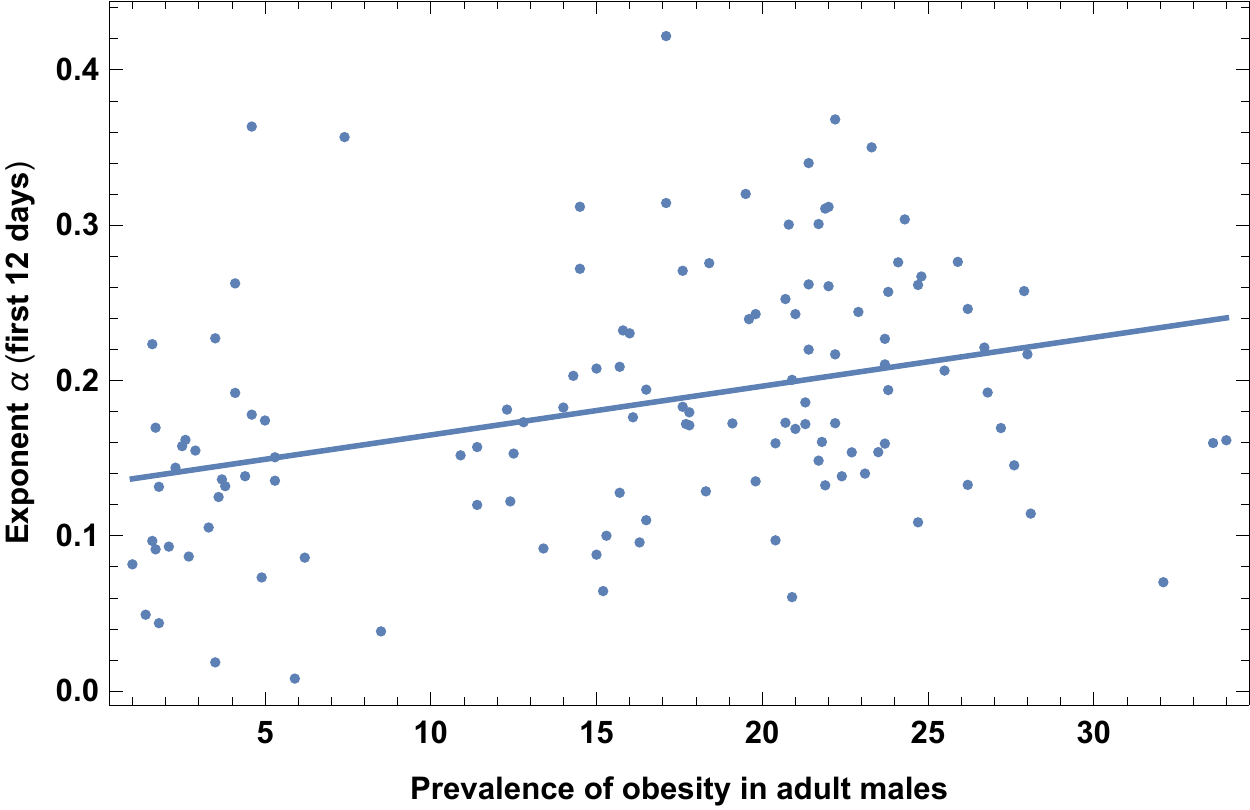} 
\caption{Exponent $\alpha$ for each country vs.~prevalence of obesity in adult males. We show the data points and the best-fit for the linear interpolation. \label{figOBE}}
\end{center}
\end{figure}

This refers to the prevalence of obesity in adult {\it males}, measured in 2014. Results are shown in fig.~\ref{figOBE} and Table~\ref{tabOBE}.  Note that this effect is mostly due to the difference between very poor countries and the rest of the world; indeed this becomes non-significant when excluding countries below 5K\$ GDP per capita. Note also that obesity in {\it females} instead is {\it not} correlated with growth rate of COVID-19 contagion in our sample. See also~\cite{obesity} for increased risk of severe COVID-19 symptoms for obese patients.

\begin{table}[H]

\begin{tabular}{cc}
    \begin{minipage}{.5\linewidth}
       \begin{tabular}{|l|c|c|c|c|c|}
       \hline
\text{} & \text{Estimate} & \text{Standard Error} & \text{t-Statistic} & \text{$p$-value} &  \text{$p$-value}, GDP>5K\$ \\
\hline
  1 & 0.134 & 0.0144 & 9.29 & $6.9\times 10^{-16}$ & \\
 \text{OBE} & 0.00314 & 0.000788 & 3.98 & 0.000115 & 0.12\\
  \hline
     \end{tabular}
     \end{minipage} 
     &
     \hspace*{12em}
    \begin{minipage}{.5\linewidth}
             \begin{tabular}{|l|c|c|c|c|}
  \hline
$R^2$ &  $0.114$ \\ \hline
$N$ & 125 \\ \hline

 \end{tabular}
    \end{minipage} 

\end{tabular}

     \vspace*{2em}

\begin{tabular}{cc}
    \begin{minipage}{.5\linewidth}
       \begin{tabular}{|l|c|c|c|c|c|}
       \hline
\text{} & \text{Estimate} & \text{Standard Error} & \text{t-Statistic} & \text{$p$-value} \\
\hline
 1 & 0.132 & 0.0148 & 8.89 & $8.35\times 10^{-15}$ \\
 \text{GDP} & $5.6\times 10^{-7}$ & $4.84\times 10^{-7}$ & 1.16 & 0.25 \\
 \text{OBE} & 0.00249 & 0.00106 & 2.35 & 0.0202 \\
  \hline
     \end{tabular}
     \end{minipage} 
     &
     \hspace*{5.5em}
    \begin{minipage}{.5\linewidth}
             \begin{tabular}{|l|c|c|c|c|}

\hline
$R^2 $ &  $0.128$ \\ \hline
$N$ & 121 \\ \hline
{\text Cross-correlation} &  -0.634\\ \hline
 \end{tabular}
    \end{minipage} 
\\

\end{tabular}

    \caption{In the left top panel: best-estimate, standard error ($\sigma$), t-statistic and $p$-value for the parameters of the linear interpolation,  for correlation of $\alpha$ with prevalence of obesity in adult males (OBE). We also show the $p$-value, excluding countries below 5 thousand \$ GDP per capita. In the left bottom panel: same quantities for correlation of $\alpha$ with  OBE and GDP per capita. In the right panels: $R^2$ for the  best-estimate and number of countries $N$. We also show the correlation coefficient between the 2 variables in the two-variable fit.}
    \label{tabOBE}
\end{table}

\subsubsection{Urbanization}

\begin{figure}[H]
\begin{center}
\vspace*{3mm}
\includegraphics[scale=0.6]{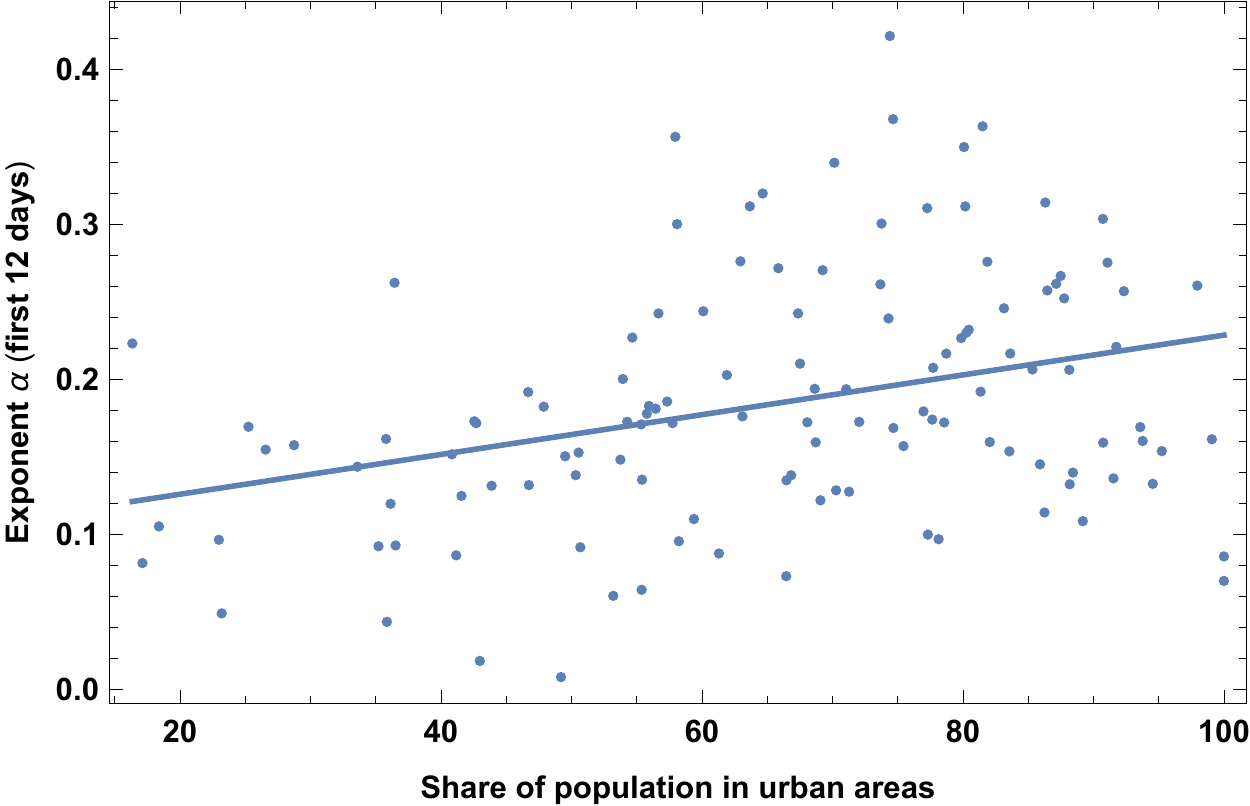} 
\caption{Exponent $\alpha$ for each country vs.~share of population in urban areas. We show the data points and the best-fit for the linear interpolation. \label{figURB}}
\end{center}
\end{figure}

This is the share of population living in urban areas, collected in year 2017. Results are shown in Fig.~\ref{figURB} and Table~\ref{tabURB}. This is an expected correlation, in agreement with~\cite{urban, urban2}.

\begin{table}[H]

\begin{tabular}{cc}
    \begin{minipage}{.5\linewidth}
       \begin{tabular}{|l|c|c|c|c|c|}
       \hline
\text{} & \text{Estimate} & \text{Standard Error} & \text{t-Statistic} & \text{$p$-value}  &  \text{$p$-value}, GDP>5K\$ \\
\hline
 1 & 0.1 & 0.0228 & 4.4 & 0.0000231 & \\
 \text{URB} & 0.00128 & 0.000331 & 3.88 & 0.000173 & 0.057 \\
\hline
     \end{tabular}
     \end{minipage} 
     &
     \hspace*{12em}
    \begin{minipage}{.5\linewidth}
             \begin{tabular}{|l|c|c|c|c|}
  \hline
$R^2$ &  $0.109$\\ \hline
$N$ & 124 \\ 
\hline 

 \end{tabular}
    \end{minipage} 

\end{tabular}

     \vspace*{2em}

\begin{tabular}{cc}
    \begin{minipage}{.5\linewidth}
       \begin{tabular}{|l|c|c|c|c|c|}
       \hline
\text{} & \text{Estimate} & \text{Standard Error} & \text{t-Statistic} & \text{$p$-value} \\
\hline
1 & 0.108 & 0.0247 & 4.37 & 0.0000267 \\
 \text{GDP} & $7.57\times 10^{-7}$ & $4.74\times 10^{-7}$ & 1.6 & 0.113 \\
 \text{URB} & 0.000916 & 0.000439 & 2.09 & 0.0391 \\
 \hline
     \end{tabular}
     \end{minipage} 
     &
     \hspace*{5em}
    \begin{minipage}{.5\linewidth}
             \begin{tabular}{|l|c|c|c|c|}

\hline
$R^2$ &  $0.133$\\ \hline
$N$ & 120 \\ \hline 
{\text Cross-correlation} &   -0.6216\\ \hline
 \end{tabular}
    \end{minipage} 

\end{tabular}

    \caption{In the left top panel: best-estimate, standard error ($\sigma$), t-statistic and $p$-value for the parameters of the linear interpolation,  for correlation of $\alpha$ with share of population in urban areas, URB, as defined in the text. We also show the $p$-value, excluding countries below 5 thousand \$ GDP per capita. In the left bottom panel: same quantities for correlation of $\alpha$ with  URB and GDP per capita. In the right panels: $R^2$ for the  best-estimate and number of countries $N$. We also show the correlation coefficient between the 2 variables in the two-variable fit.}
    \label{tabURB}
\end{table}

\subsubsection{Alcohol consumption}

\begin{figure}[H]
\begin{center}
\vspace*{3mm}
\includegraphics[scale=0.6]{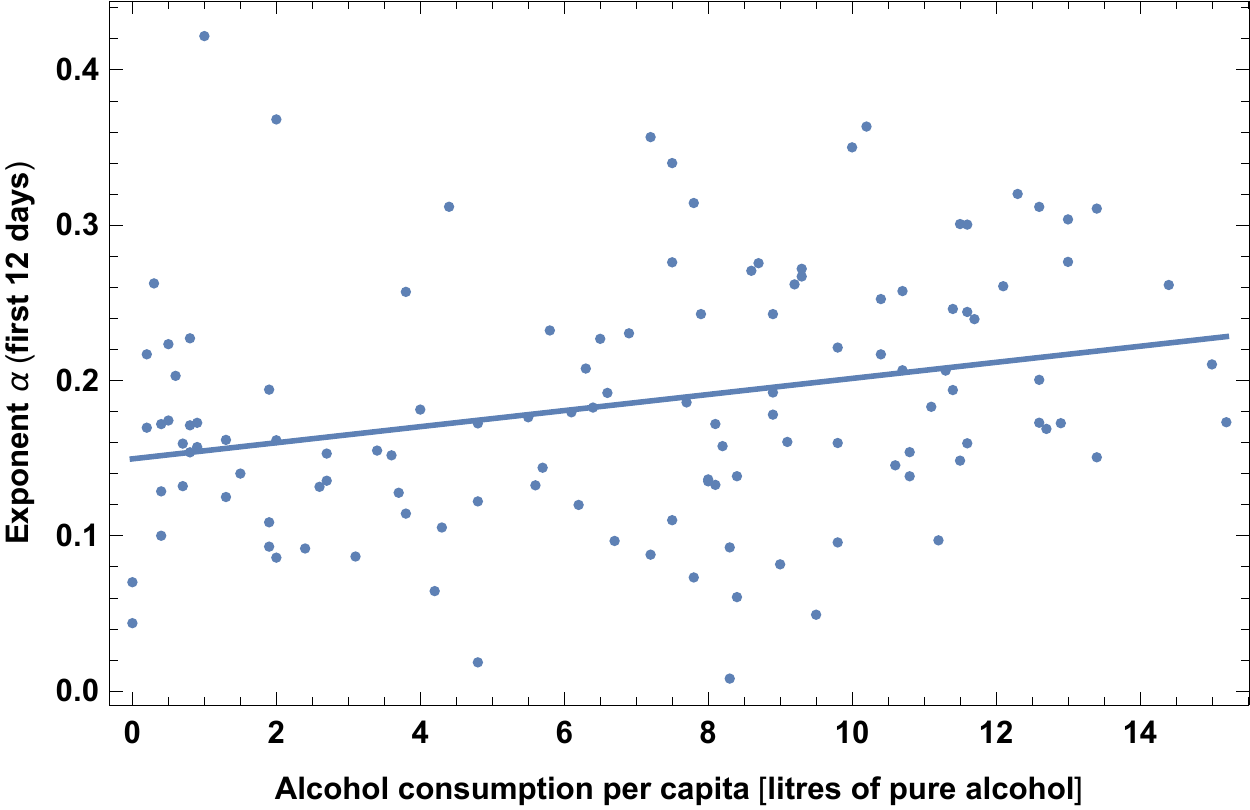} 
\caption{Exponent $\alpha$ for each country vs.~alcohol consumption. We show the data points and the best-fit for the linear interpolation. \label{figALCO}}
\end{center}
\end{figure}

This dataset refers to year 2016.  Results are shown in Fig.~\ref{figALCO} and Table~\ref{tabALCO}.
Note that this variable is highly correlated with old-age dependency ratio, as discussed in section~\ref{cross}. While the correlation with alcohol consumption may be simply due to correlation with other variables, such as old-age dependency ratio, this finding deserves anyway more research, to assess whether it may be at least partially due to the deleterious effects of alcohol on the immune system.

\begin{table}[H]

\begin{tabular}{cc}
    \begin{minipage}{.5\linewidth}
       \begin{tabular}{|l|c|c|c|c|c|}
       \hline
\text{} & \text{Estimate} & \text{Standard Error} & \text{t-Statistic} & \text{$p$-value}  &  \text{$p$-value}, GDP>5K\$  \\
\hline
 1 & 0.15 & 0.0131 & 11.4 & $7.8\times 10^{-21}$ & \\
 \text{ALCO} & 0.00518 & 0.00164 & 3.17 & 0.00195 & 0.012 \\ \hline
     \end{tabular}
     \end{minipage} 
     &
     \hspace*{12em}
    \begin{minipage}{.5\linewidth}
             \begin{tabular}{|l|c|c|c|c|}
  \hline
$R^2$ &  $0.076$\\ \hline
$N$ & 126 \\ 
\hline 

 \end{tabular}
    \end{minipage} 

\end{tabular}

     \vspace*{2em}

\begin{tabular}{cc}
    \begin{minipage}{.5\linewidth}
       \begin{tabular}{|l|c|c|c|c|c|}
       \hline
\text{} & \text{Estimate} & \text{Standard Error} & \text{t-Statistic} & \text{$p$-value} \\
\hline
 1 & 0.134 & 0.0141 & 9.48 & $3.75\times 10^{-16}$ \\
 \text{GDP} & $1.12\times 10^{-6}$ & $3.86\times 10^{-7}$ & 2.91 & 0.00437 \\
 \text{ALCO} & 0.00382 & 0.0017 & 2.25 & 0.0264 \\
 \hline
     \end{tabular}
     \end{minipage} 
     &
     \hspace*{5.5em}
    \begin{minipage}{.5\linewidth}
             \begin{tabular}{|l|c|c|c|c|}

\hline
$R^2$ &  $0.138$\\ \hline
$N$ & 120 \\ \hline 
{\text Cross-correlation} &   -0.286 \\ \hline
 \end{tabular}
    \end{minipage} 
\\
%
\end{tabular}

    \caption{In the left top panel: best-estimate, standard error ($\sigma$), t-statistic and $p$-value for the parameters of the linear interpolation,  for correlation of $\alpha$ with  alcohol consumption (ALCO). We also show the $p$-value, excluding countries below 5 thousand \$ GDP per capita. In the left bottom panel: same quantities for correlation of $\alpha$ with  ALCO and GDP per capita.  In the right panels: $R^2$ for the  best-estimate and number of countries $N$. We also show the correlation coefficient between the 2 variables in the two-variable fit.}
    \label{tabALCO}
\end{table}

\subsubsection{Smoking}

\begin{figure}[H]
\begin{center}
\vspace*{3mm}
\includegraphics[scale=0.6]{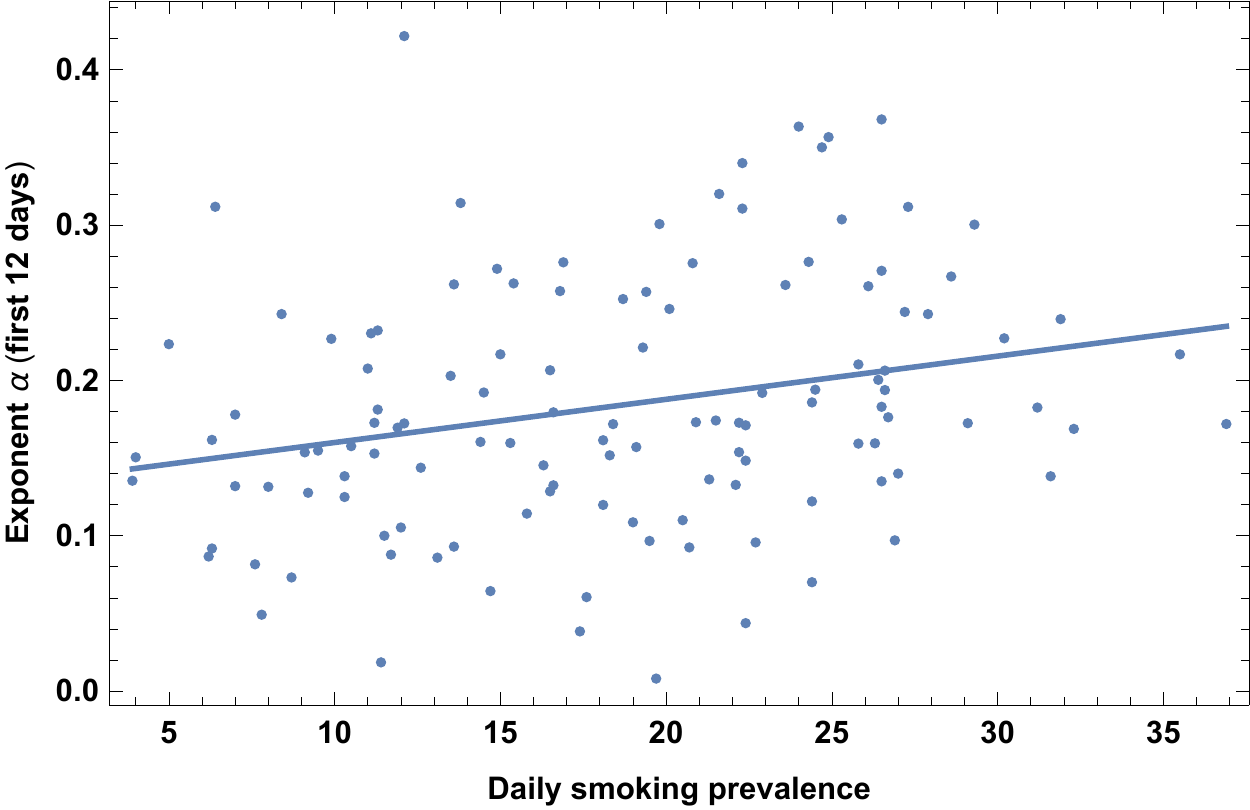} 
\caption{Exponent $\alpha$ for each country vs.~daily smoking prevalence. We show the data points and the best-fit for the linear interpolation. \label{figSMOK}}
\end{center}
\end{figure}

This dataset refers to year 2012.  Results are shown in Fig.~\ref{figSMOK} and Table~\ref{tabSMOK}.
As expected this variable is highly correlated with lung cancer, as discussed in section~\ref{cross}.
We find that COVID-19 spreads more rapidly in countries with higher daily smoking prevalence. Note however that this becomes non-significant when excluding countries below 5K\$ GDP per capita.

Correlation of $\alpha$ with smoking thus could be simply due to correlation with other variables or to a bias due to lack of testing in very poor countries. Alternative interpretations are that smoking has negative effects on conditions of lungs that facilitates contagion or that it contributes to increased transmission of virus from hand to mouth~\cite{QAsmoking}.  Interestingly, note that our finding is in contrast with claims of a possible protective effect of nicotine and smoking against COVID-19~\cite{smoking, smoking2}. 

\begin{table}[H]

\begin{tabular}{cc}
    \begin{minipage}{.5\linewidth}
       \begin{tabular}{|l|c|c|c|c|c|}
       \hline
\text{} & \text{Estimate} & \text{Standard Error} & \text{t-Statistic} & \text{$p$-value} &  \text{$p$-value}, GDP>5K\$  \\
\hline
 1 & 0.132 & 0.0187 & 7.07 & $1.07\times 10^{-10}$ & \\
 \text{SMOK} & 0.00278 & 0.000937 & 2.97 & 0.00361 & 0.19\\
 \hline
     \end{tabular}
     \end{minipage} 
     &
     \hspace*{12em}
    \begin{minipage}{.5\linewidth}
             \begin{tabular}{|l|c|c|c|c|}
  \hline
$R^2$ &  $0.067$\\ \hline
$N$ & 124 \\ 
\hline 

 \end{tabular}
    \end{minipage} 

\end{tabular}

     \vspace*{2em}

\begin{tabular}{cc}
    \begin{minipage}{.5\linewidth}
       \begin{tabular}{|l|c|c|c|c|c|}
       \hline
\text{} & \text{Estimate} & \text{Standard Error} & \text{t-Statistic} & \text{$p$-value} \\
\hline
1 & 0.121 & 0.019 & 6.38 & $3.68\times 10^{-9} $ \\
 \text{GDP} & $1.06\times 10^{-6}$ & $3.89\times 10^{-7}$ & 2.73 & 0.00729 \\
 \text{SMOK} & 0.00209 & 0.000966 & 2.16 & 0.0328 \\
 \hline
     \end{tabular}
     \end{minipage} 
     &
     \hspace*{5em}
    \begin{minipage}{.5\linewidth}
             \begin{tabular}{|l|c|c|c|c|}

\hline
$R^2$ &  $0.122$\\ \hline
$N$ & 121 \\ \hline 
{\text Cross-correlation} &   -0.2646\\ \hline
 \end{tabular}
    \end{minipage} 
\\
%
\end{tabular}

    \caption{In the left top panel: best-estimate, standard error ($\sigma$), t-statistic and $p$-value for the parameters of the linear interpolation,  for correlation of $\alpha$ with  daily smoking prevalence (SMOK). We also show the $p$-value, excluding countries below 5 thousand \$ GDP per capita. In the left bottom panel: same quantities for correlation of $\alpha$ with SMOK and GDP per capita.  In the right panels: $R^2$ for the  best-estimate and number of countries $N$. We also show the correlation coefficient between the 2 variables in the two-variable fit.}
    \label{tabSMOK}
\end{table}

\subsubsection{UV index}

\begin{figure}[H]
\begin{center}
\vspace*{3mm}
\includegraphics[scale=0.6]{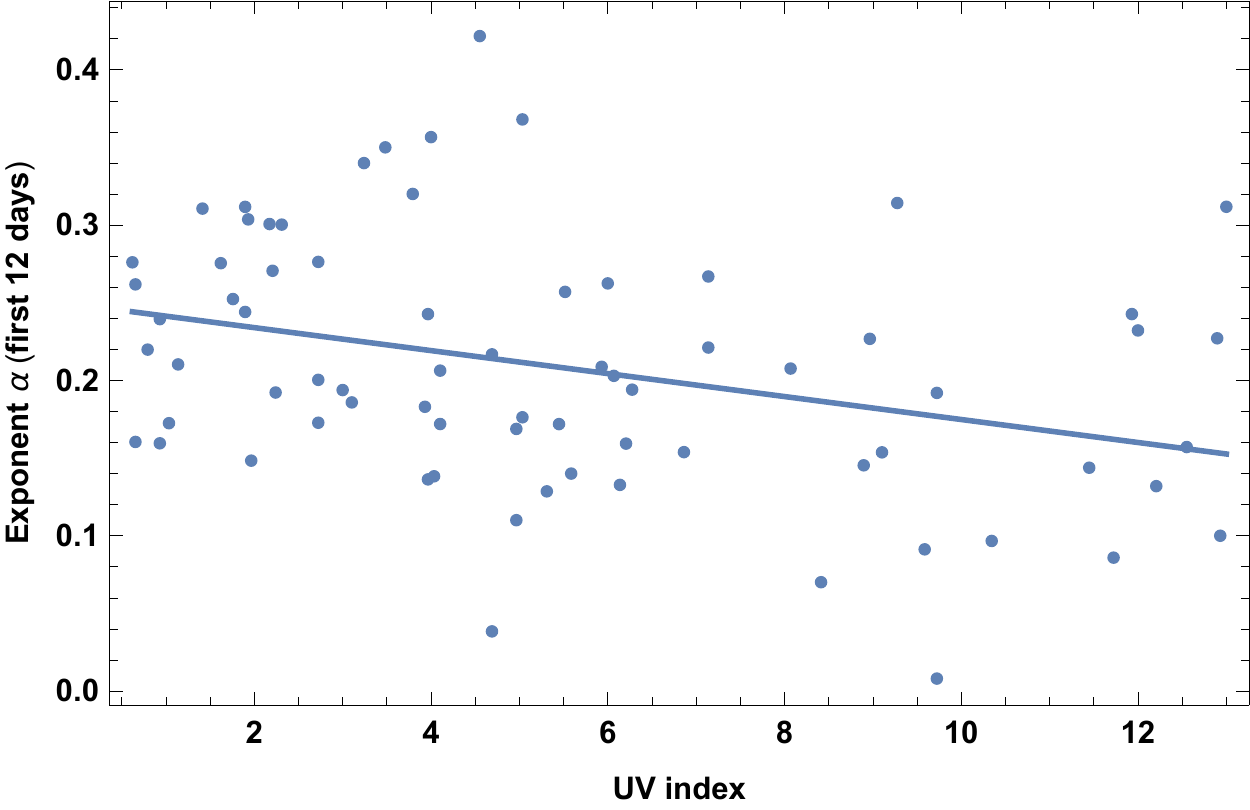} 
\caption{Exponent $\alpha$ for each country vs.~UV index for the relevant month of the epidemic. We show the data points and the best-fit for the linear interpolation. \label{figUV}}
\end{center}
\end{figure}

This is the UV index for the relevant period of time of the epidemic. In particular the UV index has been collected from~\cite{refuv}, as a monthly average, and then with a linear interpolation we have used the average value during the 12 days of the epidemic growth, for each country.  Results are shown in Fig.~\ref{figUV} and Table~\ref{tabUV}. Not surprisingly in section~\ref{cross} we will see that such quantity is very highly correlated with T (correlation coefficient $0.93$). Note also that here the sample size is smaller (73) than in the case of other variables, so it is not strange that the significance of a correlation with $\alpha$ here is not as high as in the case of $\alpha$ with Temperature. More research is required to answer more specific questions, for instance whether the virus survives less in an environment with high UV index, or whether a high UV index stimulates vitamin D production that may help the immune system, or both. Very recent results~\cite{sunlight} consistently find that indeed SARS-CoV-2 is deactivated efficiently under simulated sunlight in few minutes, due to UVB.

\begin{table}[H]

\begin{tabular}{cc}
    \begin{minipage}{.5\linewidth}
       \begin{tabular}{|l|c|c|c|c|c|}
       \hline
\text{} & \text{Estimate} & \text{Standard Error} & \text{t-Statistic} & \text{$p$-value} &  \text{$p$-value}, GDP>5K\$  \\
\hline
 1 & 0.249 & 0.0163 & 15.3 & $3.7\times 10^{-24}$ & \\
 \text{UV} & -0.0074 & 0.00249 & -2.97 & 0.00408 & 0.012\\
 \hline
     \end{tabular}
     \end{minipage} 
     &
     \hspace*{12em}
    \begin{minipage}{.5\linewidth}
             \begin{tabular}{|l|c|c|c|c|}
  \hline
$R^2$ &  $0.110$\\ \hline
$N$ & 73 \\ 
\hline 

 \end{tabular}
    \end{minipage} 

\end{tabular}

     \vspace*{2em}

\begin{tabular}{cc}
    \begin{minipage}{.5\linewidth}
       \begin{tabular}{|l|c|c|c|c|c|}
       \hline
\text{} & \text{Estimate} & \text{Standard Error} & \text{t-Statistic} & \text{$p$-value} \\
\hline
 1 & 0.242 & 0.0269 & 9.01 & $2.88\times 10^{-13}$ \\
 \text{GDP} & $1.92\times 10^{-7}$ & $5.8\times 10^{-7}$ & 0.33 & 0.742 \\
 \text{UV} & -0.00707 & 0.00274 & -2.58 & 0.0119 \\
 \hline
     \end{tabular}
     \end{minipage} 
     &
     \hspace*{6em}
    \begin{minipage}{.5\linewidth}
             \begin{tabular}{|l|c|c|c|c|}

\hline
$R^2$ &  $0.112$\\ \hline
$N$ & 73 \\ \hline 
{\text Cross-correlation} &  0.383\\ \hline
 \end{tabular}
    \end{minipage} 
\\
%
\end{tabular}

    \caption{In the left top panel: best-estimate, standard error ($\sigma$), t-statistic and $p$-value for the parameters of the linear interpolation,  for correlation of $\alpha$ with UV index for the relevant period of time of the epidemic (UV). We also show the $p$-value, excluding countries below 5 thousand \$ GDP per capita. In the left bottom panel: same quantities for correlation of $\alpha$ with UV and GDP per capita.  In the right panels: $R^2$ for the  best-estimate and number of countries $N$. We also show the correlation coefficient between the 2 variables in the two-variable fit.}
    \label{tabUV}
\end{table}

\subsubsection{Vitamin D serum concentration}

Another relevant variable is the amount of serum Vitamin D.
We collected data in the literature for the average annual level of serum Vitamin D and for the seasonal level (D$_s$). The seasonal level is defined as: the amount during the month of March or during winter for northern hemisphere,  {\it or}  during summer for southern hemisphere  {\it or}  the annual level for countries with little seasonal variation.  The dataset for the annual D was built with the available literature, which is unfortunately quite inhomogeneous as discussed in Appendix~\ref{AppD}. For many countries several studies with quite different values were found and in this case we have collected the mean and the standard error and a weighted average has been performed. The countries included in this dataset are 50, as specified in Appendix. The dataset for the seasonal levels is more restricted, since the relative literature is less complete, and we have included 42 countries.  

\begin{figure}[t!]
\begin{center}
\vspace*{3mm}
\includegraphics[scale=0.6]{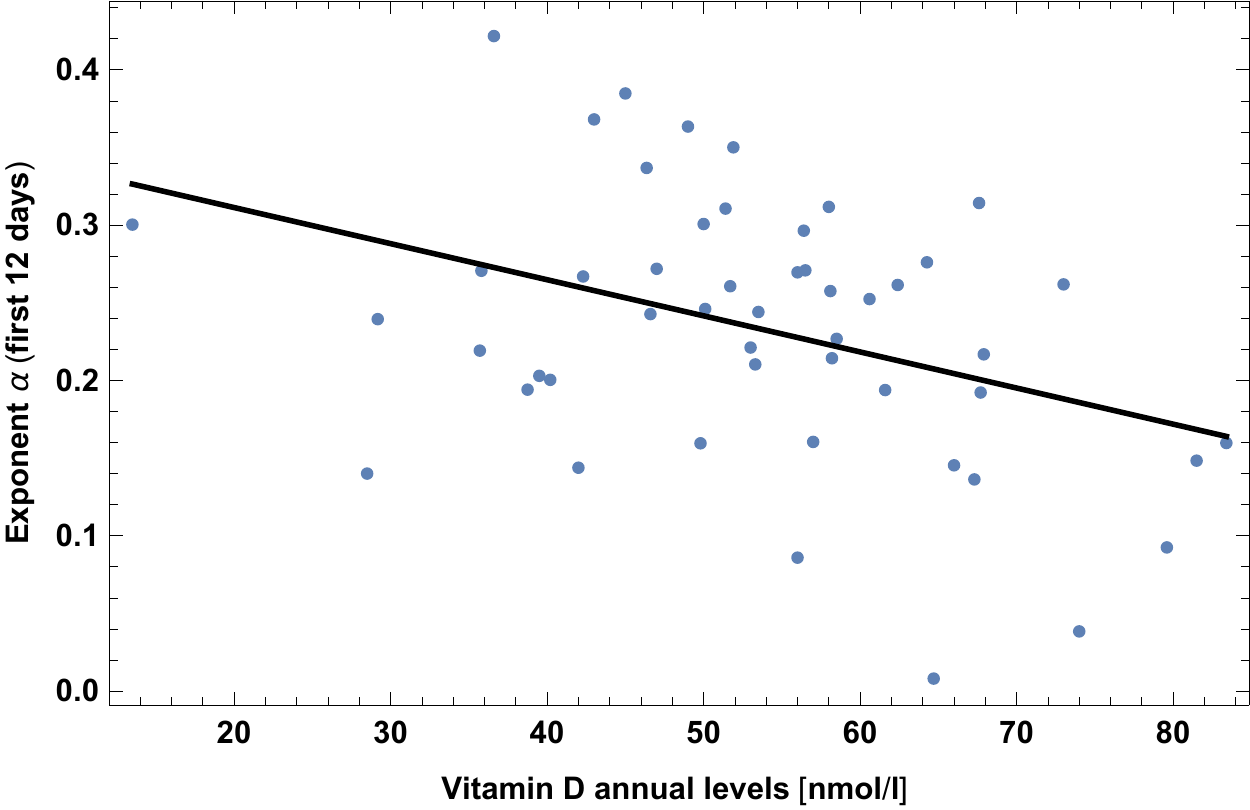} 
\caption{Exponent $\alpha$ for each country vs.~annual levels of vitamin $D$, for the relevant period of time, as defined in the text,  for the base set of 42 countries. We show the data points and the best-fit for the linear interpolation. \label{figvitD}}
\end{center}
\end{figure}

Results are shown in Fig.~\ref{figvitD} and Table~\ref{tabvitD} for the annual levels and in Fig.~\ref{figvitD} and Table~\ref{tabvitD} for the seasonal levels. Note however that such results are preliminary and based on inhomogeneous data, and have to be confirmed on a larger sample.

Interestingly, in section~\ref{cross} we will see that D is {\it not} highly correlated with T or UV index, as one naively could expect, due to different food consumption in different countries. A slightly higher correlation, as it should be, is present between T and D$_s$. Note that our results are in agreement with the fact that increased vitamin D levels have been proposed to have a protective effect against COVID-19~\cite{vitd6, vitdCOVID,vitdCOVID2}.

%
%
%
%

\begin{table}[H]

\begin{tabular}{cc}
    \begin{minipage}{.5\linewidth}
       \begin{tabular}{|l|c|c|c|c|c|}
       \hline
\text{} & \text{Estimate} & \text{Standard Error} & \text{t-Statistic} & \text{$p$-value} &  \text{$p$-value}, GDP>5K\$  \\
\hline
1 & 0.356 & 0.0443 & 8.02 & $2.02\times 10^{-10}$ & \\
 \text{D} & -0.00231 & 0.000801 & -2.88 & 0.00586 & 0.0059\\\hline
     \end{tabular}
     \end{minipage} 
     &
     \hspace*{12em}
    \begin{minipage}{.5\linewidth}
             \begin{tabular}{|l|c|c|c|c|}
  \hline
$R^2$ &  $0.147$ \\ \hline
$N$ & 50 \\ \hline

 \end{tabular}
    \end{minipage} 

\end{tabular}

     \vspace*{2em}

\begin{tabular}{cc}
    \begin{minipage}{.5\linewidth}
       \begin{tabular}{|l|c|c|c|c|c|}
       \hline
\text{} & \text{Estimate} & \text{Standard Error} & \text{t-Statistic} & \text{$p$-value} \\
\hline
1 & 0.342 & 0.0457 & 7.48 & $1.52\times 10^{-9}$ \\
 \text{GDP} & $8.29\times 10^{-7}$ & $6.99\times 10^{-7}$ & 1.19 & 0.242 \\
 \text{D} & -0.00255 & 0.000822 & -3.1 & 0.00328 \\
 \hline
     \end{tabular}
     \end{minipage} 
     &
     \hspace*{5.5em}
    \begin{minipage}{.5\linewidth}
             \begin{tabular}{|l|c|c|c|c|}

\hline
$R^2 $ &  $0.172$ \\ \hline
$N$ & 50 \\ \hline
{\text Cross-correlation} &  -0.243\\ \hline

 \end{tabular}
    \end{minipage} 

\end{tabular}

    \caption{In the left top panel: best-estimate, standard error ($\sigma$), t-statistic and $p$-value for the parameters of the linear interpolation,  for correlation of $\alpha$ with mean annual levels of vitamin D (variable name: D). We also show the $p$-value, excluding countries below 5 thousand \$ GDP per capita. In the left bottom panel: same quantities for correlation of $\alpha$ with D and GDP per capita. In the right panels: $R^2$ for the  best-estimate and number of countries $N$. We also show the correlation coefficient between the 2 variables in the two-variable fit.}
    \label{tabvitD}
\end{table}

%
%
%
%

\begin{figure}[H]
\begin{center}
\vspace*{3mm}
\includegraphics[scale=0.6]{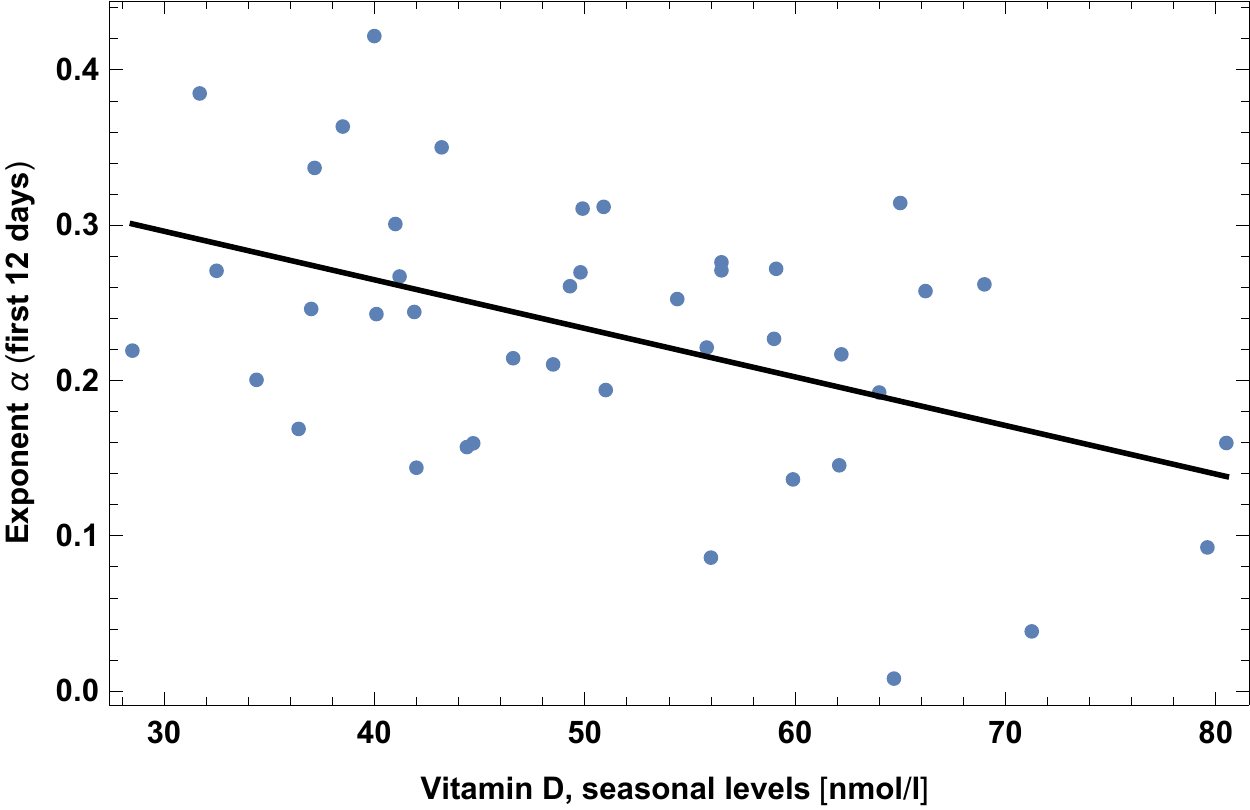} 
\caption{Exponent $\alpha$ for each country vs.~seasonal levels of vitamin D, for the relevant period of time, as defined in the text,  for the base set of 42 countries. We show the data points and the best-fit for the linear interpolation. \label{figvitDs}}
\end{center}
\end{figure}

%
%

\begin{table}[H]

\begin{tabular}{cc}
    \begin{minipage}{.5\linewidth}
       \begin{tabular}{|l|c|c|c|c|c|}
       \hline
\text{} & \text{Estimate} & \text{Standard Error} & \text{t-Statistic} & \text{$p$-value}   &  \text{$p$-value}, GDP>5K\$  \\
\hline
 1 & 0.385 & 0.0499 & 7.72 & $1.91\times 10^{-9}$ & \\
 $\text{D}_s$ & -0.00305 & 0.000949 & -3.22 & 0.00256 & 0.0024 \\
\hline
     \end{tabular}
     \end{minipage} 
     &
     \hspace*{12em}
    \begin{minipage}{.5\linewidth}
             \begin{tabular}{|l|c|c|c|c|}
  \hline
$R^2$ &  $0.206$ \\ \hline
$N$ &42 \\ \hline

 \end{tabular}
    \end{minipage} 

\end{tabular}

     \vspace*{2em}

\begin{tabular}{cc}
    \begin{minipage}{.5\linewidth}
       \begin{tabular}{|l|c|c|c|c|c|}
       \hline
\text{} & \text{Estimate} & \text{Standard Error} & \text{t-Statistic} & \text{$p$-value} \\
\hline
 1 & 0.375 & 0.0533 & 7.03 & $1.94\times 10^{-8}$ \\
 \text{GDP} & $4.66\times 10^{-7}$ & $8.02\times 10^{-7}$ & 0.581 & 0.565 \\
 $\text{D}_s$ & -0.00314 & 0.00097 & -3.24 & 0.00243 \\
 \hline
     \end{tabular}
     \end{minipage} 
     &
     \hspace*{5.5em}
    \begin{minipage}{.5\linewidth}
             \begin{tabular}{|l|c|c|c|c|}

\hline
$R^2 $ &  $0.212$ \\ \hline
$N$ & 42 \\ \hline
{\text Cross-correlation} &  -0.162\\ \hline

 \end{tabular}
    \end{minipage} 
\\
\\

\end{tabular}

    \caption{In the left top panel: best-estimate, standard error ($\sigma$), t-statistic and $p$-value for the parameters of the linear interpolation,  for correlation of $\alpha$ with seasonal levels of vitamin D (variable name: $\text{D}_s$). We also show the $p$-value, excluding countries below 5 thousand \$ GDP per capita. In the left bottom panel: same quantities for correlation of $\alpha$ with $\text{D}_s$ and GDP per capita. In the right panels: $R^2$ for the  best-estimate and number of countries $N$. We also show the correlation coefficient between the 2 variables in the two-variable fit.}
    \label{tabvitDs}
\end{table}

%
%
%

\subsubsection{$\text {CO}_2$ Emissions}

\begin{figure}[H]
\begin{center}
\vspace*{3mm}
\includegraphics[scale=0.6]{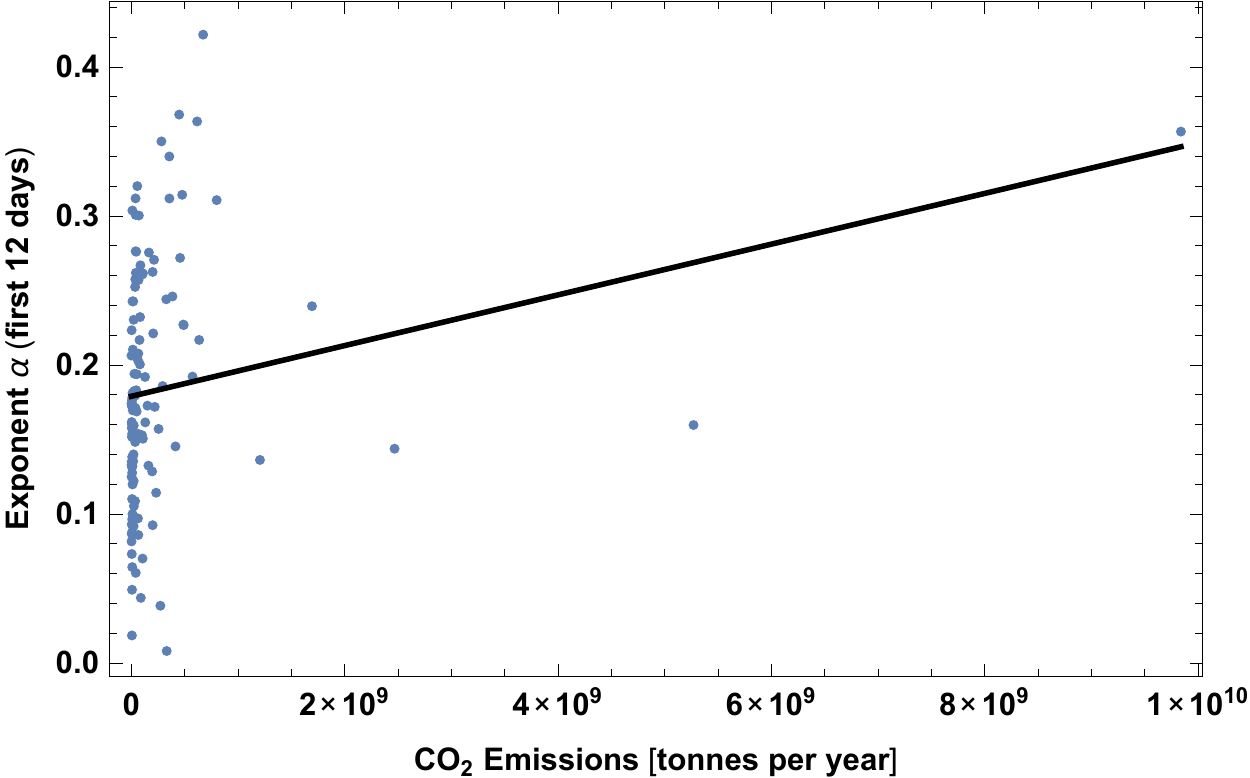} 
\caption{Exponent $\alpha$ for each country vs.~$\text{CO}_2$ emissions. We show the data points and the best-fit for the linear interpolation. \label{figCO2}}
\end{center}
\end{figure}

This is the data for year 2017. We have also checked that this has very high correlation with SO emissions (about 0.9 correlation coefficient). We show here only the case for $\text{CO}_2$, but the reader should keep in mind that a very similar result applies also to SO emissions. Note also that this is expected to have a high correlation with the number of international tourist arrivals, as we will show in section~\ref{cross}. Results are shown in Fig.~\ref{figCO2} and Table~\ref{tabCO2}. 

\begin{table}[H]

\begin{tabular}{cc}
    \begin{minipage}{.5\linewidth}
       \begin{tabular}{|l|c|c|c|c|c|}
       \hline
\text{} & \text{Estimate} & \text{Standard Error} & \text{t-Statistic} & \text{$p$-value} &  \text{$p$-value}, GDP>5K\$  \\
\hline
 1 & 0.18 & 0.0073 & 24. & $9.6\times 10^{-49}$ & \\
 $\text {CO}_2$ & $1.7\times 10^{-11}$ & $6.9\times 10^{-12}$ & 2.5 & 0.015 & 0.048 \\
\hline
     \end{tabular}
     \end{minipage} 
     &
     \hspace*{12em}
    \begin{minipage}{.5\linewidth}
             \begin{tabular}{|l|c|c|c|c|}
  \hline
$R^2$ &  $0.048$\\ \hline
$N$ & 110 \\ 
\hline 

 \end{tabular}
    \end{minipage} 

\end{tabular}

     \vspace*{2em}

\begin{tabular}{cc}
    \begin{minipage}{.5\linewidth}
       \begin{tabular}{|l|c|c|c|c|c|}
       \hline
\text{} & \text{Estimate} & \text{Standard Error} & \text{t-Statistic} & \text{$p$-value} \\
\hline
1 & 0.152 & 0.011 & 13.8 & $1.82\times 10^{-26}$ \\
 \text{GDP} & $1.23\times 10^{-6}$ & $3.75\times 10^{-7}$ & 3.29 & 0.00133 \\
 \text{CO2} & $1.57\times 10^{-11}$ & $6.74\times 10^{-12}$ & 2.33 & 0.0214 \\
  \hline
     \end{tabular}
     \end{minipage} 
     &
     \hspace*{5em}
    \begin{minipage}{.5\linewidth}
             \begin{tabular}{|l|c|c|c|c|}

\hline
$R^2$ &  $0.126$\\ \hline
$N$ & 109 \\ \hline 
{\text Cross-correlation} &  -0.0321\\ \hline
 \end{tabular}
    \end{minipage} 
\\

\end{tabular}

    \caption{In the left top panel: best-estimate, standard error ($\sigma$), t-statistic and $p$-value for the parameters of the linear interpolation,  for correlation of $\alpha$ with $\text{CO}_2$ emissions. We also show the $p$-value, excluding countries below 5 thousand \$ GDP per capita. In the left bottom panel: same quantities for correlation of $\alpha$ with $\text{CO}_2$ and GDP per capita.   In the right panels: $R^2$ for the  best-estimate and number of countries $N$. We also show the correlation coefficient between the 2 variables in the two-variable fit.}
    \label{tabCO2}
\end{table}

\subsubsection{Prevalence of type-1 Diabetes}

\begin{figure}[H]
\begin{center}
\vspace*{3mm}
\includegraphics[scale=0.6]{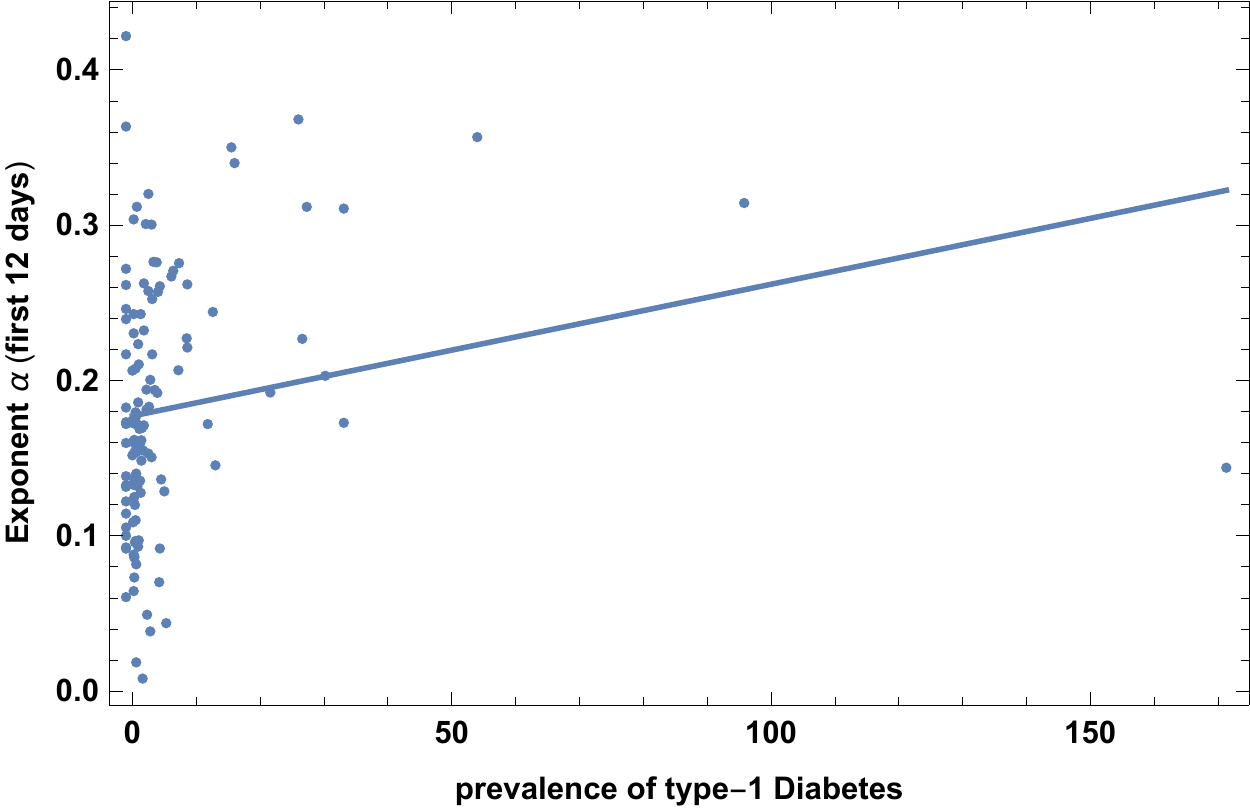} 
\caption{Exponent $\alpha$ for each country vs.~prevalence of type-1 Diabetes. We show the data points and the best-fit for the linear interpolation. \label{figDIAB}}
\end{center}
\end{figure}

This is the prevalence of  type-1 Diabetes in children, 0-19 years-old, taken from~\cite{diabetes}. Results are shown in Fig.~\ref{figDIAB} and Table~\ref{tabDIAB}.  Note however that significance becomes very small when restricting to countries with GDP per capita larger than 5K\$.  Note also that in the case of diabetes of {\it any} kind we do {\it not} find a correlation with COVID-19. Such correlation, even if not highly significant, could be non-trivial and could constitute useful information for clinical and genetic research.  See also~\cite{diabetes2}.

\begin{table}[H]

\begin{tabular}{cc}
    \begin{minipage}{.5\linewidth}
       \begin{tabular}{|l|c|c|c|c|c|}
       \hline
\text{} & \text{Estimate} & \text{Standard Error} & \text{t-Statistic} & \text{$p$-value} &  \text{$p$-value}, GDP>5K\$  \\
\hline
 1 & 0.177 & 0.00793 & 22.4 & $2.35\times 10^{-41}$ & \\
 \text{DIAB} & 0.00087 & 0.000367 & 2.37 & 0.0198 & 0.072\\
\hline
     \end{tabular}
     \end{minipage} 
     &
     \hspace*{12em}
    \begin{minipage}{.5\linewidth}
             \begin{tabular}{|l|c|c|c|c|}
  \hline
$R^2$ &  $0.052$\\ \hline
$N$ & 105 \\ 
\hline 

 \end{tabular}
    \end{minipage} 

\end{tabular}

     \vspace*{2em}

\begin{tabular}{cc}
    \begin{minipage}{.5\linewidth}
       \begin{tabular}{|l|c|c|c|c|c|}
       \hline
\text{} & \text{Estimate} & \text{Standard Error} & \text{t-Statistic} & \text{$p$-value} \\
\hline
 1 & 0.142 & 0.0116 & 12.3 & $1.52\times 10^{-21}$ \\
 \text{GDP} & $1.58\times 10^{-6}$ & $3.88\times 10^{-7}$ & 4.08 & 0.000092 \\
 \text{DIAB} & 0.000932 & 0.000349 & 2.67 & 0.00889 \\
 \hline
     \end{tabular}
     \end{minipage} 
     &
     \hspace*{5.5em}
    \begin{minipage}{.5\linewidth}
             \begin{tabular}{|l|c|c|c|c|}

\hline
$R^2$ &  $0.189$\\ \hline
$N$ & 101 \\ \hline 
{\text Cross-correlation} &  0.0415\\ \hline
 \end{tabular}
    \end{minipage} 

\end{tabular}

    \caption{In the left top panel: best-estimate, standard error ($\sigma$), t-statistic and $p$-value for the parameters of the linear interpolation,  for correlation of $\alpha$ with prevalence of type-1 Diabetes, DIA. We also show the $p$-value, excluding countries below 5 thousand \$ GDP per capita. In the left bottom panel: same quantities for correlation of $\alpha$ with DIA and GDP per capita.   In the right panels: $R^2$ for the  best-estimate and number of countries $N$. We also show the correlation coefficient between the 2 variables in the two-variable fit.}
    \label{tabDIAB}
\end{table}

\subsubsection{Tuberculosis (BCG) vaccination coverage}

\begin{figure}[H]
\begin{center}
\vspace*{3mm}
\includegraphics[scale=0.6]{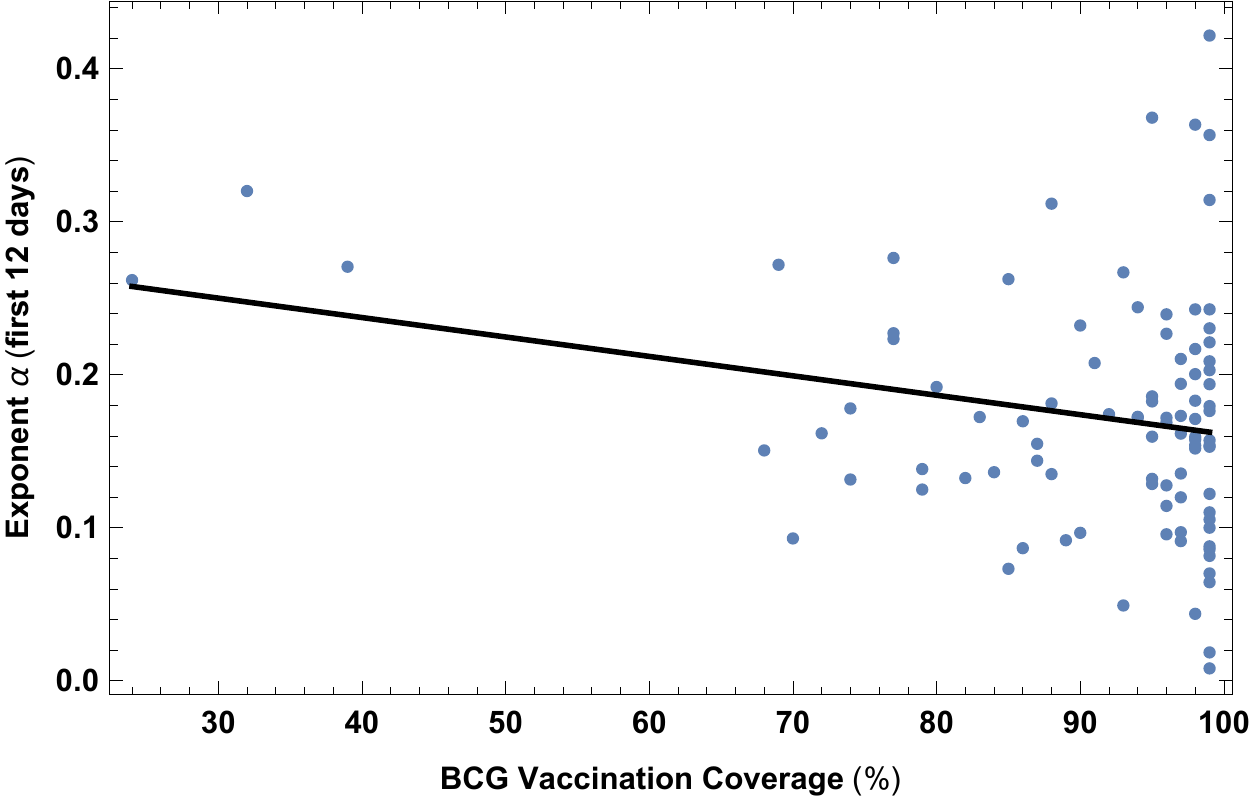} 
\caption{Exponent $\alpha$ for each country vs.~BCG vaccination coverage. We show the data points and the best-fit for the linear interpolation. \label{figBCG}}
\end{center}
\end{figure}

This dataset  is the vaccination coverage for tuberculosis for year 2015~\footnote{Countries were taken from~\cite{worlddata}, plus Taiwan added from~\cite{taiwan}.}. Results are shown in Fig.~\ref{figBCG} and Table~\ref{tabBCG}. We find a negative correlation. Note  that this depends mostly on the three countries with low coverage present in the plot: by excluding them the correlation becomes non-significant. Note also that several countries with no compulsory vaccination (such as USA or Italy) do not have an estimate for BCG coverage and were not included in the analysis. Including them, with very low coverage, would probably affect a lot the significance.  This correlation, even if not highly significant and to be confirmed by more data, is also quite non-trivial and could be useful information for clinical and genetic research (see also~\cite{bcg1,bcg2,bcg3}) and even for vaccine development~\cite{bcgvaccine,bcgvaccine2,bcgvaccine3}.

%
%
%
%

\begin{table}[H]

\begin{tabular}{cc}
    \begin{minipage}{.5\linewidth}
       \begin{tabular}{|l|c|c|c|c|c|}
       \hline
\text{} & \text{Estimate} & \text{Standard Error} & \text{t-Statistic} & \text{$p$-value} &  \text{$p$-value}, GDP>5K\$  \\
\hline
 1 & 0.289 & 0.053 & 5.46 & $3.99\times 10^{-7}$ & \\
 \text{BCG} & -0.00129 & 0.000579 & -2.22 & 0.0286 & 0.011\\
\hline
     \end{tabular}
     \end{minipage} 
     &
     \hspace*{12em}
    \begin{minipage}{.5\linewidth}
             \begin{tabular}{|l|c|c|c|c|}
  \hline
$R^2$ &  $0.051$ \\ \hline
$N$ &94 \\ \hline

 \end{tabular}
    \end{minipage} 

\end{tabular}

     \vspace*{2em}

\begin{tabular}{cc}
    \begin{minipage}{.5\linewidth}
       \begin{tabular}{|l|c|c|c|c|c|}
       \hline
\text{} & \text{Estimate} & \text{Standard Error} & \text{t-Statistic} & \text{$p$-value} \\
\hline
1 & 0.28 & 0.0534 & 5.24 & $1.07\times 10^{-6}$ \\
 \text{GDP} & $8.53\times 10^{-7} $ & $4.79\times 10^{-7} $& 1.78 & 0.0781 \\
 \text{BCG} & -0.00134 & 0.00058 & -2.3 & 0.0235 \\
 \hline
     \end{tabular}
     \end{minipage} 
     &
     \hspace*{5em}
    \begin{minipage}{.5\linewidth}
             \begin{tabular}{|l|c|c|c|c|}

\hline
$R^2 $ &  $0.084$ \\ \hline
$N$ & 92 \\ \hline
{\text Cross-correlation} &  -0.0367 \\ \hline
 \end{tabular}
    \end{minipage} 
\\

\end{tabular}

    \caption{In the left top panel: best-estimate, standard error ($\sigma$), t-statistic and $p$-value for the parameters of the linear interpolation,  for correlation of $\alpha$ with BCG vaccination coverage (BCG). We also show the $p$-value, excluding countries below 5 thousand \$ GDP per capita. In the left bottom panel: same quantities for correlation of $\alpha$ with BCG and GDP per capita. In the right panels: $R^2$ for the  best-estimate and number of countries $N$. We also show the correlation coefficient between the 2 variables in the two-variable fit.}
    \label{tabBCG}
\end{table}

%
%
%
%
%

%
%

%
%
%

\subsection{``Counterintuitive'' correlations} \label{counter}
We show here other correlations that are somehow counterintuitive, since they go in the opposite direction than from a naive expectation. We will try to interpret these results in section~\ref{cross}.

\subsubsection{Death rate from air pollution}

\begin{figure}[H]
\begin{center}
\vspace*{3mm}
\includegraphics[scale=0.6]{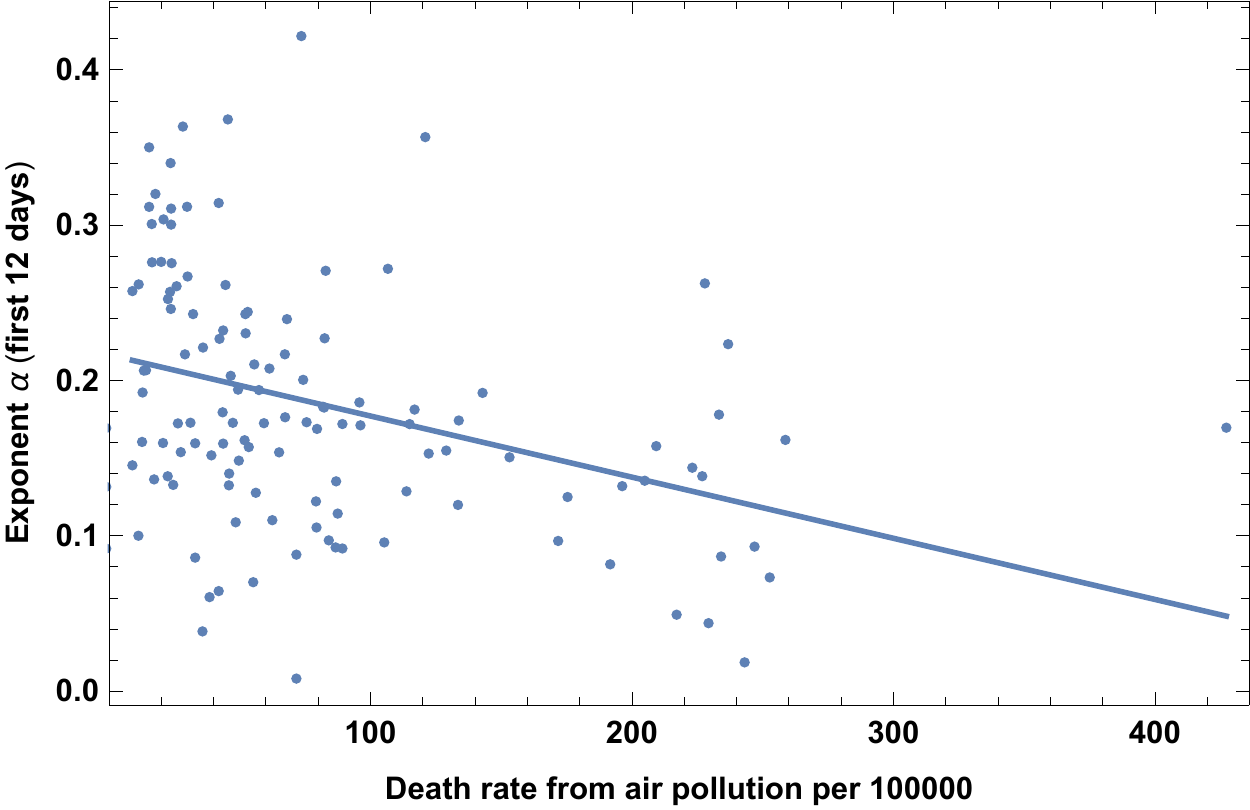} 
\caption{Exponent $\alpha$ for each country vs.~death rate from air pollution per 100000. We show the data points and the best-fit for the linear interpolation. \label{figPOLL}}
\end{center}
\end{figure}

This dataset is for year 2015.   Results are shown in Fig.~\ref{figPOLL} and Table~\ref{tabPOLL}. Contrary to naive expectations and to claims in the opposite direction~\cite{poll}, we find that countries with larger death rate from air pollution actually have {\it slower} COVID-19 contagion.

\begin{table}[H]

\begin{tabular}{cc}
    \begin{minipage}{.5\linewidth}
       \begin{tabular}{|l|c|c|c|c|c|}
       \hline
\text{} & \text{Estimate} & \text{Standard Error} & \text{t-Statistic} & \text{$p$-value}  &  \text{$p$-value}, GDP>5K\$   \\
\hline
1 & 0.216 & 0.0102 & 21.3 & $8.97\times 10^{-43} $ & \\
 \text{POLL} & -0.000394 & 0.0000917 & -4.29 & 0.0000357 & 0.040 \\
\hline
     \end{tabular}
     \end{minipage} 
     &
     \hspace*{12em}
    \begin{minipage}{.5\linewidth}
             \begin{tabular}{|l|c|c|c|c|}
  \hline
$R^2$ &  $0.132$\\ \hline
$N$ & 123 \\  
\hline 

 \end{tabular}
    \end{minipage} 

\end{tabular}

     \vspace*{2em}

\begin{tabular}{cc}
    \begin{minipage}{.5\linewidth}
       \begin{tabular}{|l|c|c|c|c|c|}
       \hline
\text{} & \text{Estimate} & \text{Standard Error} & \text{t-Statistic} & \text{$p$-value} \\
\hline
1 & 0.211 & 0.0204 & 10.3 & $3.75\times 10^{-18}$ \\
 \text{GDP} & $ 3.13\times 10^{-7}$ & $4.77\times 10^{-7}$ & 0.655 & 0.514 \\
 \text{POLL} & -0.000418 & 0.000131 & -3.19 & 0.00185 \\
 \hline
     \end{tabular}
     \end{minipage} 
     &
     \hspace*{6em}
    \begin{minipage}{.5\linewidth}
             \begin{tabular}{|l|c|c|c|c|}

\hline
$R^2$ &  $0.158$\\ \hline
$N$ & 120 \\ \hline 
{\text Cross-correlation} & 0.630 \\ \hline
 \end{tabular}
    \end{minipage} 
\\

\end{tabular}

    \caption{In the left top panel: best-estimate, standard error ($\sigma$), t-statistic and $p$-value for the parameters of the linear interpolation,  for correlation of $\alpha$ with death rate from air pollution per 100000 (POLL). We also show the $p$-value, excluding countries below 5 thousand \$ GDP per capita. In the left bottom panel: same quantities for correlation of $\alpha$ with POLL and GDP per capita. In the right panels: $R^2$ for the  best-estimate and number of countries $N$. We also show the correlation coefficient between the 2 variables in the two-variable fit.}
    \label{tabPOLL}
\end{table}

\subsubsection{High blood pressure in {\it females}}

\begin{figure}[H]
\begin{center}
\vspace*{3mm}
\includegraphics[scale=0.6]{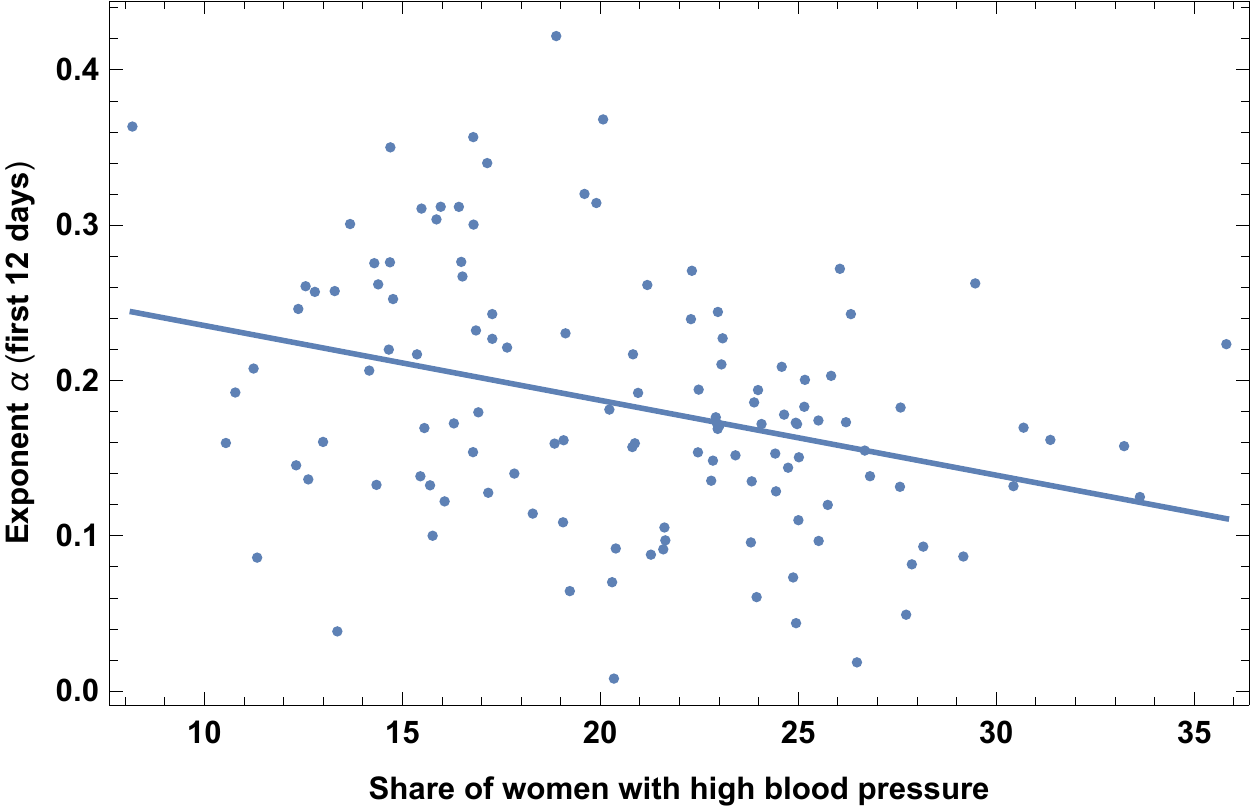} 
\caption{Exponent $\alpha$ for each country vs.~share of women with high blood pressure. We show the data points and the best-fit for the linear interpolation. \label{figPRE}}
\end{center}
\end{figure}

This dataset is for year 2015.   Results are shown in Fig.~\ref{figPRE} and Table~\ref{tabPRE}.  Countries with larger share of high blood pressure in females have {\it slower} COVID-19 contagion. Note that we do {\it not} find a significant correlation instead with high blood pressure in {\it males}.

\begin{table}[H]

\begin{tabular}{cc}
    \begin{minipage}{.5\linewidth}
       \begin{tabular}{|l|c|c|c|c|c|}
       \hline
\text{} & \text{Estimate} & \text{Standard Error} & \text{t-Statistic} & \text{$p$-value} &  \text{$p$-value}, GDP>5K\$  \\
\hline
 1 & 0.284 & 0.0265 & 10.7 & $2.76\times 10^{-19}$ & \\
 \text{PRE} & -0.00482 & 0.00124 & -3.89 & 0.000164 &  0.028 \\
\hline
     \end{tabular}
     \end{minipage} 
     &
     \hspace*{12em}
    \begin{minipage}{.5\linewidth}
             \begin{tabular}{|l|c|c|c|c|}
  \hline
$R^2$ &  $0.109$\\ \hline
$N$ & 125 \\  
\hline 

 \end{tabular}
    \end{minipage} 

\end{tabular}

     \vspace*{2em}

\begin{tabular}{cc}
    \begin{minipage}{.5\linewidth}
       \begin{tabular}{|l|c|c|c|c|c|}
       \hline
\text{} & \text{Estimate} & \text{Standard Error} & \text{t-Statistic} & \text{$p$-value} \\
\hline
 1 & 0.248 & 0.0432 & 5.75 & $7.2\times 10^{-8} $ \\
 \text{GDP} & $5.85\times 10^{-7}$ & $4.89\times 10^{-7}$ & 1.2 & 0.234 \\
 \text{PRE} & -0.00372 & 0.00168 & -2.22 & 0.0284 \\
 \hline
     \end{tabular}
     \end{minipage} 
     &
     \hspace*{6em}
    \begin{minipage}{.5\linewidth}
             \begin{tabular}{|l|c|c|c|c|}

\hline
$R^2$ &  $0.123$\\ \hline
$N$ & 121 \\ \hline 
{\text Cross-correlation} & 0.642 \\ \hline
 \end{tabular}
    \end{minipage} 
\\

\end{tabular}

    \caption{In the left top panel: best-estimate, standard error ($\sigma$), t-statistic and $p$-value for the parameters of the linear interpolation,  for correlation of $\alpha$ with share of women with high blood pressure (PRE). We also show the $p$-value, excluding countries below 5 thousand \$ GDP per capita. In the left bottom panel: same quantities for correlation of $\alpha$ with PRE and GDP per capita. In the right panels: $R^2$ for the  best-estimate and number of countries $N$. We also show the correlation coefficient between the 2 variables in the two-variable fit.}
    \label{tabPRE}
\end{table}

\subsubsection{Hepatitis B incidence rate}

\begin{figure}[H]
\begin{center}
\vspace*{3mm}
\includegraphics[scale=0.6]{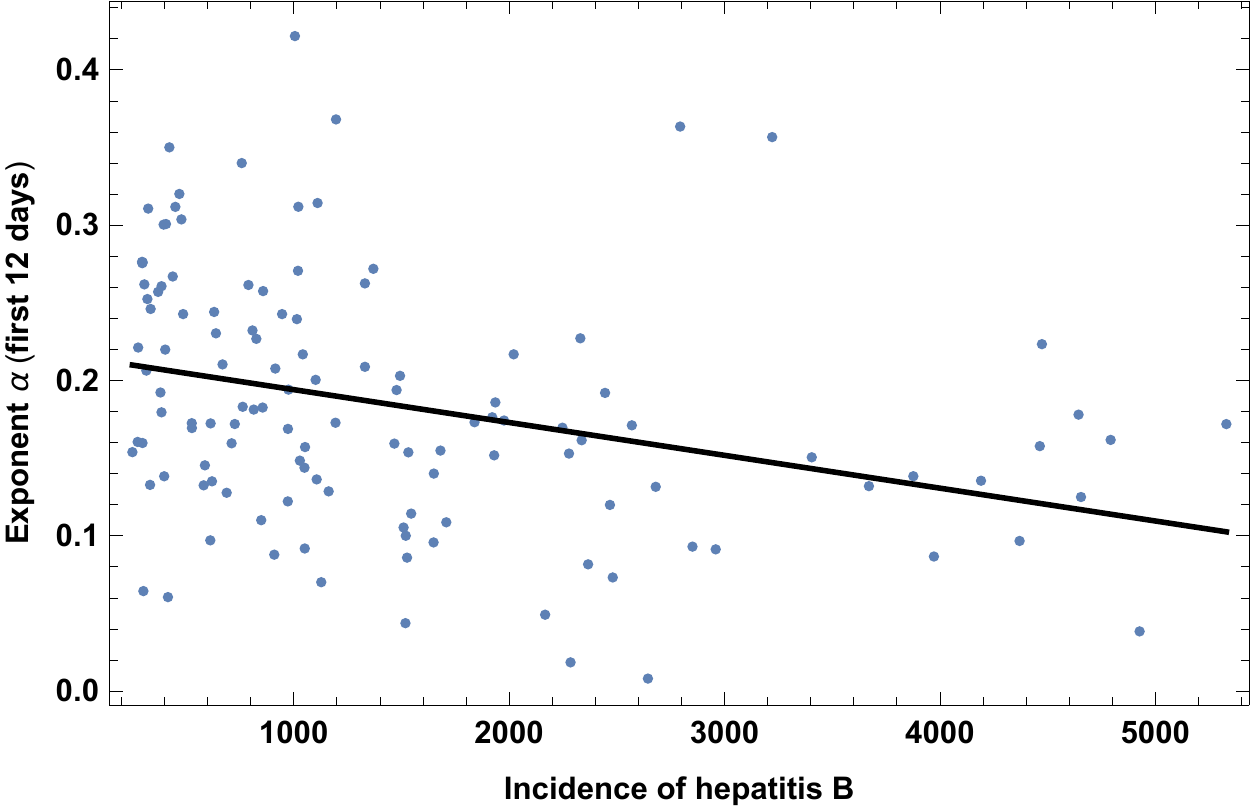} 
\caption{Exponent $\alpha$ for each country vs.~incidence of Hepatitis B, for the relevant period of time, as defined in the text,  for the base set of 42 countries. We show the data points and the best-fit for the linear interpolation. \label{figHEP}}
\end{center}
\end{figure}

This is the incidence of hepatitis B, measured as the number of new cases of hepatitis B per 100,000
individuals in a given population, for the year 2015. Results are shown in Fig.~\ref{figHEP} and Table~\ref{tabHEP}.  Countries with higher incidence of hepatitis B have {\it slower} contagion of COVID-19.  

\begin{table}[H]

\begin{tabular}{cc}
    \begin{minipage}{.5\linewidth}
       \begin{tabular}{|l|c|c|c|c|c|}
       \hline
\text{} & \text{Estimate} & \text{Standard Error} & \text{t-Statistic} & \text{$p$-value} & \text{$p$-value}, GDP>5K\$ \\
\hline
  1 & 0.215 & 0.0107 & 20.1 & $1.16\times 10^{-40}$ & \\
 \text{HEP} & -0.000021 & $5.56\times 10^{-6}$ & -3.78 & 0.000243 & 0.016\\
\hline
     \end{tabular}
     \end{minipage} 
     &
     \hspace*{12em}
    \begin{minipage}{.5\linewidth}
             \begin{tabular}{|l|c|c|c|c|}
  \hline
$R^2$ &  $0.104$\\ \hline
$N$ & 125 \\  
\hline 

 \end{tabular}
    \end{minipage} 

\end{tabular}

     \vspace*{2em}

\begin{tabular}{cc}
    \begin{minipage}{.5\linewidth}
       \begin{tabular}{|l|c|c|c|c|c|}
       \hline
\text{} & \text{Estimate} & \text{Standard Error} & \text{t-Statistic} & \text{$p$-value} \\
\hline
1 & 0.189 & 0.017 & 11.1 & $5.02\times 10^{-20}$ \\
 \text{GDP} & $8.39\times 10^{-7}$ & $4.1\times 10^{-7}$ & 2.05 & 0.0429 \\
 \text{HEP} & -0.0000161 & $6.21\times 10^{-6}$ & -2.58 & 0.011 \\
 \hline
     \end{tabular}
     \end{minipage} 
     &
     \hspace*{6em}
    \begin{minipage}{.5\linewidth}
             \begin{tabular}{|l|c|c|c|c|}

\hline
$R^2$ &  $0.136$\\ \hline
$N$ & 121 \\ \hline 
 {\text Cross-correlation} & 0.420\\ \hline
 \end{tabular}
    \end{minipage} 
\\
\\

\end{tabular}

    \caption{In the left top panel: best-estimate, standard error ($\sigma$), t-statistic and $p$-value for the parameters of the linear interpolation,  for correlation of $\alpha$ with  incidence of Hepatitis B (HEP).  We also show the $p$-value, excluding countries below 5 thousand \$ GDP per capita. In the left bottom panel: same quantities for correlation of $\alpha$ with HEP and GDP per capita. In the right panels: $R^2$ for the  best-estimate and number of countries $N$. We also show the correlation coefficient between the 2 variables in the two-variable fit.}
    \label{tabHEP}
\end{table}

\subsubsection{Prevalence of Anemia}

\begin{figure}[H]
\begin{center}
\vspace*{3mm}
\includegraphics[scale=0.6]{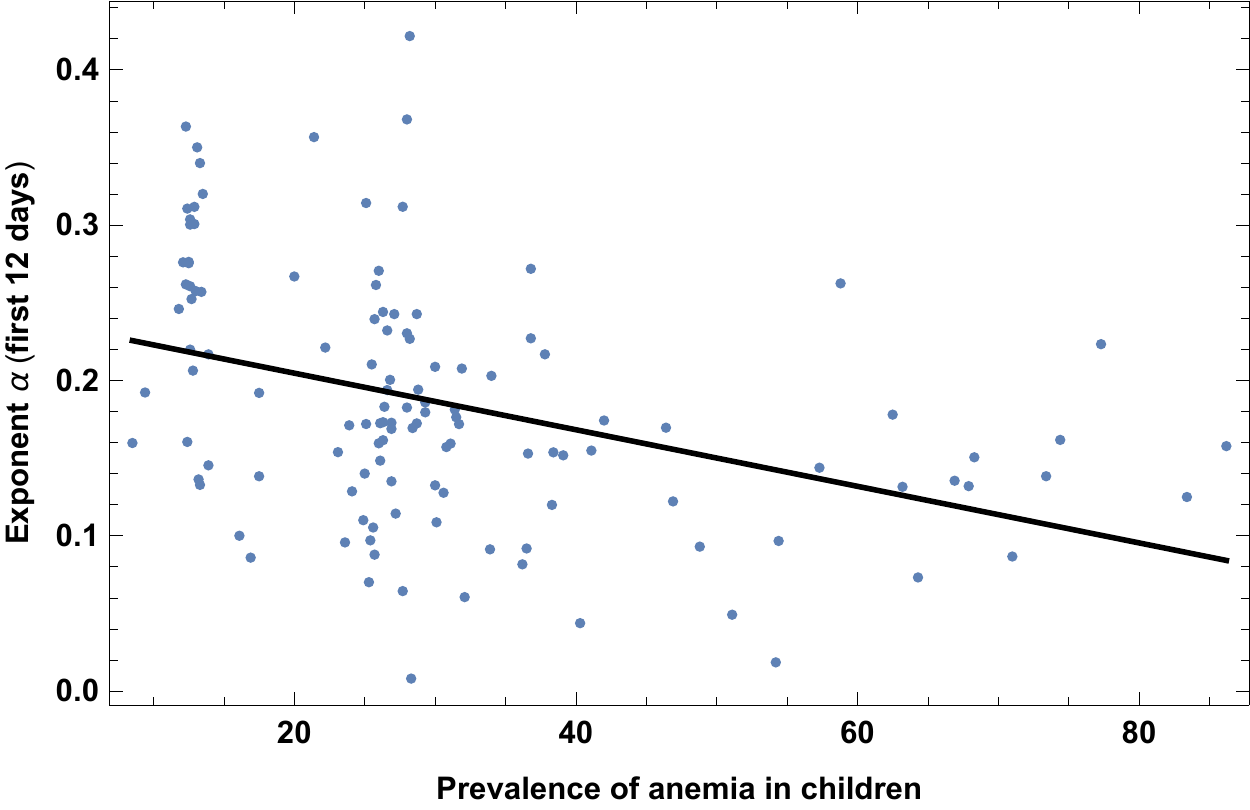} 
\caption{Exponent $\alpha$ for each country vs.~prevalence of anemia in children. We show the data points and the best-fit for the linear interpolation. \label{figANE}}
\end{center}
\end{figure}

Prevalence of anemia in children in 2016, measured as the share of children under the age of five with
hemoglobin levels less than 110 grams per liter at sea level. A similar but less significant correlation is found also with anemia in adults, which we do not report here. Results are shown in Fig.~\ref{figANE} and Table~\ref{tabANE}. The significance is quite high, but could be interpreted as due to a high correlation with life expectancy, as we explain in section~\ref{cross}. A different hypothesis is that this might be related to  genetic factors, which might affect the immune response to COVID-19.

\begin{table}[H]

\begin{tabular}{cc}
    \begin{minipage}{.5\linewidth}
       \begin{tabular}{|l|c|c|c|c|c|}
       \hline
\text{} & \text{Estimate} & \text{Standard Error} & \text{t-Statistic} & \text{$p$-value} & \text{$p$-value}, GDP>5K\$ \\
\hline
1 & 0.24 & 0.0136 & 17.7 & $1.58\times 10^{-35}$ & \\
 \text{ANE} & -0.00181 & 0.000386 & -4.7 & $6.95\times 10^{-6} $ & 0.0014 \\
\hline
     \end{tabular}
     \end{minipage} 
     &
     \hspace*{12em}
    \begin{minipage}{.5\linewidth}
             \begin{tabular}{|l|c|c|c|c|}
  \hline
$R^2$ &  $0.153$\\ \hline
$N$ & 124 \\ 
\hline 

 \end{tabular}
    \end{minipage} 

\end{tabular}

     \vspace*{2em}

\begin{tabular}{cc}
    \begin{minipage}{.5\linewidth}
       \begin{tabular}{|l|c|c|c|c|c|}
       \hline
\text{} & \text{Estimate} & \text{Standard Error} & \text{t-Statistic} & \text{$p$-value} \\
\hline
 1 & 0.22 & 0.0249 & 8.83 & $1.23\times 10^{-14}$ \\
 \text{GDP} & $4.97\times 10^{-7}$ & $4.74\times 10^{-7}$ & 1.05 & 0.297 \\
 \text{ANE} & -0.00148 & 0.000514 & -2.89 & 0.0046 \\
 \hline
     \end{tabular}
     \end{minipage} 
     &
     \hspace*{6em}
    \begin{minipage}{.5\linewidth}
             \begin{tabular}{|l|c|c|c|c|}

\hline
$R^2$ &  $0.161$\\ \hline
$N$ & 120 \\ \hline 
{\text Cross-correlation} & 0.638 \\ \hline 
 \end{tabular}
    \end{minipage} 

\end{tabular}

    \caption{In the left top panel: best-estimate, standard error ($\sigma$), t-statistic and $p$-value for the parameters of the linear interpolation,  for correlation of $\alpha$ with prevalence of anemia in children (ANE).  We also show the $p$-value, excluding countries below 5 thousand \$ GDP per capita. In the left bottom panel: same quantities for correlation of $\alpha$ with ANE and GDP per capita.  In the right panels: $R^2$ for the  best-estimate and number of countries $N$. We also show the correlation coefficient between the 2 variables in the two-variable fit.}
    \label{tabANE}
\end{table}

\subsection{Blood types}

Blood types are not equally distributed in the world and thus we have correlated them with $\alpha$. Data were taken from~\cite{wikiblood}. Very interestingly we find significant correlations, especially for blood types B+ (slower COVID-19 contagion) and A- (faster COVID-19 contagion). In general also all RH-negative blood types correlate  with faster COVID-19 contagion. It is interesting to compare with findings in clinical data: (i) our finding that blood type A is associated with a higher risk for acquiring COVID-19  is in good agreement with~\cite{bloodCOVID}, (ii) we find higher risk for group 0- and no correlation for group 0+ (while~\cite{bloodCOVID} finds lower risk for groups 0),  (iii) we have  a strong significance for lower risk for  RH+ types and in particular lower risk for group B+, which is probably a new finding, to our knowledge. These are also non-trivial findings which should stimulate further medical research on the immune response of different blood-types against COVID-19.

\subsubsection{Type A+}
\begin{figure}[H]
\begin{center}
\vspace*{3mm}
\includegraphics[scale=0.6]{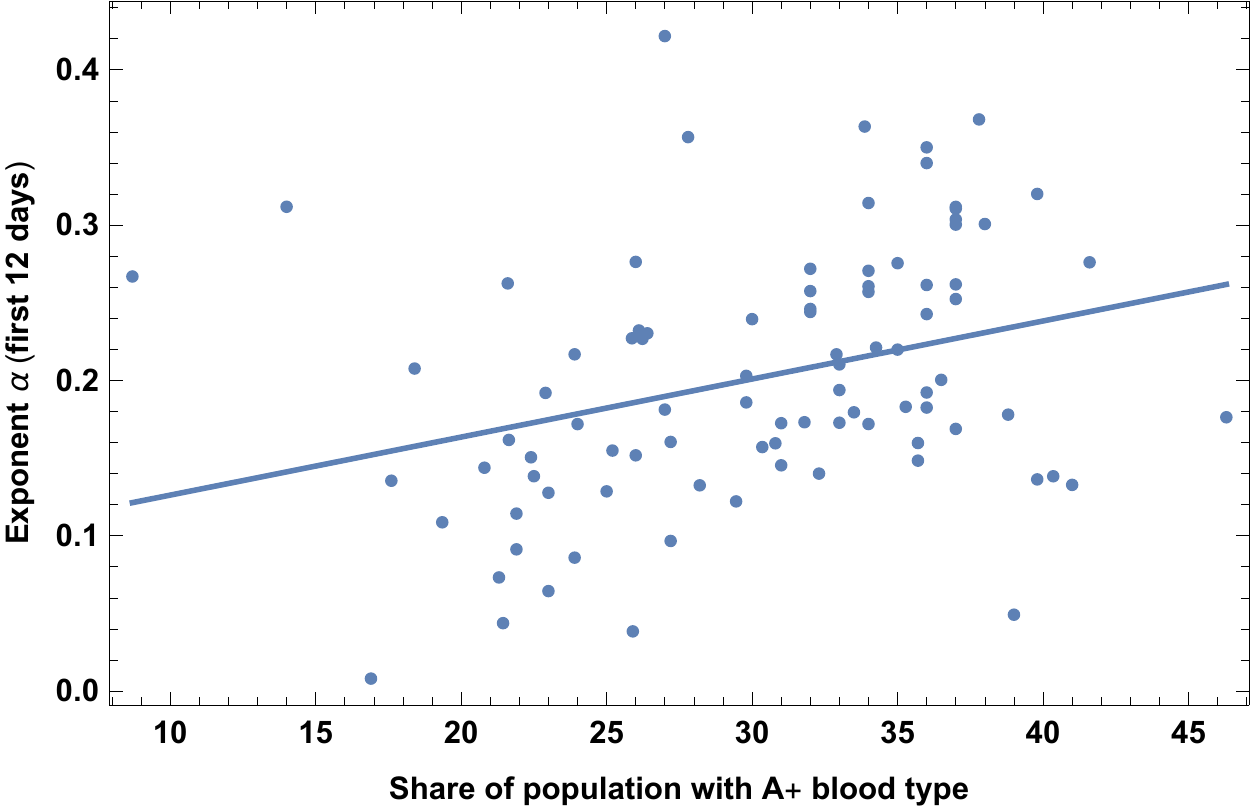} 
\caption{Exponent $\alpha$ for each country vs.~percentage of population with blood type A+. We show the data points and the best-fit for the linear interpolation. \label{figA+}}
\end{center}
\end{figure}

\begin{table}[H]

\begin{tabular}{cc}
    \begin{minipage}{.5\linewidth}
       \begin{tabular}{|l|c|c|c|c|c|}
       \hline
\text{} & \text{Estimate} & \text{Standard Error} & \text{t-Statistic} & \text{$p$-value} & \text{$p$-value}, GDP>5K\$  \\
\hline
 1 & 0.0921 & 0.0372 & 2.48 & 0.0152 & \\
 \text{A}+ & 0.00362 & 0.0012 & 3.03 & 0.0032 & 0.011  \\
   \hline
     \end{tabular}
     \end{minipage} 
     &
     \hspace*{12em}
    \begin{minipage}{.5\linewidth}
             \begin{tabular}{|l|c|c|c|c|}
  \hline
$R^2$ &  $0.093$\\ \hline
$N$ & 91 \\ 
\hline 

 \end{tabular}
    \end{minipage} 

\end{tabular}

     \vspace*{2em}

\begin{tabular}{cc}
    \begin{minipage}{.5\linewidth}
       \begin{tabular}{|l|c|c|c|c|c|}
       \hline
\text{} & \text{Estimate} & \text{Standard Error} & \text{t-Statistic} & \text{$p$-value} \\
\hline
 1 & 0.0935 & 0.0367 & 2.54 & 0.0127 \\
 \text{GDP} & $9.75\times 10^{-7}$ & $4.8\times 10^{-7}$ & 2.03 & 0.0454 \\
 \text{A}+ & 0.0028 & 0.00125 & 2.23 & 0.0282 \\
  \hline
     \end{tabular}
     \end{minipage} 
     &
     \hspace*{6em}
    \begin{minipage}{.5\linewidth}
             \begin{tabular}{|l|c|c|c|c|}

\hline
$R^2$ &  $0.136$\\ \hline
$N$ & 90 \\ \hline 
{\text Cross-correlation} &  -0.334\\ \hline
 \end{tabular}
    \end{minipage} 

\end{tabular}

    \caption{In the left top panel: best-estimate, standard error ($\sigma$), t-statistic and $p$-value for the parameters of the linear interpolation,  for correlation of $\alpha$ with percentage of population with blood type A+. We also show the $p$-value, excluding countries below 5 thousand \$ GDP per capita. In the left bottom panel: same quantities for correlation of $\alpha$ with A+ blood and GDP per capita.   In the right panels: $R^2$ for the  best-estimate and number of countries $N$. We also show the correlation coefficient between the 2 variables in the two-variable fit.}
    \label{tabDIAB}
\end{table}

\subsubsection{Type B+}
\begin{figure}[H]
\begin{center}
\vspace*{3mm}
\includegraphics[scale=0.6]{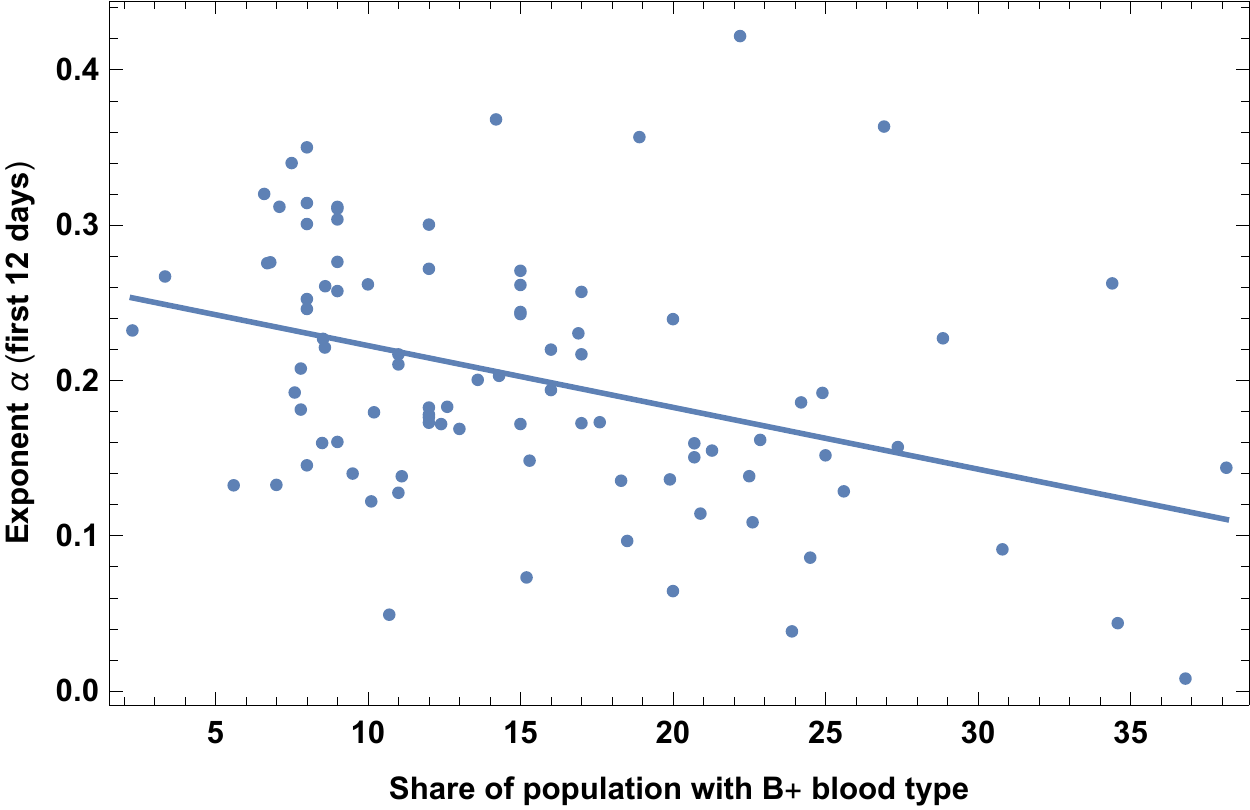} 
\caption{Exponent $\alpha$ for each country vs.~percentage of population with blood type B+. We show the data points and the best-fit for the linear interpolation. \label{figB+}}
\end{center}
\end{figure}

\begin{table}[H]

\begin{tabular}{cc}
    \begin{minipage}{.5\linewidth}
       \begin{tabular}{|l|c|c|c|c|c|}
       \hline
\text{} & \text{Estimate} & \text{Standard Error} & \text{t-Statistic} & \text{$p$-value}  & \text{$p$-value}, GDP>5K\$ \\
\hline
 1 & 0.262 & 0.0177 & 14.8 & $8.76\times 10^{-26}$ & \\
 \text{B}+ & -0.00398 & 0.00104 & -3.82 & 0.000246 & 0.00138 \\

 \hline
     \end{tabular}
     \end{minipage} 
     &
     \hspace*{12em}
    \begin{minipage}{.5\linewidth}
             \begin{tabular}{|l|c|c|c|c|}
  \hline
$R^2$ &  $0.141$\\ \hline
$N$ & 91 \\ 
\hline 

 \end{tabular}
    \end{minipage} 

\end{tabular}

     \vspace*{2em}

\begin{tabular}{cc}
    \begin{minipage}{.5\linewidth}
       \begin{tabular}{|l|c|c|c|c|c|}
       \hline
\text{} & \text{Estimate} & \text{Standard Error} & \text{t-Statistic} & \text{$p$-value} \\
\hline
 1 & 0.231 & 0.0237 & 9.76 & $1.27\times 10^{-15}$ \\
 \text{GDP} & $9.18\times 10^{-7}$ & $4.6\times 10^{-7}$ & 2. & 0.0489 \\
 \text{B}+ & -0.00341 & 0.00107 & -3.17 & 0.00211 \\
 \hline
     \end{tabular}
     \end{minipage} 
     &
     \hspace*{6em}
    \begin{minipage}{.5\linewidth}
             \begin{tabular}{|l|c|c|c|c|}

\hline
$R^2$ &  $0.181$\\ \hline
$N$ & 90 \\ \hline 
{\text Cross-correlation} &  0.284\\ \hline
 \end{tabular}
    \end{minipage} 

\end{tabular}

    \caption{In the left top panel: best-estimate, standard error ($\sigma$), t-statistic and $p$-value for the parameters of the linear interpolation,  for correlation of $\alpha$ with  percentage of population with blood type B+. We also show the $p$-value, excluding countries below 5 thousand \$ GDP per capita. In the left bottom panel: same quantities for correlation of $\alpha$ with B+ and GDP per capita.   In the right panels: $R^2$ for the  best-estimate and number of countries $N$. We also show the correlation coefficient between the 2 variables in the two-variable fit.}
    \label{tabB+}
\end{table}

\subsubsection{Type 0-}
\begin{figure}[H]
\begin{center}
\vspace*{3mm}
\includegraphics[scale=0.6]{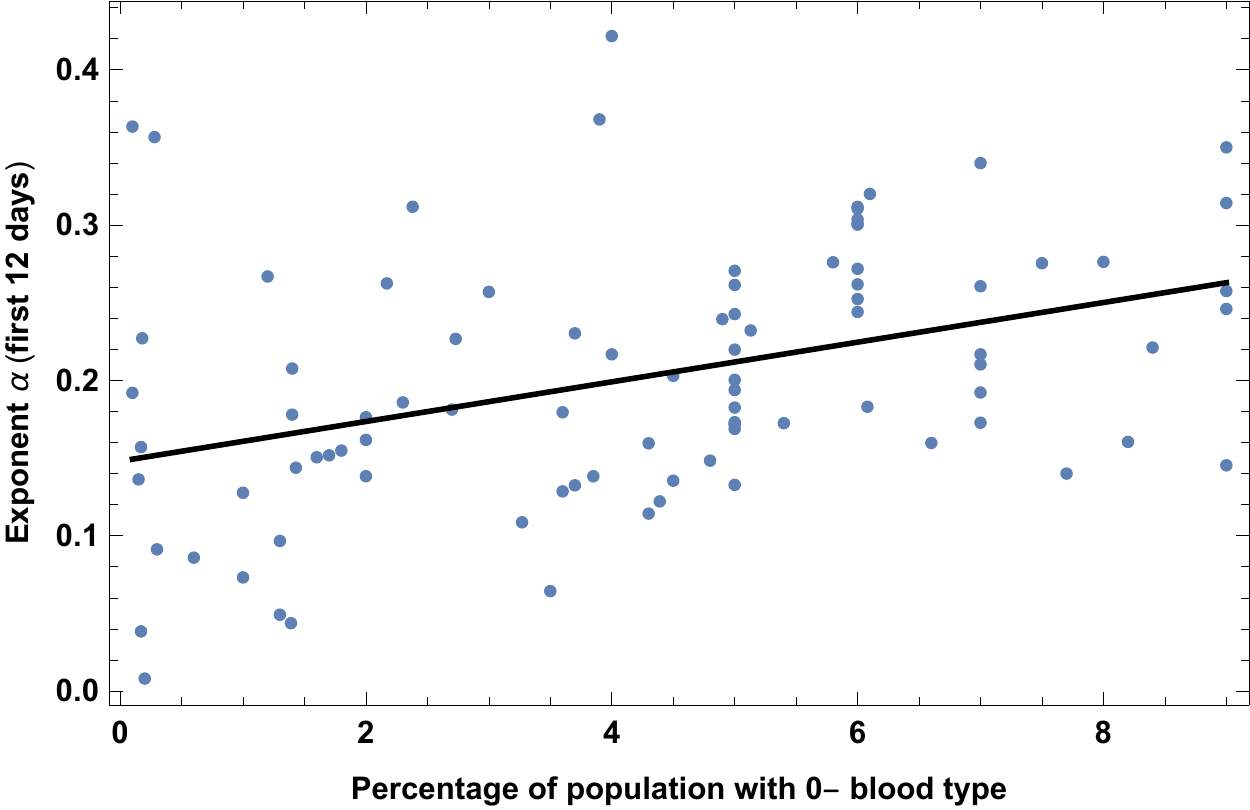} 
\caption{Exponent $\alpha$ for each country vs.~percentage of population with blood type 0-. We show the data points and the best-fit for the linear interpolation. \label{fig0-}}
\end{center}
\end{figure}

\begin{table}[H]

\begin{tabular}{cc}
    \begin{minipage}{.5\linewidth}
       \begin{tabular}{|l|c|c|c|c|c|}
       \hline
\text{} & \text{Estimate} & \text{Standard Error} & \text{t-Statistic} & \text{$p$-value} & \text{$p$-value}, GDP>5K\$  \\
\hline
 1 & 0.148 & 0.0155 & 9.55 & $2.77\times 10^{-15}$ & \\
$0$- & 0.0128 & 0.00315 & 4.08 & 0.0000991 & \\
 \hline
     \end{tabular}
     \end{minipage} 
     &
     \hspace*{12em}
    \begin{minipage}{.5\linewidth}
             \begin{tabular}{|l|c|c|c|c|}
  \hline
$R^2$ &  $0.157$\\ \hline
$N$ & 91 \\ 
\hline 

 \end{tabular}
    \end{minipage} 

\end{tabular}

     \vspace*{2em}

\begin{tabular}{cc}
    \begin{minipage}{.5\linewidth}
       \begin{tabular}{|l|c|c|c|c|c|}
       \hline
\text{} & \text{Estimate} & \text{Standard Error} & \text{t-Statistic} & \text{$p$-value} \\
\hline
 1 & 0.14 & 0.0165 & 8.47 & $5.42\times 10^{-13}$ \\
 \text{GDP} & $6.84\times 10^{-7}$ & $4.9\times 10^{-7} $ & 1.39 & 0.167 \\
 \text{0}- & 0.0107 & 0.00349 & 3.07 & 0.00285 \\
 \hline
     \end{tabular}
     \end{minipage} 
     &
     \hspace*{6em}
    \begin{minipage}{.5\linewidth}
             \begin{tabular}{|l|c|c|c|c|}

\hline
$R^2$ &  $0.175$\\ \hline
$N$ & 90 \\ \hline 
{\text Cross-correlation} &  -0.431 \\ \hline
 \end{tabular}
    \end{minipage} 

\end{tabular}

    \caption{In the left top panel: best-estimate, standard error ($\sigma$), t-statistic and $p$-value for the parameters of the linear interpolation,  for correlation of $\alpha$ with percentage of population with blood type $0-$. We also show the $p$-value, excluding countries below 5 thousand \$ GDP per capita. In the left bottom panel: same quantities for correlation of $\alpha$ with $0-$ and GDP per capita.   In the right panels: $R^2$ for the  best-estimate and number of countries $N$. We also show the correlation coefficient between the 2 variables in the two-variable fit.}
    \label{tab0-}
\end{table}

\subsubsection{Type A-}
\begin{figure}[H]
\begin{center}
\vspace*{3mm}
\includegraphics[scale=0.6]{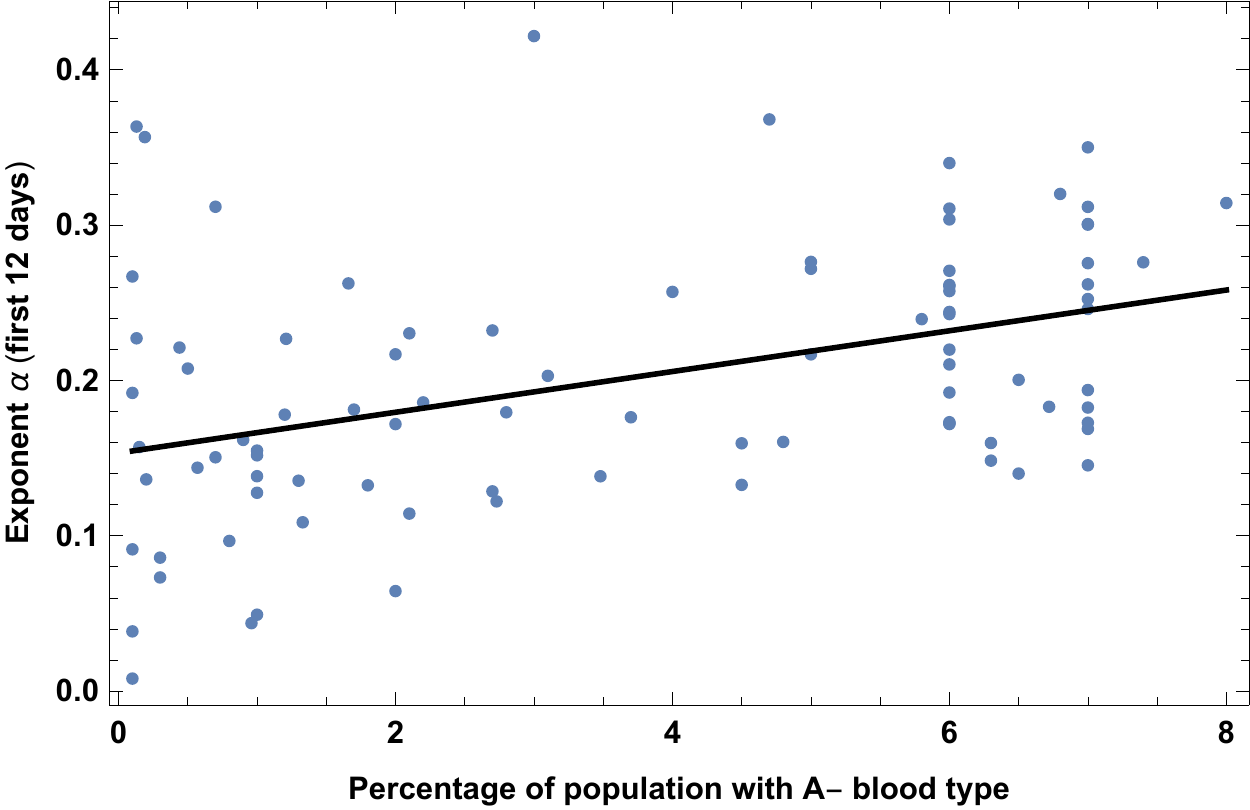} 
\caption{Exponent $\alpha$ for each country vs.~percentage of population with blood type A-. We show the data points and the best-fit for the linear interpolation. \label{figA-}}
\end{center}
\end{figure}

\begin{table}[H]

\begin{tabular}{cc}
    \begin{minipage}{.5\linewidth}
       \begin{tabular}{|l|c|c|c|c|c|}
       \hline
\text{} & \text{Estimate} & \text{Standard Error} & \text{t-Statistic} & \text{$p$-value}  & \text{$p$-value}, GDP>5K\$ \\
\hline
 1 & 0.153 & 0.0135 & 11.3 & $6.71\times 10^{-19}$ & \\
 \text{A}- & 0.0132 & 0.00298 & 4.42 & 0.0000278 & 0.0027\\
  \hline
     \end{tabular}
     \end{minipage} 
     &
     \hspace*{12em}
    \begin{minipage}{.5\linewidth}
             \begin{tabular}{|l|c|c|c|c|}
  \hline
$R^2$ &  $0.18$\\ \hline
$N$ & 91 \\ 
\hline 

 \end{tabular}
    \end{minipage} 

\end{tabular}

     \vspace*{2em}

\begin{tabular}{cc}
    \begin{minipage}{.5\linewidth}
       \begin{tabular}{|l|c|c|c|c|c|}
       \hline
\text{} & \text{Estimate} & \text{Standard Error} & \text{t-Statistic} & \text{$p$-value} \\
\hline
  1 & 0.145 & 0.0151 & 9.62 & $2.46\times 10^{-15}$ \\
 \text{GDP} & $5.99\times 10^{-7}$ & $4.87\times 10^{-7}$ & 1.23 & 0.222 \\
 \text{A}- & 0.0113 & 0.00333 & 3.4 & 0.00101 \\
  \hline
     \end{tabular}
     \end{minipage} 
     &
     \hspace*{6em}
    \begin{minipage}{.5\linewidth}
             \begin{tabular}{|l|c|c|c|c|}

\hline
$R^2$ &  $0.193$\\ \hline
$N$ & 90 \\ \hline 
{\text Cross-correlation} &  -0.442 \\ \hline
 \end{tabular}
    \end{minipage} 

\end{tabular}

    \caption{In the left top panel: best-estimate, standard error ($\sigma$), t-statistic and $p$-value for the parameters of the linear interpolation,  for correlation of $\alpha$ with percentage of population with blood type A-. We also show the $p$-value, excluding countries below 5 thousand \$ GDP per capita. In the left bottom panel: same quantities for correlation of $\alpha$ with A- and GDP per capita.   In the right panels: $R^2$ for the  best-estimate and number of countries $N$. We also show the correlation coefficient between the 2 variables in the two-variable fit.}
    \label{tabA-}
\end{table}

\subsubsection{Type B-}
\begin{figure}[H]
\begin{center}
\vspace*{3mm}
\includegraphics[scale=0.6]{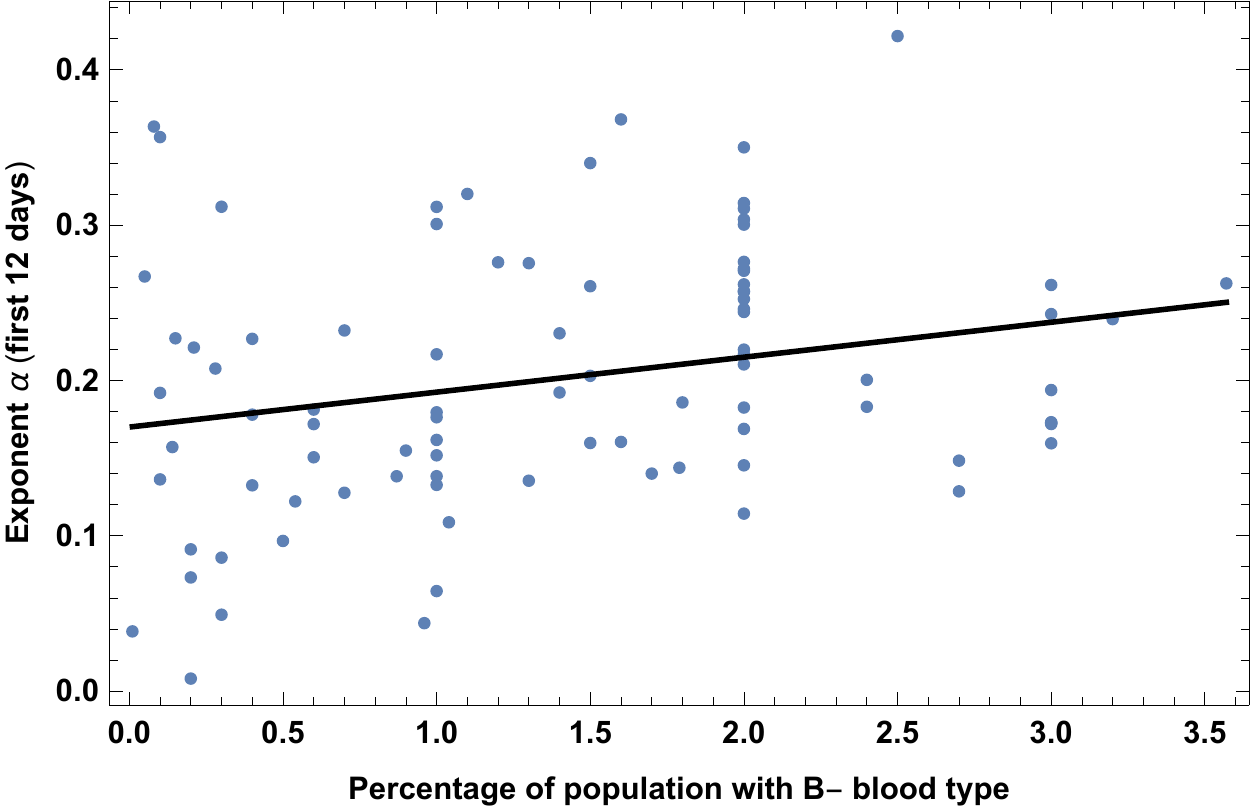} 
\caption{Exponent $\alpha$ for each country vs.~percentage of population with blood type B-. We show the data points and the best-fit for the linear interpolation. \label{figB-}}
\end{center}
\end{figure}

\begin{table}[H]

\begin{tabular}{cc}
    \begin{minipage}{.5\linewidth}
       \begin{tabular}{|l|c|c|c|c|c|}
       \hline
\text{} & \text{Estimate} & \text{Standard Error} & \text{t-Statistic} & \text{$p$-value}  & \text{$p$-value}, GDP>5K\$  \\
\hline
  1 & 0.169 & 0.0153 & 11. & $2.27\times 10^{-18}$ & \\
 \text{B}- & 0.0229 & 0.00905 & 2.54 & 0.013 & 0.146 \\
  \hline
     \end{tabular}
     \end{minipage} 
     &
     \hspace*{12em}
    \begin{minipage}{.5\linewidth}
             \begin{tabular}{|l|c|c|c|c|}
  \hline
$R^2$ &  $0.0674$\\ \hline
$N$ & 91 \\ 
\hline 

 \end{tabular}
    \end{minipage} 

\end{tabular}

     \vspace*{2em}

\begin{tabular}{cc}
    \begin{minipage}{.5\linewidth}
       \begin{tabular}{|l|c|c|c|c|c|}
       \hline
\text{} & \text{Estimate} & \text{Standard Error} & \text{t-Statistic} & \text{$p$-value} \\
\hline
 1 & 0.147 & 0.0176 & 8.36 & $8.98\times 10^{-13}$ \\
 \text{GDP} & $1.16\times 10^{-6}$ & $4.62\times 10^{-7}$ & 2.51 & 0.0139 \\
 \text{B}- & 0.0188 & 0.00899 & 2.09 & 0.0392 \\
  \hline
     \end{tabular}
     \end{minipage} 
     &
     \hspace*{6em}
    \begin{minipage}{.5\linewidth}
             \begin{tabular}{|l|c|c|c|c|}

\hline
$R^2$ &  $0.13$\\ \hline
$N$ & 90 \\ \hline 
{\text Cross-correlation} &-0.179 \\ \hline
 \end{tabular}
    \end{minipage} 

\end{tabular}

    \caption{In the left top panel: best-estimate, standard error ($\sigma$), t-statistic and $p$-value for the parameters of the linear interpolation,  for correlation of $\alpha$ with percentage of population with blood type B-. We also show the $p$-value, excluding countries below 5 thousand \$ GDP per capita. In the left bottom panel: same quantities for correlation of $\alpha$ with B- and GDP per capita.   In the right panels: $R^2$ for the  best-estimate and number of countries $N$. We also show the correlation coefficient between the 2 variables in the two-variable fit.}
    \label{tabB-}
\end{table}

\subsubsection{Type AB-}
\begin{figure}[H]
\begin{center}
\vspace*{3mm}
\includegraphics[scale=0.6]{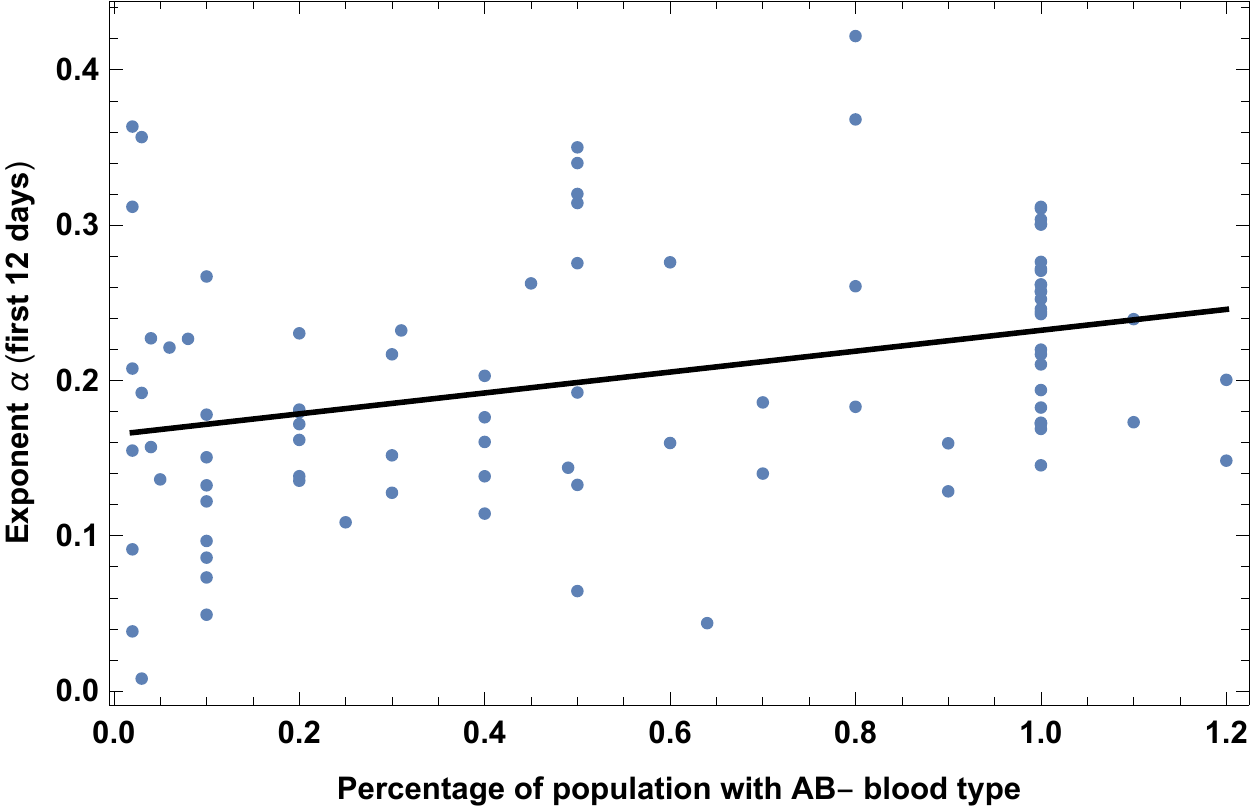} 
\caption{Exponent $\alpha$ for each country vs.~percentage of population with blood type AB-. We show the data points and the best-fit for the linear interpolation. \label{figAB-}}
\end{center}
\end{figure}

\begin{table}[H]

\begin{tabular}{cc}
    \begin{minipage}{.5\linewidth}
       \begin{tabular}{|l|c|c|c|c|c|}
       \hline
\text{} & \text{Estimate} & \text{Standard Error} & \text{t-Statistic} & \text{$p$-value} & \text{$p$-value}, GDP>5K\$ \\
\hline
 1 & 0.163 & 0.0139 & 11.8 & $7.32\times 10^{-20}$ & \\
 \text{AB}- & 0.0711 & 0.0206 & 3.46 & 0.000844 & 0.0278\\
  \hline
     \end{tabular}
     \end{minipage} 
     &
     \hspace*{12em}
    \begin{minipage}{.5\linewidth}
             \begin{tabular}{|l|c|c|c|c|}
  \hline
$R^2$ &  $0.118$\\ \hline
$N$ & 91 \\ 
\hline 

 \end{tabular}
    \end{minipage} 

\end{tabular}

     \vspace*{2em}

\begin{tabular}{cc}
    \begin{minipage}{.5\linewidth}
       \begin{tabular}{|l|c|c|c|c|c|}
       \hline
\text{} & \text{Estimate} & \text{Standard Error} & \text{t-Statistic} & \text{$p$-value} \\
\hline
 1 & 0.15 & 0.0157 & 9.53 & $3.75\times 10^{-15}$ \\
 \text{GDP} & $8.75\times 10^{-7}$ & $4.84\times 10^{-7}$ & 1.81 & 0.074 \\
 \text{AB}- & 0.0563 & 0.0221 & 2.55 & 0.0126 \\
  \hline
     \end{tabular}
     \end{minipage} 
     &
     \hspace*{6em}
    \begin{minipage}{.5\linewidth}
             \begin{tabular}{|l|c|c|c|c|}

\hline
$R^2$ &  $0.15$\\ \hline
$N$ & 90 \\ \hline 
{\text Cross-correlation} &  -0.371\\ \hline
 \end{tabular}
    \end{minipage} 

\end{tabular}

    \caption{In the left top panel: best-estimate, standard error ($\sigma$), t-statistic and $p$-value for the parameters of the linear interpolation,  for correlation of $\alpha$ with percentage of population with blood type AB-. We also show the $p$-value, excluding countries below 5 thousand \$ GDP per capita. In the left bottom panel: same quantities for correlation of $\alpha$ with AB- and GDP per capita.   In the right panels: $R^2$ for the  best-estimate and number of countries $N$. We also show the correlation coefficient between the 2 variables in the two-variable fit.}
    \label{tabAB-}
\end{table}

\subsubsection{RH-positive}
\begin{figure}[H]
\begin{center}
\vspace*{3mm}
\includegraphics[scale=0.6]{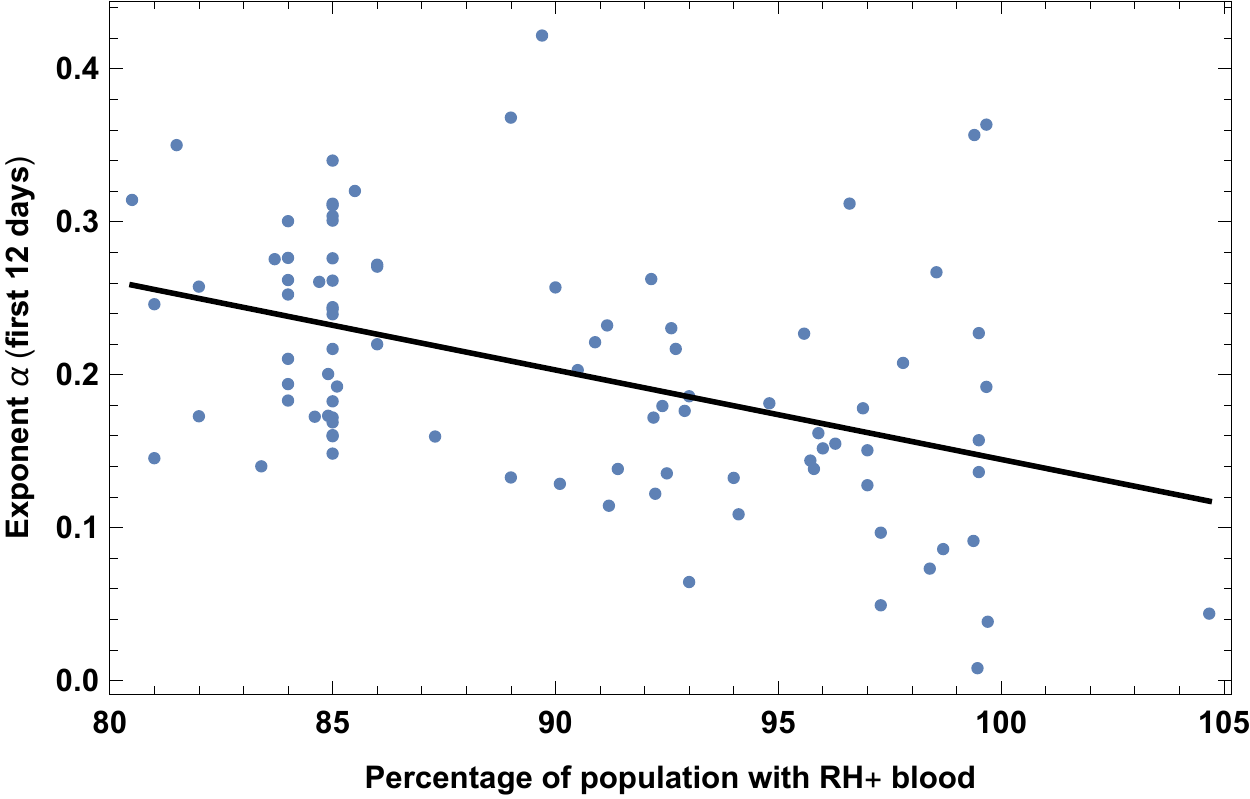} 
\caption{Exponent $\alpha$ for each country vs.~percentage of population with  RH-positive blood. We show the data points and the best-fit for the linear interpolation. \label{figRH+}}
\end{center}
\end{figure}

\begin{table}[H]

\begin{tabular}{cc}
    \begin{minipage}{.5\linewidth}
       \begin{tabular}{|l|c|c|c|c|c|}
       \hline
\text{} & \text{Estimate} & \text{Standard Error} & \text{t-Statistic} & \text{$p$-value} & \text{$p$-value}, GDP>5K\$ \\
\hline
 1 & 0.723 & 0.119 & 6.05 & $3.4\times 10^{-8}$ & \\
 \text{RH}+ & -0.00578 & 0.00132 & -4.37 & 0.0000341 & 0.00355 \\
%

 \hline
     \end{tabular}
     \end{minipage} 
     &
     \hspace*{12em}
    \begin{minipage}{.5\linewidth}
             \begin{tabular}{|l|c|c|c|c|}
  \hline
$R^2$ &  $0.176$\\ \hline
$N$ & 91  \\ \hline 

 \end{tabular}
    \end{minipage} 

\end{tabular}

     \vspace*{2em}

\begin{tabular}{cc}
    \begin{minipage}{.5\linewidth}
       \begin{tabular}{|l|c|c|c|c|c|}
       \hline
\text{} & \text{Estimate} & \text{Standard Error} & \text{t-Statistic} & \text{$p$-value} \\
\hline
 1 & 0.633 & 0.138 & 4.57 & 0.0000158 \\
 \text{GDP} & $6.27\times 10^{-7}$ & $4.85\times 10^{-7}$ & 1.29 & 0.2 \\
 \text{RH}+ & -0.00495 & 0.00147 & -3.36 & 0.00114 \\
   \hline
     \end{tabular}
     \end{minipage} 
     &
     \hspace*{6em}
    \begin{minipage}{.5\linewidth}
             \begin{tabular}{|l|c|c|c|c|}

\hline
$R^2$ &  $0.191$\\ \hline
$N$ & 90 \\ \hline 
{\text Cross-correlation} &  0.432\\ \hline
 \end{tabular}
    \end{minipage} 

\end{tabular}

    \caption{In the left top panel: best-estimate, standard error ($\sigma$), t-statistic and $p$-value for the parameters of the linear interpolation,  for correlation of $\alpha$ with percentage of population with  RH-positive blood (RH+). We also show the $p$-value, excluding countries below 5 thousand \$ GDP per capita. In the left bottom panel: same quantities for correlation of $\alpha$ with RH+ and GDP per capita.   In the right panels: $R^2$ for the  best-estimate and number of countries $N$. We also show the correlation coefficient between the 2 variables in the two-variable fit.}
    \label{tabRH+}
\end{table}

\section{Cross-correlations}\label{cross}



\begin{figure}[H]
\begin{center}
\vspace*{3mm}
\includegraphics[scale=0.32]{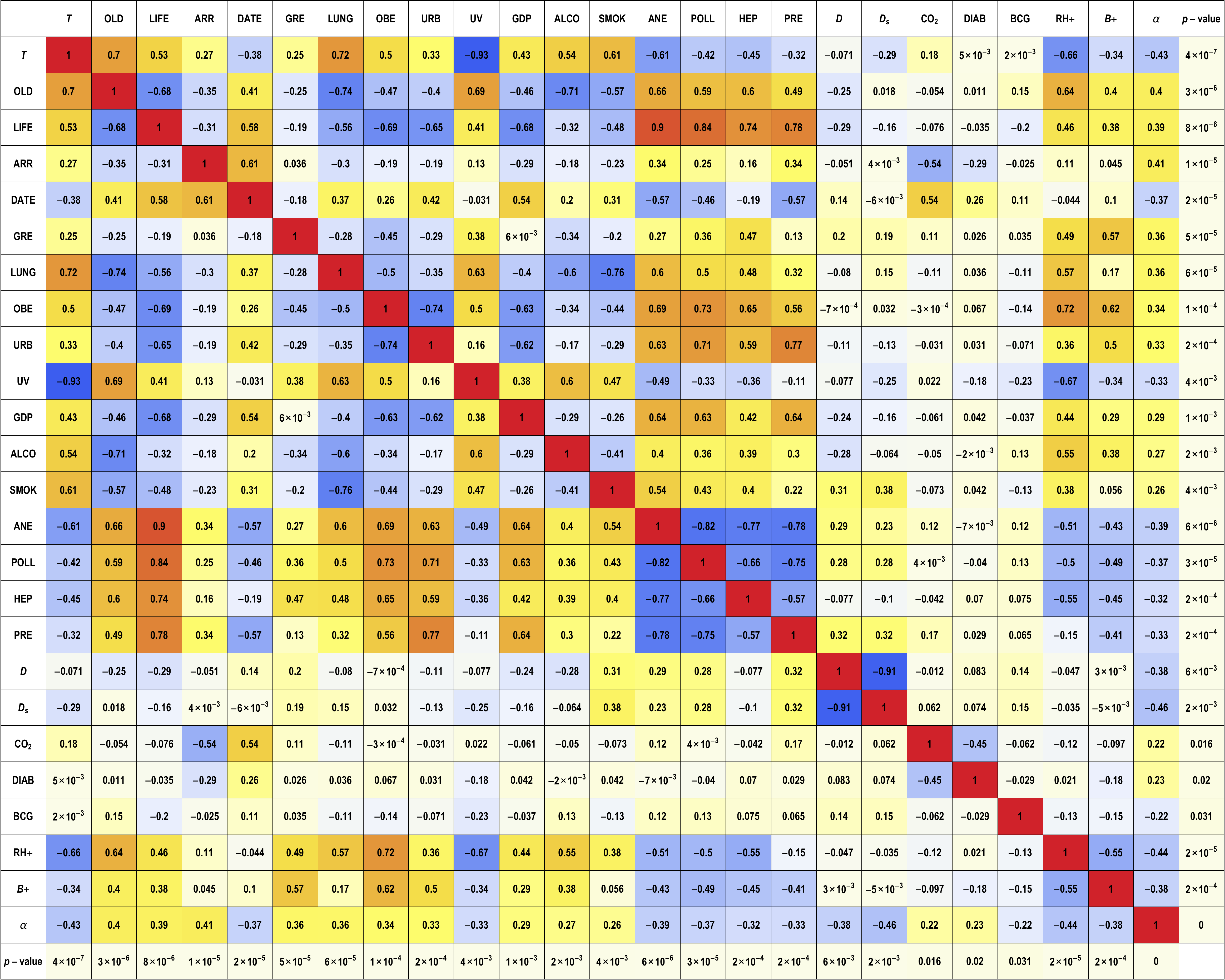} 
\caption{Correlation coefficients between each variable in a pair. Such coefficient corresponds to the off-diagonal entry of the (normalized) covariance matrix, multiplied by $-1$. In the last column and row we show the $p$-value of each variable when performing a one-variable linear fit for the growth rate $\alpha$. Note also that the fits that include vitamin D variables (D and D$_s$) and UV index are based on smaller samples than for the other fits and were collected with rather with inhomogeneous data, and so have to be confirmed on a larger sample, as explained in the text.  The variables considered here are: Temperature (T), Old age dependency ratio (OLD), Life expectancy (LIFE), Number of tourist arrivals (ARR), Starting date of the epidemic (DATE), Amount of contact in greeting habits (GRE), Lung cancer (LUNG), Obesity in males (OBE), Urbanization (URB), UV Index  (UV), GDP per capita (GDP), Alcohol consumption (ALCO), Daily smoking prevalence (SMOK), Prevalence of anemia in children (ANE), Death rate due to pollution (POLL), Prevalence of hepatitis B (HEP), High blood pressure in females (PRE), average vitamin D serum levels (D), seasonal vitamin D serum levels (D$_s$), CO$_2$ emissions (CO$_2$), type 1 diabetes prevalence (DIAB), BCG vaccination (BCG), percentage with blood of RH+ type (RH+), percentage with blood type  B+ (B+).}
 \label{heatmap}
\end{center}
\end{figure}

\begin{figure}[H]
\begin{center}
\vspace*{3mm}
\includegraphics[width=15.5cm, height=13cm]{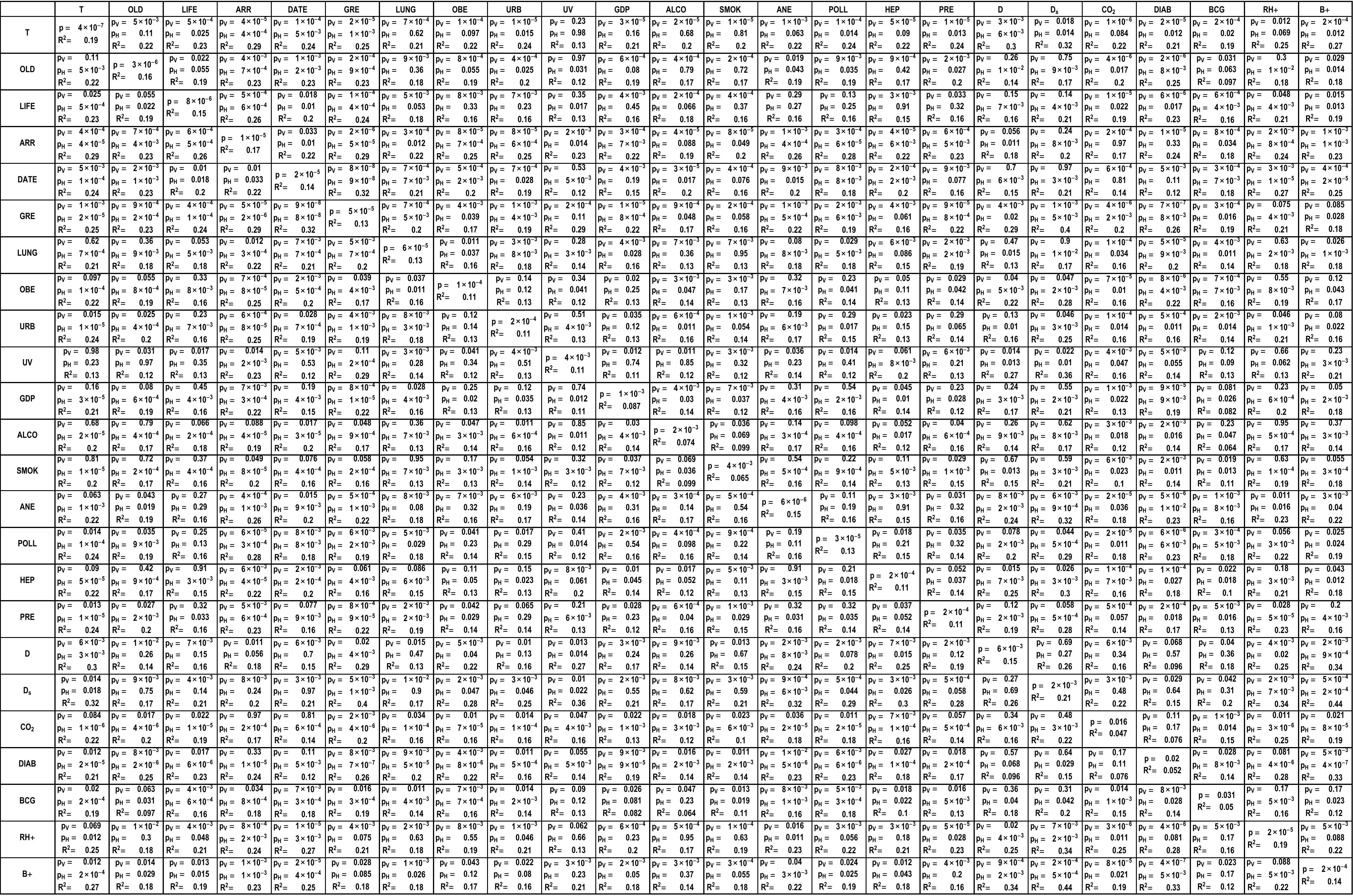} 
\caption{Significance and $R^2$ for linear fits of $\alpha$ as a function of two-variables.  Variables names are the same as in Fig.~\ref{heatmap}. Each cell in the table gives $R^2$ and the $p$-value of the each of the two variables, using $t$-statistic, labeled as p$_H$ for horizontal and p$_V$ for vertical.\label{tabpvalues}}
\end{center}
\end{figure}

In this section we first perform linear fits of $\alpha$ with each possible pair of variables (for blood types, we considered only RH+ and B+). We show the correlation coefficients between the two variables, for each pair, in Fig.~\ref{heatmap}. We also show the $p$-value of the $t$-statistic of each variable in a pair and the total $R^2$ of such fits in Fig.~\ref{tabpvalues}.

We give here below possible interpretations of the redundancy among our variables and we perform  multiple variable fits in the following subsections.

\subsection{Possible Interpretations}

The set of most significant variables, i.e. with smallest $p$-value, which correlate with faster propagation of COVID-19 are the following: low temperature, high percentage of old vs. working people and life expectancy, number of international tourist arrivals, high percentage of RH- blood types, earlier starting date of the epidemic, high physical contact in greeting habits,  prevalence of lung cancer. Such variables are however correlated with each other and we analyze them together below. Most other variables are mildly/strongly correlated with the previous set of variables, except for: prevalence of Type-I diabetes, BCG vaccination, Vitamin D levels (note however that the latter is based on more inhomogeneous data and should be confirmed on a larger sample), which might indeed be considered as almost independent factors.

From the above table one can verify that all the ``counterintuitive'' variables (death-rate due to pollution, POLL, prevalence of anemia, ANE, prevalence of hepatitis B, HEP, high blood pressure in women, PRE) have a strong negative correlation with life expectancy, LIFE. This offers a neat possible interpretation: since countries with high deaths by pollution or high prevalence of anemia or hepatitis B or high blood pressure in women have a younger population, then the virus spread is slower. Therefore one expects that when performing a fit with any of such variable and LIFE together, one of the variables will turn out to be non-significant, i.e. redundant. Indeed one may verify from Table~\ref{tabpvalues} that this happens for all of the four above variables.

Redundancy is also present when one of the following variables is used together with life expectancy in a 2 variables fit: smoking, urbanization, obesity in males. Also, old age dependency and life expectancy are obviously quite highly correlated.

Other variables instead do {\it not}  have such an interpretation:  BCG vaccination, type-1 diabetes in children and vitamin D levels. In this case other interpretations have to be looked for. Regarding the vaccination a promising interpretation is indeed that BCG-vaccinated people could be more protected against COVID-19~\cite{bcg1,bcg2,bcg3}. 

Lung cancer and alcohol consumption remain rather significant, close to a $p$-value of 0.05, even after taking into account life expectancy, but they become non-significant when combining with old age dependency. In this case it is also difficult to disentangle them from the fact that old people are more subject to COVID-19 infection.

Blood type RH+ is also quite correlated with T, however it remains moderately significant when combined with it.

Finally vitamin D, which is measured on a smaller sample, also has little correlations with the main factors. This is quite interesting, since it may open avenues for research on protective factors and health policies. It is quite possible that high Vitamin D helps the immune response to COVID-19~\cite{vitd6,vitdCOVID,vitdCOVID2}. Note however that this finding is based on more inhomogeneous data and should be confirmed on a larger and more homogeneous sample.

\subsection{Multiple variable fits}

It is not too difficult to identify redundant variables, looking at very strongly correlated pairs in Table~\ref{heatmap}. It is generically harder, instead, to extract useful information when combining more than 2 or 3 variables, since we have many variables with comparable predictive power (individual $R^2$ are at most around 0.2) and several of them exhibit mild/strong correlations.
In the following we perform examples of fits with some of the most predictive variables, trying to keep small correlation between them. 

\subsubsection{Temperature+Arrivals+Greetings}

Here we show an example of a fit with 3 parameters.

\begin{table}[H]

\begin{tabular}{cc}
    \begin{minipage}{.5\linewidth}
       \begin{tabular}{|l|c|c|c|c|c|}
       \hline
 \text{} & \text{Estimate} & \text{Standard Error} & \text{t-Statistic} & \text{$p$-value} \\
\hline
 1 & 0.149 & 0.0247 & 6.02 & $2.67\times 10^{-8}$ \\
 T & -0.00247 & 0.000709 & -3.48 & 0.000741 \\
 \text{ARR} & $1.72\times 10^{-9}$ & $4.24\times 10^{-10}$ & 4.05 & 0.0000976 \\
 \text{GRE} & 0.0972 & 0.0276 & 3.52 & 0.000651 \\
  \hline
      \end{tabular}
     \end{minipage} 
&
\hspace*{10em}
    \begin{minipage}{.5\linewidth}
             \begin{tabular}{|l|c|c|c|c|}
 \hline
$R^2$ &  $0.36$\\ \hline
$N$ & 107 \\ \hline 
 \end{tabular}
    \end{minipage} 
    
\end{tabular}

\begin{center}
\vspace*{1em}
$
\begin{pmatrix}
  1. & -0.65 & -0.37 & -0.83 \\
 -0.65 & 1. & 0.28 & 0.24 \\
 -0.37 & 0.28 & 1. & 0.1 \\
 -0.83 & 0.24 & 0.1 & 1. \\
 \end{pmatrix}
$
\end{center}

    \caption{In the left upper panel: best-estimate, standard error ($\sigma$), t-statistic and $p$-value for the parameters of the linear fit. In the right panels: $R^2$ for the  best-estimate and number of countries $N$. Below we show the correlation matrix for all variables.}
    \label{tabtourists}
\end{table}

\subsubsection{Temperature+Arrivals+Greetings+Starting date}

Here we show an example of a fit with 4 parameters. As we combine more than 3 parameters, typically at least one of them becomes less significant.

\begin{table}[H]

\begin{tabular}{cc}
         \hspace*{4em}
         \begin{minipage}{.5\linewidth}
       \begin{tabular}{|l|c|c|c|c|c|}
       \hline
 \text{} & \text{Estimate} & \text{Standard Error} & \text{t-Statistic} & \text{$p$-value} \\
\hline
 1 & 0.299 & 0.0519 & 5.77 & $8.66\times 10^{-8}$ \\
 T & -0.00169 & 0.000718 & -2.36 & 0.0203 \\
 \text{ARR} & $7.69\times 10^{-10}$ & $4.99\times 10^{-10}$ & 1.54 & 0.126 \\
 \text{GRE} & 0.13 & 0.0283 & 4.59 & 0.0000127 \\
 \text{DATE} & -0.00237 & 0.000728 & -3.26 & 0.00154 \\
  \hline
      \end{tabular}
     \end{minipage} 
&
\hspace*{6em}
    \begin{minipage}{.5\linewidth}
             \begin{tabular}{|l|c|c|c|c|}
 \hline
$R^2$ &  $0.42$\\ \hline
$N$ & 107 \\ \hline 
 \end{tabular}
    \end{minipage} 
    
\end{tabular}

\begin{center}
\vspace*{1em}
$
\begin{pmatrix}
 1. & 0.016 & -0.66 & -0.04 & -0.89 \\
 0.016 & 1. & 0.025 & 0.33 & -0.33 \\
 -0.66 & 0.025 & 1. & -0.13 & 0.58 \\
 -0.04 & 0.33 & -0.13 & 1. & -0.35 \\
 -0.89 & -0.33 & 0.58 & -0.35 & 1. \\
 \end{pmatrix}
$
\end{center}

    \caption{In the left upper panel: best-estimate, standard error ($\sigma$), t-statistic and $p$-value for the parameters of the linear interpolation. In the right panels: $R^2$ for the  best-estimate and number of countries $N$. Below we show the correlation matrix for all variables.}
    \label{tabtourists}
\end{table}

\subsubsection{Many variables fit}

One may think of combining {\it all} our variables. This is however not a straightforward task, because we do not have data on the same number $N$ of countries for all variables. 
As a compromise we may restrict to a large number of variables, but still keeping a large number of countries. For instance we may choose the following set: T, OLD, LIFE, ARR, DATE, GRE, LUNG, OBE, URB, GDP, ALCO, SMOK, ANE, POLL, HEP, PRE, CO$_2$. These variables are defined for a sample of $N=103$ countries.
By combining all of them we get $R^2=0.48$, which tells us that only about half of the variance is described by these variables. However, clearly, many of these variables are redundant.
 We perform thus now a Principal Component Analysis (PCA), i.e. we look for linear combinations of such variables that diagonalize the covariance matrix.

%
%
%
%
%
%
%
%
%
%

We perform the PCA analysis in two different ways:
\begin{enumerate}
\item  First, we we exclude the variables with a counterintuitive behavior, see section~\ref{counter}. Moreover we also exclude LIFE, since its meaning is also captured by the similar variable OLD.  We performed thus a Principal Component Analysis, fitting with linear combinations of: 1, T, OLD, ARR, GRE, OBE, URB, SMOK, CO$_2$, DATE, LUNG, GDP, ALCO.
\item Second, we include also the ``counterintuitive'' variables, using thus linear combinations of: 1, T, OLD, ARR, GRE, OBE, URB, SMOK, CO$_2$, DATE, LUNG, GDP, ALCO, ANE, HEP, POLL, PRE; 
\end{enumerate}

In both cases we fit with
\begin{eqnarray}
\alpha= \sum_i \beta_i v_i \label{PCA1} \, ,
\end{eqnarray}
where $v_i$ are linear combinations of the above variables.

In case (1)  we find a fit with total $R^2=0.46$, with $N=103$. There are however only 6 significant independent orthogonal linear combinations. Such 6 combinations, which have $R^2=0.42$, are:

\begin{eqnarray}
v_1 &=&  \text{GRE} \nonumber \, , \\
 v_2 &=& 0.11 \, \text{OBE}+0.11  \,  \text{OLD}+0.18  \,  \text{SMOK}+0.18  \,  T+0.2  \,  \text{LUNG}+0.51  \,  \text{URB}+0.78 \,   \text{DATE} \nonumber  \, , \\
 v_3 &= & \text{ARR} \nonumber \, , \\
v_4 &=& 0.1  \,  \text{ALCO}+0.11  \,  \text{OBE}+0.14  \,  \text{URB}+0.26  \,  \text{SMOK}+0.29  \text{OLD}-0.32  \,  \text{DATE}-0.4  \,  T+0.74   \,  \text{LUNG} \nonumber \, , \\
 v_5 &=& 0.12  \,  \text{OBE}-0.29  \,  \text{LUNG}-0.45  \, \text{DATE}+0.82  \,  \text{URB} 
\nonumber \, , \\
v_6 &=& -0.23  \, \text{DATE}+0.28  \,  \text{OLD}+0.45  \,  \text{SMOK}+0.56  \,  T-0.59  \,  \text{OBE} \, , \label{PCAvars}
\end{eqnarray}
where we omitted variables whose coefficients (``loadings'') on $v_1,v_2,v_3,v_4,v_6$ are less than 0.1. The variables with loadings of at least 0.3 are: GRE, URB, DATE, ARR, T, LUNG, SMOK, OBE. The significance of the principal components is given in Table~\ref{signPCA}.

\begin{table}[H]

\begin{tabular}{cc}
        \hspace*{5em}
         \begin{minipage}{.5\linewidth}
       \begin{tabular}{|l|c|c|c|c|c|}
       \hline
\text{} & \text{Estimate} & \text{Standard Error} & \text{t-Statistic} & \text{$p$-value} \\
\hline  $\beta_1$ & -0.14 & 0.034 & -4.1 & 0.000075 \\
 $\beta_2 $ & -0.00038 & 0.000099 & -3.9 & 0.00021 \\
 $\beta_3$ & $1.5\times 10^{-9}$ & $3.6\times 10^{-10}$ & 4.2 & 0.000057 \\
  $\beta_4$ & 0.0011 & 0.00033 & 3.3 & 0.0012 \\
  $\beta_5$ & 0.0011 & 0.00043 & 2.5 & 0.014 \\
 $\beta_6$  & -0.002 & 0.001 & -2. & 0.051 \\
\hline
     \end{tabular}
     \end{minipage} 
     &
     \hspace*{6em}
    \begin{minipage}{.5\linewidth}
             \begin{tabular}{|l|c|c|c|c|}
  \hline
$R^2$ &  $0.42$ \\ \hline
$N$ & 103 \\ \hline

 \end{tabular}
 
    \end{minipage}

\end{tabular}

    \caption{Best-estimate, standard error, t-statistic and $p$-value for the parameters of the Principal Components, see eqs.~(\ref{PCA1}-\ref{PCAvars}). In the right panel: $R^2$ for the  best-estimate and number of countries $N$.}
    \label{signPCA}
\end{table}

In case (2)  we find a fit with total $R^2=0.48$, with $N=103$. There are again only 6 significant independent orthogonal linear combinations. Such 6 combinations, which have $R^2=0.42$, are:
\begin{eqnarray}
v_1 &=&  -\text{HEP} \nonumber \, , \\
v_2 &=&  -\text{GRE} \nonumber \, , \\
v_3 &=&  -\text{ARR} \nonumber \, , \\
 v_4 &=& -0.49 \, \text{ANE}+0.27 \, \text{LUNG}-0.2 \, \text{OBE}-0.35 \, \text{OLD}+0.17 \, \text{SMOK}+0.69 \, T \nonumber \, , \\
 v_5 &= & \text{CO}_2 \nonumber \, ,\\
v_6 &=& -0.25 \, \text{DATE}-0.24 \, \text{LUNG}-0.15 \, \text{OBE}-0.12 \, \text{OLD}+0.74 \, \text{POLL}-0.15 \, \text{SMOK}-0.52
 \,  \text{URB}  \label{PCAvars2} \nonumber \, , \\ 
\end{eqnarray}
where, again, we omitted variables whose loadings are less than 0.1. The variables with loadings of at least 0.3 are now: HEP, GRE, ARR, ANE, OLD, T, CO$_2$, URB. The significance of the principal components is given in Table~\ref{signPCA2}.

\begin{table}[H]

\begin{tabular}{cc}
        \hspace*{5em}
         \begin{minipage}{.5\linewidth}
       \begin{tabular}{|l|c|c|c|c|c|}
       \hline
\text{} & \text{Estimate} & \text{Standard Error} & \text{t-Statistic} & \text{$p$-value} \\
\hline   
$\beta_1$ & 0.000019 & $4.1\times 10^{-6}$ & 4.6 & 0.000015 \\
$\beta_2$ & -0.16 & 0.038 & -4.3 & 0.000045 \\
$\beta_3$  & $-1.4\times 10^{-9}$ & $4.1\times 10^{-10}$ & -3.4 & 0.001 \\
$\beta_4$  & -0.003 & 0.0011 & -2.8 & 0.0054 \\
$\beta_5$ & $-1.5\times 10^{-11}$ & $5.5\times 10^{-12}$ & -2.7 & 0.0083 \\
$\beta_6$  & -0.00042 & 0.00019 & -2.2 & 0.032 \\
\hline
     \end{tabular}
     \end{minipage} 
     &
     \hspace*{6em}
    \begin{minipage}{.5\linewidth}
             \begin{tabular}{|l|c|c|c|c|}
  \hline
$R^2$ &  $0.42$ \\ \hline
$N$ & 103 \\ \hline

 \end{tabular}
 
    \end{minipage}

\end{tabular}

    \caption{Best-estimate, standard error, t-statistic and $p$-value for the parameters of the Principal Components, see eqs.~(\ref{PCA1}-\ref{PCAvars2}). In the right panel: $R^2$ for the  best-estimate and number of countries $N$.}
    \label{signPCA2}
\end{table}

\section{Conclusions} \label{conclusions}

We have collected data for countries that had at least 12 days of data after a starting point, which we fixed to be at the threshold of 30 confirmed cases. We considered a dataset of 126 countries, collected on April 15th. We have fit the data for each country with an exponential and extracted the exponents $\alpha$, for each country. Then we have correlated such exponents with several variables, one by one.

We found a  positive correlation with {\it high confidence} level with the following variables, with respective $p$-value:  
low temperature (negative correlation, $p$-value $4\cdot10^{-7}$),
high ratio of old people vs. people in the
 working-age (15-64 years)  ($p$-value $3\cdot10^{-6}$), 
life expectancy  ($p$-value $8\cdot10^{-6}$),
international tourism: number of arrivals ($p$-value $1\cdot10^{-5}$),
earlier start of the epidemic ($p$-value $2\cdot 10^{-5}$), high amount of contact in greeting habits (positive correlation, $p$-value $5\cdot 10^{-5}$), lung cancer death rates ($p$-value $6\cdot 10^{-5}$), obesity in males ($p$-value $1\cdot10^{-4}$), share of population in urban areas ( $p$-value $2\cdot10^{-4}$),
 share of population with cancer ($p$-value $2.8\cdot10^{-4}$),
 alcohol consumption ($p$-value $0.0019$),
 daily smoking prevalence ($p$-value $0.0036$),
 low UV index ($p$-value $0.004$; smaller sample, 73 countries),
 low vitamin D serum levels (annual values $p$-value $0.006$, seasonal values $0.002$; smaller sample, $\sim 50$ countries).

%

We also find strong evidence for correlation with blood types:
  RH + blood group system (negative correlation, $p$-value $3\cdot 10^{-5}$);
  A+ (positive correlation, $p$-value $3\cdot 10^{-3}$);
 B + (negative correlation, $p$-value $2\cdot 10^{-4}$);
 A-  (positive correlation, $p$-value $3\cdot 10^{-5}$);
 0-  (positive correlation, $p$-value $8\cdot 10^{-4}$);
 AB-  (positive correlation, $p$-value $0.028$);
We find moderate evidence for correlation with:
 B-  (positive correlation, $p$-value $0.013$).

We find {\it moderate} evidence for positive correlation with:
 CO$_2$ (and SO) emissions  ($p$-value $0.015$), type-1 diabetes in children ($p$-value $0.023$), vaccination coverage for Tuberculosis (BCG)  ($p$-value $0.028$).

Counterintuitively we also find negative correlations, in a direction opposite to a naive expectation, with:
death rate from air pollution ($p$-value $3\cdot10^{-5}$), prevalence of anemia, adults and children, ($p$-value $1\cdot10^{-4}$ and $7 \cdot 10^{-6}$, respectively), share of women with high-blood-pressure ($p$-value $2\cdot10^{-4}$), incidence of Hepatitis B ($p$-value $2\cdot10^{-4}$), PM2.5 air pollution ($p$-value $0.029$).

As is clear from the figures, the data present a high amount of dispersion, for all fits that we have performed.  This is of course unavoidable, given the existence of many systematic effects.   One obvious factor is that the data are collected at {\em country} level, whereas many of the factors considered are regional.   This is obvious from empirical data (see for instance the difference between the epidemic development in Lombardy vs. other regions in Italy, or New York vs.~more rural regions), and also sometimes has obvious explanations (climate, health factors vary a lot region by region) as well as not so obvious ones.   Because of this, we consider $R^2$ values as at least as important as $p$-values and correlation coefficients: 
an increase of the $R^2$ after a parameter is included means that the parameter has a systematic effect in reducing the dispersion (``more data points are explained''). 

Several of the above variables are correlated with each other and so they are likely to have a common interpretation and it is not easy to disentangle them.
The correlation structure is quite rich and non-trivial, and we encourage interested readers to study the tables in detail, giving both $R^2$, $p$-values and correlation estimates.
Note that some correlations are ``obvious'', for example between temperature and UV radiation.  Others are accidental, historical and sociological.  For instance, social habits like alcohol consumption and smoking are correlated with climatic variables.  In a similar vein correlation of smoking and lung cancer is very high, and this is likely to contribute to the correlation of the latter with climate.  Historical reasons also correlate climate with GDP per capita. 

Other variables are found to have a counterintuitive {\it negative} correlation, which can be explained due their strong negative correlation with life expectancy:  death-rate due to pollution, prevalence of anemia, Hepatitis B and high blood pressure for women.

We also analyzed the possible existence of a bias: countries with low GDP-per capita, typically located in warm regions, might have less intense testing and we discussed the correlation with the above variables, showing that most of them remain significant, even after taking GDP into account. %
In this respect, note that in countries where testing is not prevalent, registration of the illness is dependent on the development of severe symptoms.
Hence, while this study is about {\em infection rates} rather than {\em mortality}, in quite a few countries we are actually measuring a proxy of mortality rather than infection rate.   Hence, effects affecting mortality will be more relevant.  Pre-existing lung conditions, diabetes, smoking and health indicators in general as well as pollution are likely to be important in this respect,   perhaps  not affecting $\alpha$ per se but the detected amount of $\alpha$. These are in turn generally correlated with GDP and temperature for historical reasons.
 Other interpretations, which may be complementary, are that co-morbidities and old age  affect immune response and thus  may directly increase the growth rate of the contagion. Similarly it is likely that individuals with co-morbidities and old age, developing a more severe form of the disease, are also more contagious than younger or asymptomatic individuals, producing thus an increase in $\alpha$.  In this regard, we wish to point the reader's attention to the relevant differences in correlations once we apply a threshold on GDP per capita.  It has long been known that human wellness (we refer to a psychological happiness study~\cite{happy}, but the point is more general) depends non-linearly on material resources, being strongly correlated when resources are low and reaching a plateau after a critical limit.   The biases described above (weather comorbidity, testing facilities, pre-existing conditions and environmental factors) seem to reflect this, changing considerably in the case our sample has a threshold w.r.t. a more general analysis without a threshold.

About pollution our findings are mixed. We find no correlation with generic air pollution (``Suspended particulate matter (SPM), in micrograms per cubic metre''). We find higher contagion to be moderately correlated only with and CO$_2$/SO emissions.
Instead we find a {\it negative} correlation with death rates due to air pollution and PM2.5 concentration (in contrast with~\cite{poll}, see also~\cite{pansini}). Note however that correlation with PM2.5 becomes non significant when combined with GDP per capita, while CO$_2$/SO becomes non significant when combined with tourist arrivals. Finally death rates due to air pollution is also redundant when correlating with life expectancy.

We also performed a Principal Component Analysis, in a sample of N=103 countries, where: (1) we omitted all variables with a counterintuitive behavior, (2) we omitted Vitamin D, BCG, UV and Blood variables, as we did not have data for so many countries, (3) we also omitted LIFE, since its meaning is already described by the similar variable OLD. 
Therefore we used the following variables: 1, T, OLD, ARR, GRE, OBE, URB, SMOK, CO2, DATE, LUNG, GDP, ALCO. 
As a result we found that only 6 linear orthogonal combinations are significant, and the variables with a loading of at least 0.3 on any of such 6 combinations are:  GRE, URB, DATE, ARR, T, LUNG, SMOK, OBE.

Including also the counterintuitive variables we find again only 6 significant linear orthogonal combinations, but the variables with a loading of at least 0.3 are now:  HEP, GRE, ARR, ANE, OLD, T, CO$_2$, URB.

Some of the variables that we have studied cannot be arbitrarily changed, but can be taken into account  by public health policies, such as temperature, amount of old people and life expectancy, by implementing stronger testing and tracking policies, and possibly lockdowns, both with the arrival of the cold seasons and for the old aged population. 

Other variables instead can be controlled by governments: testing and isolating international travelers and reducing number of flights in more affected regions; promoting social distancing habits as long as the virus is spreading, such as campaigns for reducing physical contact in greeting habits; campaigns against vitamin D deficiency, decrease smoking and obesity.

We also emphasize that some variables are useful to inspire and support medical research, such as correlation of contagion with: lung cancer, obesity, low vitamin D levels, blood types (higher risk for all RH- types, A types, lower risk for B+ type), type 1 diabetes. This definitely deserves further study, also of correlational type using data from patients.

In conclusion, our findings could thus be very useful both for policy makers and for further experimental research.

\begin{acknowledgments}
  GT acknowledges support from FAPESP proc. 2017/06508-7,
partecipation in FAPESP tematico 2017/05685-2 and CNPQ bolsa de
 produtividade 301432/2017-1.
We would like to acknowledge  Alberto Belloni,  Jordi Miralda and Miguel Quartin  for useful discussions and comments.
\end{acknowledgments}

\appendix

\section{Vitamin D} \label{AppD}
We collected most data on vitamin D from~\cite{vitdtable1,vitdtable2,vitdtable3,vitdtable4,vitdtable5} and from references therein. For a first dataset of 50 countries we have collected annual averages. For many countries several studies with different values were found and in this case we have collected the mean and the standard error (when available) and a weighted average has been performed. The resulting values that we have used are listed in Table~\ref{tabvitD}. The sample size here is smaller and such dataset is based on quite inhomogeneous research and thus should be confirmed by a more complete dataset.

\begin{table}[H]

\begin{tabular}{cc}
    \begin{minipage}{.5\linewidth}
       \begin{tabular}{|l|c|c|c|c|c|}
       \hline
 \text{Country} & \text{Vit D (annual) } & \text{Vit D (seasonal)} \\
\hline
  \text{Argentina} & 53 & 55.8 \\
 \text{Australia} & 66 & 62.1 \\
 \text{Austria} & 13.5 & \text{N/A} \\
 \text{Belgium} & 51.7 & 49.3 \\
 \text{Brazil} & 67.6 & 65. \\
 \text{Canada} & 67.7 & 64. \\
 \text{Chile} & 42.3 & 41.2 \\
 \text{China} & 45 & 31.7 \\
 \text{Croatia} & 46.6 & 40.1 \\
 \text{Czech Republic} & 62.4 & \text{N/A} \\
 \text{Denmark} & 60.6 & 54.4 \\
 \text{Estonia} & 49.8 & 44.7 \\
 \text{Finland} & 58.2 & 46.6 \\
 \text{France} & 58 & 50.9 \\
 \text{Germany} & 51.4 & 49.9 \\
 \text{Greece} & 67.9 & 62.2 \\
 \text{Hungary} & 61.6 & 51. \\
 \text{Iceland} & 57 & \text{N/A} \\
 \text{India} & 42 & 42. \\
 \text{Iran} & 36.6 & 40. \\
 \text{Ireland} & 56.4 & \text{N/A} \\
 \text{Israel} & 56.5 & 56.5 \\
 \text{Italy} & 46.4 & 37.2 \\
 \text{Japan} & 67.3 & 59.9 \\
 \text{Lebanon} & 28.5 & \text{N/A} \\

  \hline
      \end{tabular}
     \end{minipage} 
&
\hspace*{1.6em}

    \begin{minipage}{.5\linewidth}
       \begin{tabular}{|l|c|c|c|c|c|}
       \hline
 \text{Country} & \text{Vit D (annual) } & \text{Vit D (seasonal)} \\
\hline
  \text{Lithuania} & 53.3 & 48.5 \\
 \text{Mexico} & 58.5 & 59. \\
 \text{Morocco} & 39.5 & \text{N/A} \\
 \text{Netherlands} & 56 & 49.8 \\
 \text{New Zealand} & 58.1 & 66.2 \\
 \text{Norway} & 64.27 & 56.5 \\
 \text{Poland} & 53.5 & 41.9 \\
 \text{Romania} & 40.2 & 34.4 \\
 \text{Russia} & 29.2 & \text{N/A} \\
 \text{Arabia Saudi} & 35.7 & 28.5 \\
 \text{Singapore} & 56 & 56. \\
 \text{Slovakia} & 81.5 & \text{N/A} \\
 \text{South Africa} & 47 & 59.1 \\
 \text{South Korea} & 49 & 38.5 \\
 \text{Spain} & 51.9 & 43.2 \\
 \text{Sweden} & 73 & 69. \\
 \text{Switzerland} & 50 & 41. \\
 \text{Taiwan} & 74 & 71.2 \\
 \text{Thailand} & 64.7 & 64.7 \\
 \text{Tunisia} & 38.8 & \text{N/A} \\
 \text{Turkey} & 43 & \text{N/A} \\
 \text{Ukraine} & 35.8 & 32.5 \\
 \text{UK} & 50.1 & 37. \\
 \text{USA} & 83.4 & 80.5 \\
 \text{Vietnam} & 79.6 & 79.6 \\
  \hline
      \end{tabular}
     \end{minipage}

%
%
\end{tabular}

    \caption{Vitamin D serum levels (in nmol/l) obtained with a weighted average from refs.~\cite{vitdtable1,vitdtable2,vitdtable3,vitdtable4,vitdtable5} and references therein. The ``annual'' level refers to an average over the year. The ``seasonal'' level refers to the value present in the literature, which is closer to the months of January-March: either the amount during such months or during winter for northern hemisphere,  {\it or}  during summer for southern hemisphere  {\it or}  the annual level for countries with little seasonal variation.}
    \label{tabvitD}
\end{table}

\renewcommand{\bibsection}{\section*{References}}

\bibliographystyle{unsrt}

\begin{thebibliography}{}
\bibitem{ref2} In practice we choose, as the first day, the one in which the number of cases $N_i$ is closest to 30. In some countries, such a number $N_i$ is repeated for several days; in such cases we choose the last of such days as the starting point. For the particular case of China, we started from January 16th, with 59 cases, since the number before that day was essentially frozen.
\end{thebibliography}
%

\end{document}